\documentclass[aps,prd,twocolumn,nofootinbib,superscriptaddress]{revtex4-2}
\usepackage{mathrsfs}
\usepackage{amsmath}
\usepackage{amsfonts}
\usepackage{amssymb}
\usepackage{array}
\usepackage{verbatim}
\usepackage{epsfig}
\usepackage{graphicx} 
\usepackage[usenames, dvipsnames]{color}
\usepackage{marginnote}
\usepackage{slashed}
\usepackage{bm}
\usepackage{todonotes}
\usepackage{tikz}
\usepackage{tikz-feynman}
\usepackage{dsfont}
\usepackage{ulem}

\newcommand{\dhd}{{\textstyle d}
\lower.03ex\hbox{\kern-0.38em$^{\scriptstyle-}$}\kern-0.05em{}}

\begin{document}

\title{DIS dijet production in Background Field Approach: General formalism and methods}
\author{Tiyasa~Kar}
\email{tkar@ncsu.edu}
\affiliation{Department of Physics and Astronomy, North Carolina State University, Raleigh, NC 27695, USA}

\author{Andrey~Tarasov}
\email{ataraso@ncsu.edu}
\affiliation{Department of Physics and Astronomy, North Carolina State University, Raleigh, NC 27695, USA}
\affiliation{Center for Frontiers in Nuclear Science (CFNS) at Stony Brook University, Stony Brook, NY 11794, USA}

\author{Vladimir~V.~Skokov}
\email{vskokov@ncsu.edu}
\affiliation{Department of Physics and Astronomy, North Carolina State University, Raleigh, NC 27695, USA}




\begin{abstract}
We develop a general formalism for computing physical observables within the background field approach, based on representing propagators of the Feynman diagrams in the background fields as path-ordered exponents. This representation allows systematic expansion of the background fields onto arbitrary linear piecewise contours in coordinate space, yielding gauge-covariant QCD operators to any required order of the expansion. We apply this formalism to DIS dijet production and derive a general form of the cross section in terms of (anti)quark propagators in the background fields, valid in arbitrary kinematics. To demonstrate the versatility of our approach, we consider two kinematic limits. In the back-to-back limit, the expansion contour reduces to that of TMD operators. In this limit we recover the known leading-power results. In the small-$x$ regime, defined by the high-energy power counting for boosted background fields, the expansion contour assumes a staple-like shape. We find that, at the leading eikonal order, the transverse component of the background field $B_i$, though parametrically suppressed relative to the light-cone component, contributes non-trivially through the field-strength tensor $F_{-i}$ and the transverse gauge links. Setting $B_i = 0$ recovers the standard CGC result. We also demonstrate matching between the eikonal and back-to-back expansions, providing a quantitative dictionary between these two distinct kinematic regimes.
\end{abstract}


\maketitle


\section{Introduction}
One of the main goals of modern nuclear science is understanding how the characteristic properties of hadrons arise from the interactions of their most elementary constituents --- quarks and gluons. In essence, this problem was introduced with the discovery of Quantum Chromodynamics (QCD) more than 50 years ago, yet a complete and satisfactory solution remains elusive. The fundamental complication is the phenomenon of confinement: because of the small value of a typical hadron scale $\sim \Lambda_{\rm QCD}$ and the running of the strong coupling constant, hadron structure is governed by the dynamics of a dense QCD medium. Analyzing the properties of this medium, which ultimately define the characteristic qualities of hadrons, remains a formidable task for both theory and experiment.

The dense QCD medium can be probed in high-energy scattering experiments. At high energies, when a large external scale is present, the QCD medium admits a factorization description in which different partonic interaction modes, characterized by widely separated scales, can be identified~\cite{Collins:2011zzd}. At the simplest level, one can distinguish a perturbative mode associated with a large energy scale from a non-perturbative mode defined by a small hadron scale. In more complicated scattering scenarios, however, a richer structure of factorized modes can emerge, e.g.\ Transverse Momentum Dependent (TMD) factorization~\cite{Collins:1981uk,Collins:1984kg,Collins:1987pm,Collins:1989gx,Meng:1995yn,Ji:2004wu,Ji:2004xq,Boussarie:2023izj}.

The perturbative mode can be systematically studied using perturbative methods. Crucially, different partonic modes of the hadron are not independent but interact with each other. By computing the perturbative mode in the background of the non-perturbative fields of the hadron, one can extract information about the non-perturbative structure of the dense QCD medium through comparison of theoretical results with experimental data.

In this approach, the dense QCD medium is characterized by matrix elements of QCD operators, which can be parametrized in terms of various parton distribution functions (PDFs)~\cite{Collins:2011zzd,Diehl:2015uka}. These distribution functions describe properties of the dense QCD medium and are crucial for understanding the origin of the hadronic properties. The form of the QCD operators is determined by the structure of interactions between partonic modes, which is specific to a given scattering reaction and its kinematic regime. Consequently, precise determination of the operators is an essential part of the analysis within the QCD factorization approach.

At the same time, determining the operators that define the scattering is not usually straightforward. While this can be done in some simple cases --- e.g.\ deeply inelastic scattering (DIS) in the Bjorken limit of large $Q^2\to \infty$~\cite{Politzer:1974fr,Anselmino:1994gn,Jaffe:1996zw,Blumlein:2012bf} with leading-twist collinear operators~\cite{Brandt:1970kg,Anikin:1978tj,Aoyama:1981ev,Ellis:1982wd,Shuryak:1981kj,Ellis:1982cd,Jaffe:1983hp,Karchev:1983ai,Geyer:1985vw,Braunschweig:1985nr}, or DIS in the Regge limit of small $x_B\to 0$ with dipole operators~\cite{Balitsky:1995ub,Jalilian-Marian:1997qno,Jalilian-Marian:1997ubg,Balitsky:1997mk,Balitsky:1998ya,Kovchegov:1999yj,Kovchegov:2012mbw} --- once higher-order corrections are to be calculated, or non-trivial effects such as the interplay between large and small $x$ are to be quantitatively described, the problem becomes considerably more involved \cite{Kutak:2004ym,Motyka:2009gi,Mueller:2013wwa,Mueller:2012uf,Balitsky:2015qba,Iancu:2015vea,Iancu:2015joa,Balitsky:2016dgz,Boussarie:2020fpb,Caucal:2021ent,Caucal:2022ulg,Caucal:2023fsf,Altinoluk:2023hfz,Duan:2024qck,Duan:2024qev}. Examples include the computation of power corrections in the TMD framework \cite{Balitsky:2017flc,Balitsky:2017gis,Ebert:2021jhy,Vladimirov:2021hdn,Rodini:2023plb,Piloneta:2025jjb,Jaarsma:2025ksf,Balitsky:2026nux}, sub-eikonal corrections for describing scattering in the region of moderate $x$~\cite{Altinoluk:2014oxa,Balitsky2015,Balitsky:2016dgz,Altinoluk:2015gia,Agostini:2024xqs,Chirilli:2018kkw,Chirilli:2021lif}, spin effects at small $x$~\cite{Kovchegov:2015pbl,Kovchegov:2018znm,Kovchegov:2021lvz,Cougoulic:2022gbk,Borden:2024bxa}, and the role of the anomaly in high-energy scattering~\cite{Tarasov:2020cwl,Tarasov:2021yll,Bhattacharya:2022xxw,Bhattacharya:2023wvy,Bhattacharya:2024geo,Tarasov:2025mvn}.





Currently, the most efficient approach for deriving the operators that define the scattering is the background field method~\cite{Abbott:1980hw,Abbott:1981ke,Balitsky:2025bup}. In this approach, the QCD fields corresponding to different partonic modes are separated at the level of the QCD Lagrangian. As a result, the perturbative mode --- defined by ``perturbative'' partons propagating in the non-perturbative background --- can be conveniently described in terms of propagators in the background fields~\cite{Shuryak:1981kj,Balitsky:1987bk,Balitsky:1995ub}. These propagators possess a gauge-covariant structure, which allows derivation of gauge-covariant QCD operators defining the scattering, an essential requirement for the problem.

This procedure involves expanding the propagators in a parameter defined by the scattering kinematics, e.g.\ $q_\perp/Q \ll 1$ in TMD factorization, where $q_\perp$ is a small transverse momentum of the final state and $Q^2$ is the hard scale of the problem. Remarkably, this expansion can be formulated as an expansion of the background fields of the propagators onto a contour of a certain shape in coordinate space, such that the resulting QCD operators are ordered along the contour --- e.g.\ a straight-line contour along the light-cone direction in collinear factorization~\cite{Balitsky:1988fi}, or a dipole contour in small-$x$ scattering~\cite{Balitsky1996}. Hence, computing the operators in the background field approach requires developing methods for constructing such expansions.

At present, little is known about the general structure of these expansions. Typically, a brute-force approach is applied in which the expansion for a particular scattering problem is constructed to only the first few orders~\cite{Balitsky2015,Altinoluk:2021lvu,Chirilli:2018kkw}. This approach requires considerable effort, since the resulting operator structure is not obvious from the form of the propagators.

In this paper, we develop a general framework for deriving such expansions, applicable to any given order of the expansion onto an arbitrary linear piecewise contour. The key element of our formalism is the representation of the ``quantum" propagator in the background field in terms of a path-ordered exponent, which describes propagation of the quantum parton over all possible trajectories in the background field, providing a completely general formulation of the scattering problem.

The background fields of the path-ordered exponents can be expanded onto a given contour. We derive a general form of the expansion onto the light-cone direction; the result is constructed from gauge-covariant operators, which allows unambiguous determination of the operators to any given order. We subsequently derive an equation for a parallel shift of the path-ordered exponent in the transverse direction; this gives rise to transverse gauge links with various insertions of the transverse field-strength tensors. Using these formal operator transformations, the path-ordered exponents representing quantum propagators in the background field can be expanded onto an arbitrary piecewise contour constructed from light-cone segments connected by transverse gauge links.

Importantly, we formulate these transformations in a gauge-covariant form, which makes the determination of the QCD operators direct and straightforward. This is one of the main advantages of our approach and can be carried out to any required order of the expansion in a given kinematic limit.

To illustrate the approach, we apply it to DIS dijet production~\cite{Altinoluk:2015dpi,Caucal:2021ent,delCastillo:2020omr,delCastillo:2021znl,Caucal_2022,Caucal:2023fsf,Altinoluk:2023qfr}. The motivation behind this choice is that dijet production will be one of the main channels for probing the dense QCD medium inside hadrons at the future Electron-Ion Collider (EIC) at Brookhaven National Laboratory (BNL), see Refs.~\cite{Accardi:2012qut,AbdulKhalek:2021gbh}. By leveraging the dependence of this process on various transverse momentum scales and polarization, this scattering reaction opens broad possibilities for precision tomography of the hadron structure, e.g.\ extraction of transverse momentum dependent parton distribution functions (TMDPDFs)~\cite{Altinoluk:2024zom,Caucal:2023nci,Mukherjee:2026cte,Echevarria:2026vca}, probing the orbital angular momentum (OAM)~\cite{Kovchegov:2024wjs} contribution to the hadron spin, and searching for signatures of the saturation regime~\cite{Gribov:1983ivg,Iancu:2003xm,Weigert:2005us,Gelis:2010nm,Albacete:2014fwa,Kovchegov:2012mbw}. 


We begin our analysis by computing DIS dijet production in the general framework of the background field approach, deriving the dijet production cross section in terms of quantum (anti)quark propagators in the non-perturbative background field. We limit our computation of the quantum modes to the leading order, without loop contributions. The resulting expression is completely general and valid in arbitrary kinematics. By specifying the kinematics of the problem, we construct an expansion of the background fields onto a contour of a given choice.

First, we consider dijet production in the back-to-back regime, where the transverse momentum imbalance between the jets is much smaller than the transverse momenta of the jets. The ratio of the imbalance to the transverse momenta serves as the expansion parameter, which we apply to our general expression for the dijet production amplitude. Expanding in this parameter corresponds to expanding the background fields of the propagators onto a contour of TMD operators~\cite{Mulders:2000sh,Boer:2016xqr}. Our approach allows one to straightforwardly derive the gauge-covariant form of this expansion. We explicitly provide the leading-power result and outline an algorithm for computing higher-power corrections, which within our framework can be obtained to any necessary order.


Second, to demonstrate the universality of our approach, we compute dijet production in the small-$x$ regime. To introduce an expansion parameter for this analysis, we employ the high-energy power counting for the background fields, defined by a boost of the fields~\cite{Jackiw:1991ck,Verlinde:1993te}:
\begin{eqnarray}
&&B_-(x^+, x^-, x_\perp) \sim \lambda \tilde{B}_-(\lambda^{-1} x^+, \lambda x^-, x_\perp);
\label{eq:pw-couting-sx}\\
&&B_i(x^+, x^-, x_\perp) \sim \tilde{B}_i(\lambda^{-1} x^+, \lambda x^-, x_\perp);
\nonumber\\
&&B_+(x^+, x^-, x_\perp) \sim \lambda^{-1} \tilde{B}_-(\lambda^{-1} x^+, \lambda x^-, x_\perp)\,,
\nonumber
\end{eqnarray}
where $\lambda$ is a large boost parameter and $\tilde{B}_\mu$ is the field before the boost. The high-energy power counting~\eqref{eq:pw-couting-sx} has previously been used in computing power corrections in the TMD framework~\cite{Balitsky:2017flc} and in deriving sub-eikonal corrections in the small-$x$ formalism~\cite{Chirilli:2018kkw,Altinoluk:2021lvu}.

The power counting~\eqref{eq:pw-couting-sx} is distinct from the Color Glass Condensate (CGC) Effective Field Theory~\cite{McLerran:1993ni,McLerran:1993ka,McLerran:1994vd}, which is defined by the shock-wave structure of the background field~\cite{Balitsky1996,Gelis:2010nm,Blaizot:2016qgz}:
\begin{eqnarray}
&&B_-(x^-, x_\perp) \sim \delta(x^-)\tilde{B}_-(x_\perp);
\label{eq:cgc-frm}\\
&& B_i(x^-, x_\perp) = B_+(x^-, x_\perp) = 0\,.
\nonumber
\end{eqnarray}

In both cases the field is dominated by a large longitudinal component $B_-(x)$. However, the high-energy power counting~\eqref{eq:pw-couting-sx} retains the other background field components, which are suppressed compared to the $B_-(x)$ contribution but are nevertheless non-trivial. 

Besides, the high-energy power counting~\eqref{eq:pw-couting-sx} does not assume the shock-wave picture of interaction between quantum partons and background field and implies a smooth distribution of the field in the longitudinal direction.

We compute the dijet production amplitude within the high-energy power counting~(\ref{eq:pw-couting-sx}) using the boost parameter $\lambda$ as the expansion parameter. We demonstrate that the corresponding expansion of the path-ordered exponents is onto a staple-like contour, typical of high-energy computations at small $x$. We describe the structure of the expansion, which in our approach is explicitly gauge covariant, and explain how the corresponding QCD operators defining the scattering can be readily derived to any given order of the expansion.

We illustrate the derivation by computing the dijet production cross section at the leading eikonal order $\sim \lambda^0$. We show that gauge covariance of the derivation requires the contribution of the transverse background field component $B_i(x)$, see Eq.~\eqref{eq:pw-couting-sx}. This contribution gives rise to various operators, which are systematically derived in our analysis. In particular, we reconstruct the contribution of the transverse gauge links.


While there are certain ambiguities in the choice of the operator basis for dijet production, which we discuss in the main text, the transverse background field component also contributes through field-strength tensor insertions $\sim \sigma^{\mu\nu}F_{\mu\nu}$. These insertions describe interactions of the quantum (anti)quark with the background field via Pauli vertices and arise naturally in our formalism. 


In our computation, however, the contribution of the transverse background field component to the field-strength tensors appears already at the eikonal order, as dictated by gauge covariance. Indeed, according to the high-energy power counting~(\ref{eq:pw-couting-sx}), all terms in the field-strength tensor contribute at the eikonal order: $F_{-i} = - \partial_i B_- + \partial_- B_i - ig [B_-, B_i] \sim \lambda^0 + \lambda^0 + \lambda^0$. That is, while the transverse component itself is suppressed, $B_i \ll B_-$, its longitudinal derivative is enhanced, $\partial_- B_i \sim \partial_i B_-$, due to the Lorentz boost.


The non-trivial contribution of the transverse background field component at the eikonal order becomes particularly evident when matching the eikonal result with the dijet production cross section in the correlation/back-to-back limit~\cite{Dominguez_2011_uni,Caucal:2023fsf,Altinoluk:2024zom,Mukherjee:2026cte,Echevarria:2026vca}. Owing to the gauge covariance of our computation, matching between different types of expansions for a physical observable is straightforward. The matching is constructed by re-expanding a given term of one expansion --- e.g.\ the leading eikonal contribution $\sim \lambda^0$ within the high-energy power counting~(\ref{eq:pw-couting-sx}) --- using a different expansion parameter, e.g.\ the ratio of transverse momentum scales in the back-to-back limit. In this way, one can trace how any given term of the first expansion contributes to various orders of the second. The matching thus provides a dictionary between different types of expansions of a physical observable, which can be systematically constructed within our approach.

Specifically, as in Refs.~\cite{Dominguez_2011_uni,Dominguez:2011br} where CGC counting was assumed, we match the leading-order contribution of the eikonal expansion of dijet production withing the high-energy counting scheme with the back-to-back limit. At the leading eikonal order, we find complete agreement between the two results, up to an exponential factor in the definition of the TMD operator that does not appear at the eikonal order of the high-energy expansion~(\ref{eq:pw-couting-sx}). At the same time, we observe that the contribution of the transverse component in our eikonal result reduces to the contribution of this component in the TMD operator in the back-to-back limit, indicating a non-trivial role of this component at the eikonal order within the high-energy power counting~(\ref{eq:pw-couting-sx}). We argue that the contribution of this component should not be neglected and should be included in the analysis of physical observables at the eikonal level.

The paper is organized as follows. In Sec.~\ref{sec:cross-section} we introduce the general framework for computing the dijet production cross section using the background field approach, expressing it in terms of (anti)quark propagators in the background field of the target. In Sec.~\ref{sec:prop-path-ordered-exp} we derive a representation of these propagators in terms of path-ordered exponents, which is crucial for expanding the propagators in a given kinematic regime. We discuss the structure of this result; in particular, in Sec.~\ref{sec:tr-link} we explain how the transverse gauge link contributions arise in this formalism. Using these results, in Sec.~\ref{sec:cross-section-gen-sec} we present a general expression for the dijet production cross section at the leading order, valid in arbitrary kinematics and, in particular, for arbitrary values of the Bjorken-$x_B$ variable.

To derive an expansion of the cross section in a specific kinematic regime, in Sec.~\ref{sec:gen-exp} we develop a formalism for expanding the path-ordered exponent representation of the propagators onto an arbitrary linear piecewise contour. This involves expansion onto the light-cone direction in Sec.~\ref{sec:generalBineq0} and a transverse shift in Sec.~\ref{sec:pshift}. We present the results in a general form that allows one to construct the expansion to any required order.

Using these results, we consider the dijet production cross section in different kinematic limits. In Sec.~\ref{sec:power-cop} we discuss the back-to-back limit and present the derivation of the leading term of the expansion, which corresponds to expanding the path-ordered exponents onto a contour of TMD operators. In Sec.~\ref{sec:djprod-hepc} we then construct the expansion of the dijet production amplitude within the high-energy power counting~(\ref{eq:pw-couting-sx}), which involves expanding the path-ordered exponents onto a staple-like contour. We discuss the general structure of this expansion and provide explicit results at the leading eikonal order. We identify distinct operators contributing to the cross section at the eikonal order and analyze the relations between them.

In Sec.~\ref{sec:cgc-eik} we compare our results with the dijet production cross section at the eikonal order computed within the CGC framework~(\ref{eq:cgc-frm}). As expected, the difference between the two results is due to the contribution of the transverse background field component. To highlight the role of this component, in Sec.~\ref{sec:btb-eik} we match our eikonal result with the back-to-back limit developed in Sec.~\ref{sec:power-cop}. We find that the eikonal result matches the leading contribution in the back-to-back power counting, with the transverse background field component contributing to the field-strength tensors of the standard bilocal gluon TMD operator.

\section{Dijet production in lepton-hadron scattering\label{sec:cross-section}}
\begin{figure}[tb]
 \begin{center}
\includegraphics[width=0.9\linewidth]{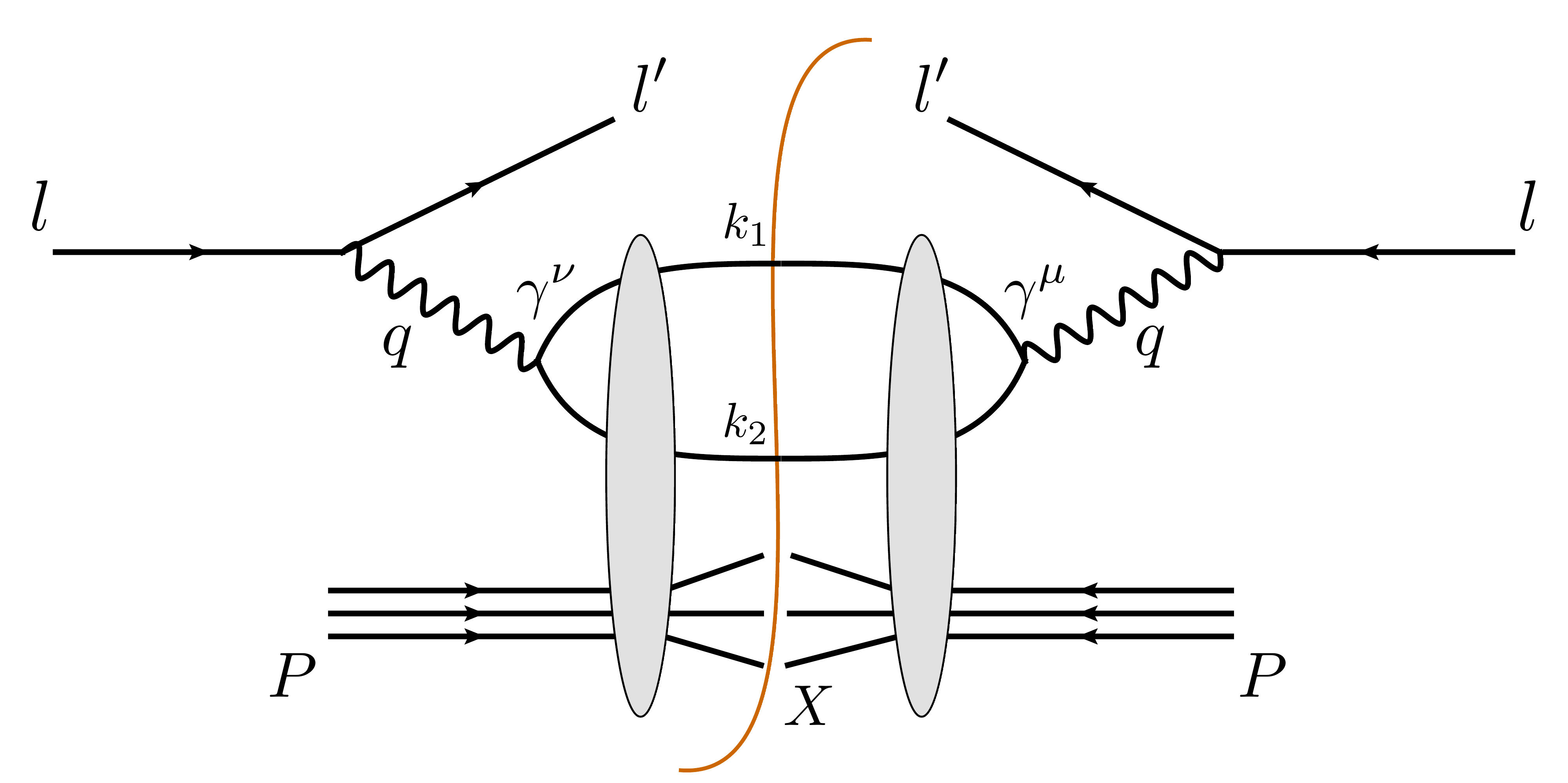}
 \end{center}
\caption{\label{fig:gif1}Dijet production in lepton-hadron scattering.}
 \end{figure}

We begin by computing the dijet production cross section in lepton-hadron collisions using the background-field approach: 
\begin{eqnarray}
&&E'\frac{d\sigma}{d^3l'} = \frac{2M_p\alpha^2_{\rm EM}}{sQ^4}L_{\mu\nu}(l, l')W^{\mu\nu}(q, P, S)\,,
\label{eq:cross-section-basic}
\end{eqnarray}
where $\alpha_{\rm EM} = e^2/(4\pi)$ is the fine-structure constant, $l$ and $l'$ are the incoming and outgoing lepton momenta, $P$ is the momentum of the hadron  with mass $M$ and spin $S^\mu = S_L P^\mu/M + S^\mu_T - M S_L \bar{n}^\mu /P^+$,\footnote{The light-like vector $\bar{n}^2=0$ is defined by $\bar{n}^- = 1$, and $P\cdot S = 0$.} see Fig.~\ref{fig:gif1}. As usual, $q$ denotes the virtual photon momentum, and the standard invariants are $Q^2 = -q^2$ and $s = (P + l)^2$.

The leptonic and hadronic tensors are given by (see e.g. Ref.~\cite{osti_6014092}) 
\begin{eqnarray}
L^{\mu\nu}(l,l')= 2(l^\mu l'^\nu + l^\nu l'^\mu - g^{\mu\nu}l\cdot l'+i \sigma_e \epsilon^{\mu\nu\rho\sigma} l_\rho l'_\sigma)\,,
\end{eqnarray}
and
\begin{eqnarray}
&&2M W^{\mu\nu}(q, P, S) = \frac{1}{2\pi}\int d^4z e^{iqz} \langle P, S| j^\mu(z)j^\nu(0)|P, S\rangle
\nonumber\\
&&=(2\pi)^3\sum_X \delta^{(4)}(p_X - P - q)
\notag \\ && \qquad \qquad \qquad \times 
\langle P, S|j^\mu(0)|X\rangle \langle X| j^\nu(0)|P, S\rangle\,,
\label{eq:hadronic-def}
\end{eqnarray}
where electromagnetic current $j^\mu = \sum_f e_f \bar{\psi}_f\gamma^\mu\psi_f$. In this work we consider inelastic dijet production, so that the sum over final states $X$ includes two jets accompanied by any other possible hadronic final states. 

We work in a reference frame in which the transverse momentum of the virtual photon vanishes:
\begin{eqnarray}
&&q^\mu = \Big(-\frac{Q^2}{2q^-}, q^-, 0_\perp\Big)\,.
\label{eq:ref-fr}
\end{eqnarray}
It is convenient to introduce transverse and longitudinal polarizations of the virtual photon. We adopt the convention
\begin{eqnarray}
&&\epsilon^{\lambda}_\mu = \frac{1}{\sqrt{2}} (0, 0, \lambda, -i),\ \ \ \epsilon^0_\mu = \frac{Q}{q^-}n_\mu\,,
\label{eq:gauge-vec}
\end{eqnarray}
where $\lambda=\pm1$ and $n_\mu$ is a light-like vector, $n^2 = 0$, with $n^+ = 1$. With this choice of polarization vectors, the metric tensor can be decomposed as\footnote{Alternatively, one could choose $$\epsilon^0_\mu = \Big(\frac{Q^2}{2q^-}, \frac{q^-}{Q}, 0_\perp\Big)\,,$$so that the metric tensor takes the form $$g_{\mu\nu} = - \sum_{\lambda = \pm1}\epsilon^{\lambda\ast}_{\mu}\epsilon^\lambda_{\nu} + \epsilon^{0\ast}_{\mu}\epsilon^0_{\nu} + \frac{q_\mu q_\nu}{q^2}\,.$$ The two choices differ by $q_\mu/Q$ and therefore yield the same cross section~(\ref{eq:cross-section-basic}). Our choice~(\ref{eq:gauge-vec}), however, significantly simplifies the calculation.}

\begin{eqnarray}
&&g_{\mu\nu} = - \sum_{\lambda = \pm1}\epsilon^{\lambda\ast}_{\mu}\epsilon^\lambda_{\nu} + \epsilon^{0\ast}_{\mu} \epsilon^0_{\nu}+ \frac{n_\mu q_\nu + q_\mu n_\nu}{n\cdot q}\,.
\label{eq:gdecomp}
\end{eqnarray}
The contribution of the last term to the cross section~(\ref{eq:cross-section-basic}) vanishes by the Ward identity~\cite{Peskin:1995ev}. Using Eq.~(\ref{eq:gdecomp}), the tensor contraction in Eq.~(\ref{eq:cross-section-basic}) can therefore be rewritten as
\begin{eqnarray}
&&E'\frac{d\sigma}{d^3 l'} = \frac{2M_p\alpha^2_{\rm EM}}{sQ^4}\sum_{\lambda,\lambda'=-1}^{1}(-1)^{\lambda + \lambda'}L_{\lambda\lambda'}W^{\lambda\lambda'}\,,
\label{eq:crosssecpol}
\end{eqnarray}
where $L_{\lambda\lambda'}\equiv L^{\mu\nu}\epsilon^\lambda_\mu \epsilon^{\lambda'\ast}_\nu$, $W^{\lambda\lambda'}\equiv W^{\mu\nu}\epsilon^{\lambda\ast}_\mu \epsilon^{\lambda'}_\nu$.

Equation~(\ref{eq:crosssecpol}) can be further rewritten in terms of the virtual photon cross section, defined through its relation to the hadronic tensor:
\begin{eqnarray}
&&\sigma^{\gamma^\ast p}_{\lambda\lambda'} = \frac{4\pi^2\alpha_{\rm EM}x}{Q^2} 2M  W^{\lambda\lambda'}\,,
\label{eq:vpcs-def}
\end{eqnarray}
where $x = Q^2/(2P\cdot q)$ is the Bjorken variable. Substituting into Eq.~(\ref{eq:crosssecpol}), one obtains
\begin{eqnarray}
&&E'\frac{d\sigma}{d^3 l'} = \frac{\alpha_{\rm EM}y}{4\pi^2Q^4}\sum_{\lambda,\lambda'=-1}^1(-1)^{\lambda + \lambda'}L^{\lambda\lambda'} \sigma^{\gamma^\ast p}_{\lambda\lambda'}\,,
\label{eq:crosssecvirt}
\end{eqnarray}
where $y=P\cdot q/P\cdot l$, which reduces to $y = Q^2/(sx)$ when hadron and lepton masses are neglected.

The explicit form of the leptonic tensor $L^{\lambda\lambda'}$ in Eq.~(\ref{eq:crosssecvirt}) can be found, for instance, in Refs.~\cite{Ananikyan:2006kz,Mantysaari:2020lhf,Kovchegov:2024wjs}. The problem of computing dijet production in lepton-hadron scattering thus reduces to computing the virtual photon cross section $\sigma^{\gamma^\ast p}_{\lambda\lambda'}$. Using Eqs.~(\ref{eq:hadronic-def}) and~(\ref{eq:vpcs-def}), the cross section can be written as
\begin{align}
&\sigma^{\gamma^\ast p}_{\lambda\lambda'} = \frac{2\pi\alpha_{\rm EM}x}{V_4Q^2} \epsilon^{\lambda\ast}_\mu \epsilon^{\lambda'}_\nu \sum_X \int d^4x e^{iqx} \langle P, S|  j^\mu(x)|X\rangle\nonumber\\
&\times \int d^4y e^{-iqy} \langle X| j^\nu(y)|P, S\rangle\,,
\label{eq:hadronic-def2}
\end{align}
where $V_4$ is a four-dimensional volume. Applying the LSZ reduction formulas for the quark and antiquark jets, this expression becomes 
\begin{widetext}
\begin{eqnarray}
&&d\sigma^{\gamma^\ast p}_{\lambda\lambda'} = \sum_{s_1,s_2}\frac{2\pi\alpha_{\rm EM}x}{V_4Q^2} \epsilon^{\lambda\ast}_\mu \epsilon^{\lambda'}_\nu \frac{dk^-_1d^2k_{1\perp}}{(2\pi)^3 2k^-_1}\Big|_{k^2_1=0} \frac{dk^-_2d^2k_{2\perp}}{(2\pi)^3 2k^-_2}\Big|_{k^2_2=0} \sum_X Z_2^{-1} \int d^4x e^{iqx} 
\label{eq:hadronic-LSZ}\\
&&\times \int d^4x_1 e^{-ik_1 x_1} \int d^4x_2 e^{-ik_2x_2} \bar{v}_{s_2}(k_2)\slashed{k}_2 \langle P, S| \tilde{T}\{\psi(x_2) j^\mu(x)\bar{\psi}(x_1)\} |X\rangle \slashed{k}_1 u_{s_1}(k_1) Z_2^{-1}
\nonumber\\
&&\times \int d^4y e^{-iqy} \int d^4y_1 e^{ik_1y_1} \int d^4y_2 e^{ik_2 y_2} \bar{u}_{s_1}(k_1)\slashed{k}_1 \langle X|  T\{\psi(y_1)j^\nu(y)\bar{\psi}(y_2)\}|P, S\rangle \slashed{k}_2v_{s_2}(k_2)\,,\nonumber
\end{eqnarray}
where $k_1$ and $k_2$ are the momenta of the quark and antiquark jets satisfying $k^2_1 = k^2_2 = 0$, and the sum runs over all possible final states $X$. Here $T$ and $\tilde{T}$ denote time ordering and anti-time ordering of operators, respectively.\footnote{Since the fields in Eq.~(\ref{eq:hadronic-def}) commute or anticommute depending on whether they are bosonic or fermionic fields respectively, the operator products in that expression can be replaced by time-ordered products.}

To introduce the background field approach, we rewrite the time-ordered products in Eq.~(\ref{eq:hadronic-LSZ}) as a double functional integral\footnote{For brevity, $C$ denotes collectively the quark and gluon fields.}
\begin{eqnarray}
&& \sum_X \langle P, S| \tilde{T}\{\psi(x_2) j^\mu(x)\bar{\psi}(x_1)\} |X\rangle \langle X|  T\{\psi(y_1)j^\nu(y)\bar{\psi}(y_2)\}|p, S\rangle
\label{eq:hadronic-func}\\
&&= \mathcal{N}^{-1}\int \mathcal{D}C \int \mathcal{D} \tilde{C} \Psi^\ast_{P, S}(\tilde{\vec{C}}(t_i)) \tilde{\psi}(x_2) \tilde{j}^\mu(x)\tilde{\bar{\psi}}(x_1)   \psi(y_1)j^\nu(y)\bar{\psi}(y_2) \Psi_{P,S}(\vec{C}(t_i))
\: e^{iS_{QCD}(C)-iS_{QCD}(\tilde{C})}
\nonumber
\end{eqnarray}
\end{widetext}
with the boundary condition $\tilde{C}(\vec{x}, t=\infty) = C(\vec{x}, t=\infty)$, which implements the sum over all possible final states $X$. Here $\Psi_{P,S}$ is the hadron wave function at $t_i\to-\infty$, and the normalization constant is chosen so that the target state satisfies $\langle P|P'\rangle = (2\pi)^3\, 2P^+\, \delta(P^+ - P'^+)\delta^{(2)}(P_\perp - P'_\perp)$.

Computation of the functional integrals in Eq.~(\ref{eq:hadronic-func}) proceeds by assuming and then proving the consistency of a factorization regime for the scattering of the virtual photon off the hadron target. Specifically, we consider high-energy scattering with a wide separation of scales, $(q+P)^2 \gg P^2$, in the photon-hadron system. This allows one to separate the functional integrals in Eq.~(\ref{eq:hadronic-func}) into integrations over perturbative and non-perturbative field modes, where the latter corresponds to the evaluation of a matrix element between hadron states.


This strategy can be made precise within the background field approach~\cite{Balitsky1996,Scimemi:2019gge,PhysRevD.109.034035}. We split the fields as
\begin{eqnarray}
&&C = A + B\,,
\label{eq:split-bfields}
\end{eqnarray}
where the $A$ mode is associated with the perturbative component of the scattering, and the $B$ mode, or background field, with the non-perturbative dynamics.

Using Eq. (\ref{eq:split-bfields}), we can rewrite Eq. (\ref{eq:hadronic-func}) as
\begin{widetext}
\begin{eqnarray}
&& \sum_X \langle P, S| \tilde{T}\{\psi(x_2) j^\mu(x)\bar{\psi}(x_1)\} |X\rangle \langle X|  T\{\psi(y_1)j^\nu(y)\bar{\psi}(y_2)\}|P, S\rangle
\label{eq:hadronic-func-split}\\
&&= \tilde{\mathcal{N}}^{-1}\int \mathcal{D}B \int \mathcal{D} \tilde{B} \Psi^\ast_{P, S}(\tilde{\vec{B}}(t_i)) \tilde{T}\{\psi(x_2) j^\mu(x)\bar{\psi}(x_1)\}_{\tilde{B}} T\{\psi(y_1)j^\nu(y)\bar{\psi}(y_2)\}_B 
\: 
\Psi_{P,S}(\vec{B}(t_i)) e^{iS_{QCD}(B)-iS_{QCD}(\tilde{B})}\,,
\nonumber
\end{eqnarray}
where
\begin{eqnarray}
&&T\{\mathcal{O}(A)\}_B
\equiv \frac{\int \mathcal{D}A e^{iS_{bg}(A, B)}\mathcal{O}(A)}{\int \mathcal{D}A e^{iS_{bg}(A, B)}} - ({\rm same\ at\ }B=0)\,,
\label{eq:TinB}
\end{eqnarray}
and similarly for $\tilde{T}\{\mathcal{O}\}_{\tilde{B}}$.
Here, the background QCD action
\begin{eqnarray}
&&S_{bg}(A, B)\equiv S_{QCD}(A+B) - S_{QCD}(B)
\end{eqnarray}
defines the action for $A$ fields in the background of $B$ fields. Subtraction in Eq. (\ref{eq:TinB}) removes disconnected diagrams. In Eqs. (\ref{eq:hadronic-func-split}) and (\ref{eq:TinB}) we assume that the integration satisfies the boundary conditions $\tilde{A}(\vec{x}, t=\infty) = A(\vec{x}, t=\infty)$ and $\tilde{B}(\vec{x}, t=\infty) = B(\vec{x}, t=\infty)$.

Integrating over the $B$ fields in Eq.~(\ref{eq:hadronic-func-split}) yields a matrix element between the hadron states:
\begin{eqnarray}
&& \sum_X \langle P, S| \tilde{T}\{\psi(x_2) j^\mu(x)\bar{\psi}(x_1)\} |X\rangle \langle X|  T\{\psi(y_1)j^\nu(y)\bar{\psi}(y_2)\}|P, S\rangle
\nonumber\\
&&= \langle P, S| \tilde{T}\{\psi(x_2) j^\mu(x)\bar{\psi}(x_1)\}_{\tilde{B}} T\{\psi(y_1)j^\nu(y)\bar{\psi}(y_2)\}_B |P, S\rangle\,.
\label{eq:hadronic-func-split-mel}
\end{eqnarray}

Substituting Eq.~(\ref{eq:hadronic-func-split-mel}) into Eq.~(\ref{eq:hadronic-LSZ}), we find that the cross section reduces to computing the time-ordered operators~(\ref{eq:TinB}) in the background of the $B$ fields:
\begin{eqnarray}
&&d\sigma^{\gamma^\ast p}_{\lambda\lambda'} = \sum_{s_1,s_2} \frac{2\pi\alpha_{\rm EM}x}{V_4Q^2} \epsilon^{\lambda\ast}_\mu \epsilon^{\lambda'}_\nu \frac{dk^-_1d^2k_{1\perp}}{(2\pi)^3 2k^-_1} \Big|_{k^2_1=0}\frac{dk^-_2d^2k_{2\perp}}{(2\pi)^3 2k^-_2} \Big|_{k^2_2=0} Z_2^{-1} \int d^4x e^{iqx}
\label{eq:hadronic-LSZ-bfm}\\
&&\times  \int d^4x_1 e^{-ik_1 x_1} \int d^4x_2 e^{-ik_2x_2} \bar{v}_{s_2}(k_2)\slashed{k}_2 \langle P, S| \tilde{T}\{\psi(x_2) j^\mu(x)\bar{\psi}(x_1)\}_{\tilde{B}} \slashed{k}_1 u_{s_1}(k_1) Z_2^{-1}
\nonumber\\
&&\times \int d^4y e^{-iqy} \int d^4y_1 e^{ik_1y_1} \int d^4y_2 e^{ik_2 y_2} \bar{u}_{s_1}(k_1)\slashed{k}_1 T\{\psi(y_1)j^\nu(y)\bar{\psi}(y_2)\}_B |P, S\rangle \slashed{k}_2v_{s_2}(k_2)\,.\nonumber
\end{eqnarray}
\end{widetext}

This expression corresponds to integration over the quantum modes, $A$. The result takes the form of matrix elements of the $B$-field operators, which encode the interaction of the perturbative component, $A$, with the non-perturbative fields of the hadron, $B$.
\begin{figure*}[htb]
 \begin{center}
\includegraphics[width=0.8\linewidth]{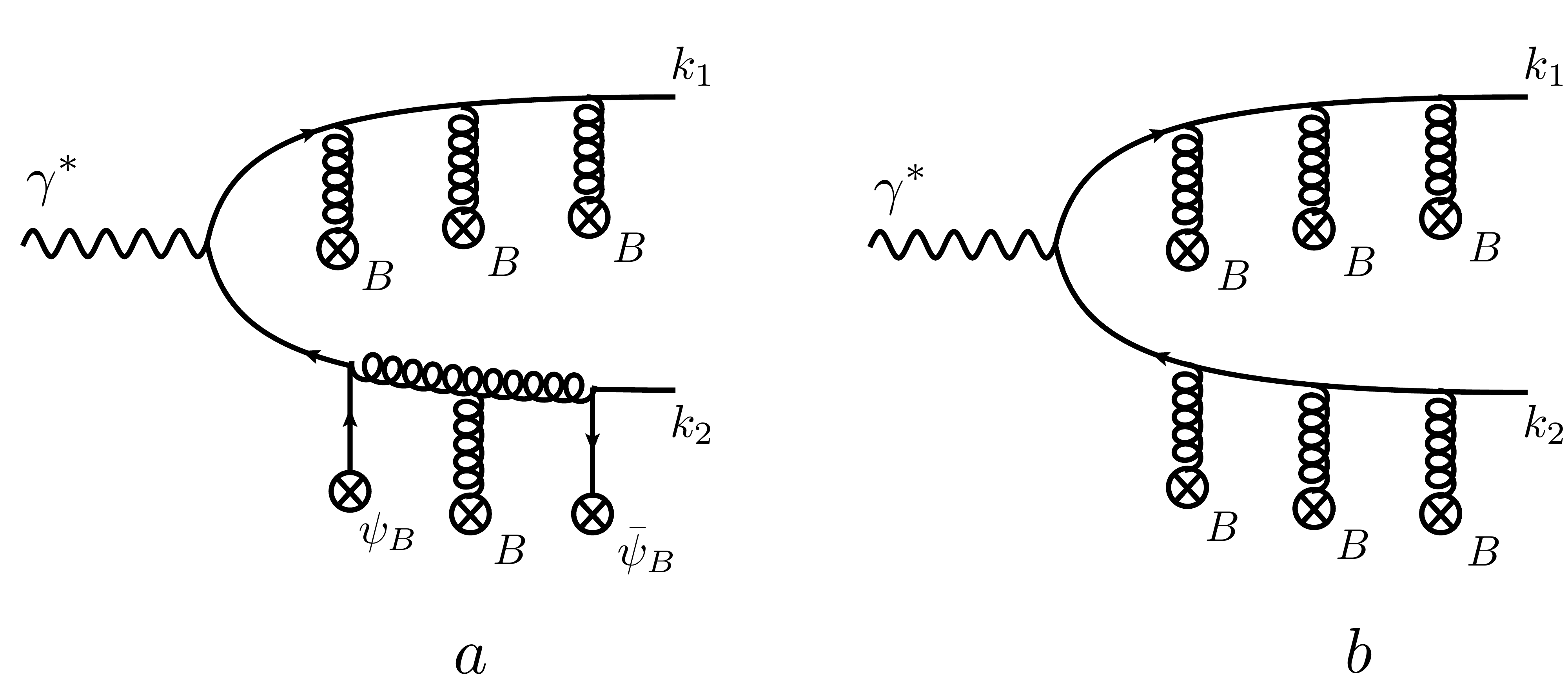}
 \end{center}
\caption{\label{fig:gif2}a) Dijet production at the leading order of quantum fields $A$ in the background fields of the target $B$; b) dijet production in the gluon background field.}
 \end{figure*}

In work, we perform the integration over the quantum fields $A$ at leading order, i.e.\ without quantum loops; the extension of our formalism to loop computations is straightforward. At leading order, a typical contribution to dijet production is shown in Fig.~\ref{fig:gif2}a, where the quantum (anti)quark propagates in the background of the $B$ fields. We focus on the gluon background field contribution, see Fig.~\ref{fig:gif2}b, leaving the analysis of the quark background field for a separate publication. Some elements of our approach can be found in Ref.~\cite{Mukherjee:2025aiw}.


For the gluon background field, the time-ordered products in Eq.~(\ref{eq:hadronic-LSZ-bfm}) take the form
\begin{widetext}
\begin{eqnarray}
&&T\{\psi(y_1)j^\nu(y)\bar{\psi}(y_2)\}_B = e(y_1|\frac{i\slashed{\hat{p}}}{\hat{p}^2+i\epsilon} + \frac{i\slashed{\hat{p}}}{\hat{p}^2+i\epsilon}(ig\slashed{\hat{B}})\frac{i\slashed{\hat{p}}}{\hat{p}^2+i\epsilon} + \dots|y)\gamma^\nu
\nonumber\\
&&\times (y|\frac{i\slashed{\hat{p}}}{\hat{p}^2+i\epsilon}+\frac{i\slashed{\hat{p}}}{\hat{p}^2+i\epsilon}(ig\slashed{\hat{B}})\frac{i\slashed{\hat{p}}}{\hat{p}^2+i\epsilon}+\dots|y_2) - (B=0)\,,
\label{eq:Tprod-init}
\end{eqnarray}
\end{widetext}
where the ellipses stand for higher-order terms in the expansion in the strong coupling constant.

In what follows, we adopt the Schwinger notation~\cite{Schwinger:1951nm}, which provides a compact and efficient formulation for computations in the background field.
We introduce coherent states $|x)$ and $|p)$, which are eigenstates of the position and momentum operators,
\begin{eqnarray}
&&\hat{x}_\mu|x) = x_\mu|x);\ \ \ \hat{p}_\mu|p) = p_\mu|p)\,,
\label{eq:sch-eihen}
\end{eqnarray}
with the normalization
\begin{eqnarray}
&&\int d^4x\, |x)( x| = 1\,;\\ && \int \dhd^4p \, |p) (p| = 1;\\ && ( x|p) = e^{-ipx}\,.
\label{eq:bosonort}
\end{eqnarray}
The background field operator acts on the coordinate states as $\hat{B}|x) = B(x)|x)$. For brevity, in the following we drop the ``hat'' notation for operators.

The infinite sums in Eq. (\ref{eq:Tprod-init}) can be formally resummed as
\begin{eqnarray}
&&T\{\psi(y_1)j^\nu(y)\bar{\psi}(y_2)\}_B\nonumber\\
&&= e (y_1|\frac{i}{\slashed{P}+i\epsilon}|y)\gamma^\nu (y|\frac{i}{\slashed{P}+i\epsilon}|y_2) - (B=0)\,,
\label{eq:Tprod-resum}
\end{eqnarray}
where the covariant derivative $P_\mu = p_\mu + gB_\mu$. And similarly,
\begin{eqnarray}
&&\tilde{T}\{\psi(x_2)j^\mu(x)\bar{\psi}(x_1)\}_{\tilde{B}}\nonumber\\
&&= e(x_2|\frac{-i}{\slashed{\tilde{P}}-i\epsilon} |x)\gamma^\mu (x|\frac{-i}{\slashed{\tilde{P}}-i\epsilon}|x_1) - (\tilde{B}=0)\,,
\label{eq:Tprprod-resum}
\end{eqnarray}
where $\tilde{P}_\mu = p_\mu + g\tilde{B}_\mu$.

In the cross section~(\ref{eq:hadronic-LSZ-bfm}), the time-ordered products can be further simplified. Using $\slashed{P}^2 = P^2 + \frac{g}{2}\sigma^{\mu\nu}F_{\mu\nu}$, we obtain
\begin{widetext}
\begin{eqnarray}
&&\int d^4y_1 e^{ik_1y_1} \int d^4y_2 e^{ik_2 y_2} \bar{u}_{s_1}(k_1)\slashed{k}_1 T\{\psi(y_1)j^\nu(y)\bar{\psi}(y_2)\}_B\slashed{k}_2v_{s_2}(k_2) 
\label{eq:Tprod-cross}\\
&&= e \bar{u}_{s_1}(k_1)\slashed{k}_1 (k_1|\slashed{P}\frac{i}{\slashed{P}^2+i\epsilon}|y)\gamma^\nu (y|\frac{i}{\slashed{P}^2+i\epsilon}\slashed{P}|-k_2)\slashed{k}_2v_{s_2}(k_2) - (B=0)
\nonumber\\
&&= e \bar{u}_{s_1}(k_1) k^2_1 (k_1|\frac{i}{P^2 + \frac{g}{2}\sigma^{\mu\nu}F_{\mu\nu}+i\epsilon}|y)\gamma^\nu (y|\frac{i}{P^2 + \frac{g}{2}\sigma^{\mu\nu}F_{\mu\nu}+i\epsilon}|-k_2)(-k^2_2) v_{s_2}(k_2) 
\nonumber\\
&&- (B=0)\,.\nonumber
\end{eqnarray}
Similarly, for the anti-time-ordered product:
\begin{eqnarray}
&&\int d^4x_1 e^{-ik_1 x_1} \int d^4x_2 e^{-ik_2x_2} \bar{v}_{s_2}(k_2)\slashed{k}_2 \tilde{T}\{\psi(x_2)j^\mu(x)\bar{\psi}(x_2)\}_{\tilde{B}} \slashed{k}_1 u_{s_1}(k_1)
\label{eq:Tprprod-cross}\\
&&= e \bar{v}_{s_2}(k_2) (-k^2_2) (-k_2|\frac{-i}{\tilde{P}^2 + \frac{g}{2}\sigma^{\mu\nu}\tilde{F}_{\mu\nu}-i\epsilon} |x)\gamma^\mu (x|\frac{-i}{\tilde{P}^2 + \frac{g}{2}\sigma^{\mu\nu}\tilde{F}_{\mu\nu}-i\epsilon}|k_1) k^2_1 u_{s_1}(k_1)\nonumber\\
&&- (\tilde{B}=0)\,.\nonumber
\nonumber
\end{eqnarray}

Substituting Eqs.~(\ref{eq:Tprod-cross}) and~(\ref{eq:Tprprod-cross}) into Eq.~(\ref{eq:hadronic-LSZ-bfm}), we obtain the cross section at leading order in the quantum fields:
\begin{eqnarray}
&&d\sigma^{\gamma^\ast p}_{\lambda\lambda'} = \sum_f\sum_{s_1,s_2}\frac{2\pi e^2_f \alpha_{\rm EM}x}{V_4Q^2}   \epsilon^{\lambda\ast}_\mu \epsilon^{\lambda'}_\nu \frac{dk^-_1d^2k_{1\perp}}{(2\pi)^3 2k^-_1} \Big|_{k^2_1=0}\frac{dk^-_2d^2k_{2\perp}}{(2\pi)^3 2k^-_2} \Big|_{k^2_2=0}
\label{eq:hadronic-bfm-LO}\\
&&\times \int d^4x e^{iqx}  \langle P, S| \Big\{ \bar{v}_{s_2}(k_2) (-k^2_2) (-k_2|\frac{-i}{\tilde{P}^2 + \frac{g}{2}\sigma^{\mu\nu}\tilde{F}_{\mu\nu}-i\epsilon} |x)\gamma^\mu (x|\frac{-i}{\tilde{P}^2 + \frac{g}{2}\sigma^{\mu\nu}\tilde{F}_{\mu\nu}-i\epsilon}|k_1)\nonumber\\
&&\times k^2_1 u_{s_1}(k_1)
- (\tilde{B}=0)\Big\}
\int d^4y e^{-iqy} \Big\{\bar{u}_{s_1}(k_1) k^2_1 (k_1|\frac{i}{P^2 + \frac{g}{2}\sigma^{\mu\nu}F_{\mu\nu}+i\epsilon}|y)\gamma^\nu\nonumber\\
&&\times (y|\frac{i}{P^2 + \frac{g}{2}\sigma^{\mu\nu}F_{\mu\nu}+i\epsilon}|-k_2) (-k^2_2) v_{s_2}(k_2)
- (B=0)\Big\} |P, S\rangle \,.
\nonumber
\end{eqnarray}
\end{widetext}

We thus find that computing the dijet production cross section at leading order reduces to evaluating the (anti)quark propagator in the background field:
\begin{eqnarray}
&&\lim_{k^2\to 0}k^2 (k|\frac{i}{P^2 + \frac{g}{2}\sigma^{\mu\nu}F_{\mu\nu}+i\epsilon}|y)\,.
\label{eq:prop-def}
\end{eqnarray}

The structure of the quark propagator~(\ref{eq:prop-def}) explicitly reflects the separation of scalar and spin degrees of freedom. This can be understood from the functional integral representation of the theory. The one-loop QCD effective action in the gluon background field $B_\mu$ can be written in terms of functional integrals over quark trajectories as (see e.g.\ Ref.~\cite{Schubert:2001he})\footnote{Here the spacetime metric is taken to be Euclidean.}
\begin{eqnarray}
&&\Gamma_{QCD}[B]\nonumber\\
&&= -\frac{1}{2} \int^T_0 \frac{dT}{T}\, 
\int_P \mathcal{D}x \exp\Big\{-\int^T_0 d\tau \Big(\frac{1}{4} \dot{x}^2 + ig\dot{x}^\mu B_\mu \Big)\Big\} 
\nonumber\\
&&\times \int_A \mathcal{D} \psi \exp\Big\{-\int^T_0 d\tau \Big( \frac{1}{2}\psi_\mu\dot{\psi}^\mu - ig \psi^\mu \psi^\nu F_{\mu\nu}\Big)\Big\}\,.
\label{MLag}
\end{eqnarray}

The first functional integral in Eq.~(\ref{MLag}) is over the coordinate trajectory $x_\mu(\tau)$, parametrized by the proper-time variable $\tau$:
\begin{eqnarray}
&&\int_P \mathcal{D}x \exp\Big\{-\int^T_0 d\tau \Big(\frac{1}{4} \dot{x}^2 + ig\dot{x}^\mu B_\mu \Big)\Big\}\,.
\label{MLag-scal}
\end{eqnarray}

\begin{figure*}[htb]
 \begin{center}
\includegraphics[width=150mm]{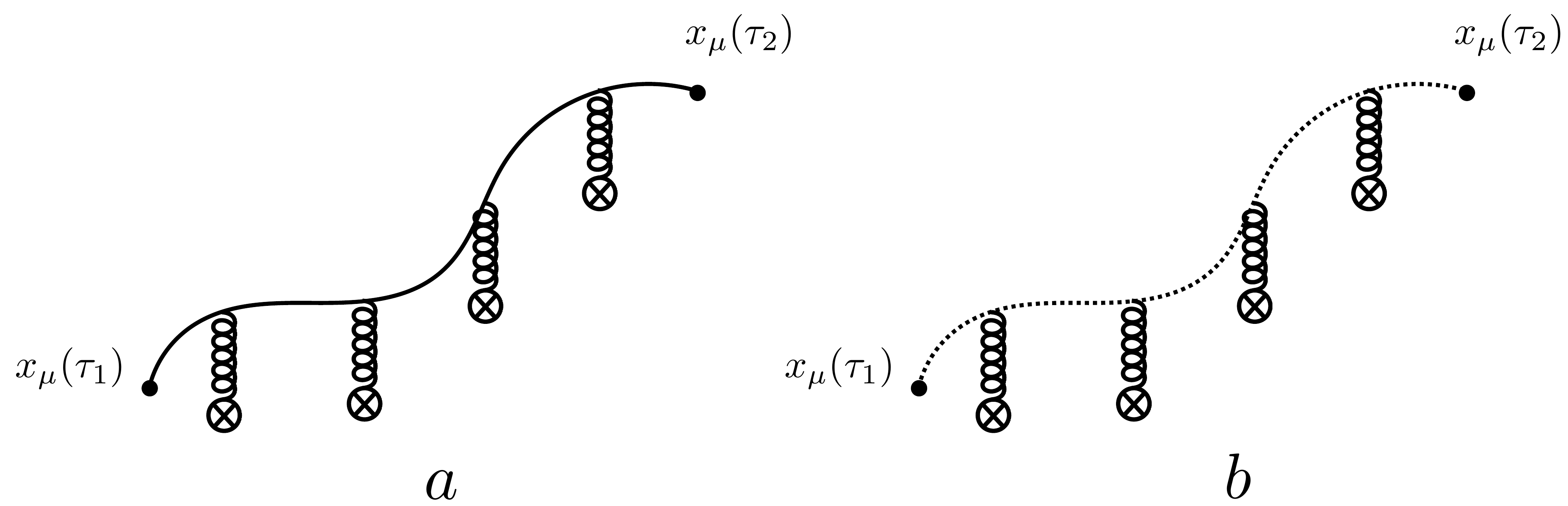}
 \end{center}
\caption{\label{fig:gif3}
a) A quark propagating along a coordinate trajectory $x_\mu(\tau)$ in the background field $B_\mu$; b) the quark acquires the gauge phase~(\ref{eq:scalar-phase}) as a result of this propagation.
}
 \end{figure*}

This integral has periodic boundary conditions and describes the scalar degrees of freedom of the quark. An integral of this form defines the one-loop effective action of scalar QED with the background photon field $B_\mu$. Propagating along a coordinate trajectory $x_\mu(\tau)$, the quark acquires a scalar phase
\begin{eqnarray}
&&\exp\Big(- ig \int d\tau \dot{x}^\mu B_\mu \Big)\,,
\label{eq:scalar-phase}
\end{eqnarray}
which is simply a Wilson line along the path $x_\mu(\tau)$, see Fig.~\ref{fig:gif3}.

 We note that in Eq.~(\ref{MLag-scal}) the sum is over all possible trajectories $x_\mu(\tau)$. Each trajectory yields a particular scalar phase~(\ref{eq:scalar-phase}), which makes the analysis of the background field contribution in Eq.~(\ref{MLag-scal}) highly non-trivial. One can study the functional integral either by applying a semiclassical approximation or by expanding the background fields onto a fixed contour in coordinate space, which yields a representation of the functional integral as an infinite series. This series can be further analyzed provided that the process contains an expansion parameter defined by the kinematics of the problem --- an approach well suited for the QCD factorization framework. This expansion method will be our main tool throughout the rest of this paper.


The scalar functional integral~(\ref{MLag-scal}) corresponds to the $1/P^2$ term of the quark propagator~(\ref{eq:prop-def}), see Refs.~\cite{Morgan:1995te,Tarasov:2019rfp}. 


The fermionic nature of the degrees of freedom of the quark, on the other hand, are described by a functional integral over the Grassmann trajectory $\psi_\mu(\tau)$ with anti-periodic boundary conditions. In this case, the interaction of the quark with the background field is governed by the term $\sim \psi^\mu \psi^\nu F_{\mu\nu}$, which maps exactly onto the $\sim \sigma^{\mu\nu}F_{\mu\nu}$ term of the propagator~(\ref{eq:prop-def}), see Ref.~\cite{Tarasov:2019rfp}. The quark's interaction with the background field via genuine fermionic part is thus determined by the field-strength tensor $F_{\mu\nu}$. In particular, as we will see, within the high-energy power counting~(\ref{eq:pw-couting-sx}) all terms of this field-strength tensor are of the same order, so this operator will naturally appear in our final results.


While the propagator~(\ref{eq:prop-def}) is completely general, its direct application is problematic since, as noted above, it incorporates gauge factors over all possible quark trajectories. To make this structure tractable, one needs to expand the gauge factors onto a given contour, the form of which is determined by the kinematics of the problem. In the next section, we compute the propagator~(\ref{eq:prop-def}) and derive its path-ordered exponent representation, which renders this expansion straightforward.

\section{Quark propagator in the background field\label{sec:prop-path-ordered-exp}}
In the previous section, we derived a general expression for the dijet production cross section~(\ref{eq:hadronic-bfm-LO}). The structure of this result is transparent: the incoming virtual photon splits into a quark-antiquark pair that propagates in the background field of the target until it produces the final dijet state. The parton propagation in the background field is governed by the (anti)quark propagator~(\ref{eq:prop-def}).

The goal of this section is to compute the (anti)quark propagator~(\ref{eq:prop-def}) in a general background field of the form
\begin{eqnarray}
&&B_\mu(x) = B_\mu(x^-, x_\perp)\,.
\label{eq:bfield-form}
\end{eqnarray}
To simplify the computation, we fix the gauge of the background field to $B_+(x) = 0$ (the so-called wrong light-cone gauge) and impose the boundary condition $\lim_{x^-\to\infty}B_i(x) = 0$. In Eq.~(\ref{eq:bfield-form}), we neglect the dependence on the $x^+$ coordinate. As we will demonstrate later, this is sufficient for computing dijet production at the level of precision considered in this paper. Specifically, in the back-to-back limit this dependence contribute to the higher-power TMD operators, while in the high-energy limit (\ref{eq:pw-couting-sx}) it leads to the sub-eikonal corrections.

We emphasize that, given the background field~(\ref{eq:bfield-form}), our computation is completely general and involves no additional assumptions. In particular, our analysis with Eq.~(\ref{eq:bfield-form}) includes the derivation of the transverse gauge link contribution to the operator for dijet production, which is one of the main results of this paper.

\subsection{Scalar propagator in the background field}
The most involved part of computing the (anti)quark propagator~(\ref{eq:prop-def}) is the derivation of its scalar part. This is due to the necessity of manipulating the gauge factors~(\ref{eq:scalar-phase}) appearing in this propagator. Our approach provides a solution to this problem by constructing a gauge-covariant expansion of these factors within the (anti)quark propagator. To construct such an expansion, we first need to derive a path-ordered exponent representation of the propagator, which simplifies handling  the non-trivial background-field dependence  and enables a systematic expansion. We proceed to construct this representation below.


Starting with a scalar propagator in the background field,
\begin{eqnarray}
&&\lim_{k^2\to0}k^2(k| \frac{1}{P^2 + i\epsilon}|y)\,,
\label{eq:scalar-prop}
\end{eqnarray}
we rewrite it as\footnote{Here we use the fact that, since the background field is independent of $x^+$, the operator $\hat{p}^-$ commutes with the field operators.}
\begin{eqnarray}
&&\lim_{k^2\to0}k^2(k| \frac{1}{P^2 + i\epsilon}|y)\label{eq:scal-prop-first}\\
&&= \lim_{k^2\to0}k^2 \int d^4z  \frac{ e^{ikz} }{2k^-}(z|\frac{1}{p^+ + gB_- - \frac{P^2_\perp}{2p^-} + i\epsilon p^-}|y)\,,
\nonumber
\end{eqnarray}
where the transverse covariant derivative $P_k = p_k + gB_k$.

We formally expand the scalar propagator~(\ref{eq:scal-prop-first}) in powers of the coupling to the $B_-$ field component:
\begin{widetext}
\begin{eqnarray}
&&\lim_{k^2\to0}k^2(k| \frac{1}{P^2 + i\epsilon}|y) 
\label{eq:scalar-exp-Amin}\\
&&= \lim_{k^2\to0}k^2 \int d^4z  \frac{ e^{ikz} }{2k^-} \Big\{ (z| \frac{1}{p^+  - \frac{P^2_\perp}{2p^-} + i\epsilon p^-}|y)
- \int d^4z_1 (z| \frac{1}{p^+  - \frac{P^2_\perp}{2p^-} + i\epsilon p^-}|z_1) gB_-(z_1) (z_1|\frac{1}{p^+  - \frac{P^2_\perp}{2p^-} + i\epsilon p^-} |y)
\nonumber\\
&&+ \int d^4z_1 \int d^4z_2 (z| \frac{1}{p^+  - \frac{P^2_\perp}{2p^-} + i\epsilon p^-}|z_1) gB_-(z_1) (z_1|\frac{1}{p^+  - \frac{P^2_\perp}{2p^-} + i\epsilon p^-}|z_2) gB_-(z_2)
 (z_2|\frac{1}{p^+  - \frac{P^2_\perp}{2p^-} + i\epsilon p^-} |y)  
+ \dots \Big\}\,,
\nonumber
\end{eqnarray}
where the ellipsis denotes terms with higher-order $B_-$ insertions.

Equation~(\ref{eq:scalar-exp-Amin}) has a simple meaning: successive interactions of the scalar quark with the $B_-$ field component are connected by the propagator
\begin{eqnarray}
(x|\frac{1}{p^+  - \frac{P^2_\perp}{2p^-} + i\epsilon p^-}|y)\,,
\label{eq:transv-prop}
\end{eqnarray}
which describes the quark dynamics in the transverse plane, including interactions with the transverse field component $B_i$. From Eq.~(\ref{eq:scalar-exp-Amin}) we therefore conclude that computing the scalar propagator~(\ref{eq:scalar-prop}) reduces to evaluating the propagator~(\ref{eq:transv-prop}) and dressing it with an infinite number of $B_-$ insertions.


We perform a detailed computation of Eq.~(\ref{eq:transv-prop}) in Appendix~\ref{app:tr-prop}, where we derive the following path-ordered exponent representation:
\begin{eqnarray}
&&(x| \frac{1}{p^+  - \frac{P^2_\perp}{2p^-} + i\epsilon p^-}|y)
= \Big( \frac{-i}{2\pi} \theta(x^- - y^-) \int^\infty_0 dp^- + \frac{i}{2\pi} \theta(y^- - x^-) \int^0_{-\infty} dp^- \Big)
 e^{-ip^- (x^+ - y^+)}\label{eq:tr-prop-Pexp-main}\\ 
&&\times (x_\perp| e^{-i\frac{ P^2_\perp(x^-)}{2p^-}x^- }\mathcal{P}\exp\Big\{ i\int^{x^-}_{y^-} dz^- \Big(e^{i\frac{ P^2_\perp(z^-)}{2p^-}z^-}i\partial_- e^{-i\frac{ P^2_\perp(z^-)}{2p^-}z^-} - \frac{P^2_\perp(z^-)}{2p^-}\Big) \Big\}
e^{i\frac{P^2_\perp(y^-)}{2p^-}y^-} |y_\perp)\,,
 \nonumber
\end{eqnarray}
where $\partial_- = \partial/\partial z^-$ acts on the corresponding $z^-$ variable.

Substituting Eq.~(\ref{eq:tr-prop-Pexp-main}) into Eq.~(\ref{eq:scalar-exp-Amin}) and performing the integrations over intermediate coordinate variables, we obtain
\begin{eqnarray}
&&\lim_{k^2\to0}k^2(k| \frac{1}{P^2 + i\epsilon}|y) 
\label{eq:scalar-exp-Amin-exp-form}\\
&&= \lim_{k^2\to0}k^2 \int d^4z e^{ikz} \frac{1}{2k^-} \Big( \frac{-i}{2\pi} \theta(z^- - y^-) \int^\infty_0 dp^- + \frac{i}{2\pi} \theta(y^- - z^-) \int^0_{-\infty} dp^- \Big) 
\nonumber\\
&&\times e^{-ip^- (z^+ - y^+)}  (z_\perp| e^{-i\frac{ P^2_\perp(z^-)}{2p^-}z^- } \Big[ \mathcal{P}\exp\Big\{ i\int^{z^-}_{y^-} dz^-_1 \Big(e^{i\frac{ P^2_\perp(z^-_1)}{2p^-}z^-_1}i\partial_- e^{-i\frac{ P^2_\perp(z^-_1)}{2p^-}z^-_1} - \frac{P^2_\perp(z^-_1)}{2p^-}\Big) \Big\}  
\nonumber\\
&&+ i \int^{z^-}_{y^-} dz^-_1 \mathcal{P}\exp\Big\{ i\int^{z^-}_{z^-_1} dz^-_3 \Big(e^{i\frac{ P^2_\perp(z^-_3)}{2p^-}z^-_3}i\partial_- e^{-i\frac{ P^2_\perp(z^-_3)}{2p^-_1}z^-_3} - \frac{P^2_\perp(z^-_3)}{2p^-_1}\Big) \Big\} e^{i\frac{P^2_\perp(z^-_1)}{2p^-_1}z^-_1} B_-(z^-_1)\nonumber\\
&&\times e^{-i\frac{ P^2_\perp(z^-_1)}{2p^-}z^-_1 }
\mathcal{P}\exp\Big\{ i\int^{z^-_1}_{y^-} dz^-_4 \Big(e^{i\frac{ P^2_\perp(z^-_4)}{2p^-}z^-_4}i\partial_- e^{-i\frac{ P^2_\perp(z^-_4)}{2p^-}z^-_4} - \frac{P^2_\perp(z^-_4)}{2p^-}\Big) \Big\} + \dots \Big]\nonumber\\
&&\times e^{i\frac{P^2_\perp(y^-)}{2p^-}y^-}|y_\perp)\,,
\nonumber
\end{eqnarray}
where the ellipsis denotes higher-order terms in the number of $B_-$ insertions. These terms have a form analogous to the leading-order terms shown explicitly in Eq.~(\ref{eq:scalar-exp-Amin-exp-form}).


The infinite series of $B_-$ insertions in Eq.~(\ref{eq:scalar-exp-Amin-exp-form}) can be exponentiated into a path-ordered exponent:
\begin{eqnarray}
&&\lim_{k^2\to0}k^2(k| \frac{1}{P^2 + i\epsilon}|y) = \lim_{k^2\to0}k^2 \int d^4z e^{ikz} \frac{1}{2k^-} \Big( \frac{-i}{2\pi} \theta(z^- - y^-) \int^\infty_0 dp^-
\label{eq:scalar-exp-Amin-exp-form-full-S}\\
&& + \frac{i}{2\pi} \theta(y^- - z^-) \int^0_{-\infty} dp^- \Big) e^{-ip^- (z^+ - y^+)} (z_\perp| e^{-i\frac{ P^2_\perp(z^-)}{2p^-}z^- } \mathcal{S}(z^-, y^-) e^{i\frac{P^2_\perp(y^-)}{2p^-}y^-}|y_\perp)\,,
\nonumber
\end{eqnarray}
where
\begin{eqnarray}
&&\mathcal{S}(x^-, y^-)\equiv \mathcal{P}\exp\Big\{ i\int^{x^-}_{y^-} dz^- \Big(e^{i\frac{ P^2_\perp(z^-)}{2p^-}z^-}iD_- e^{-i\frac{ P^2_\perp(z^-)}{2p^-}z^-} - \frac{P^2_\perp(z^-)}{2p^-}\Big) \Big\}\,,
\label{eq:scalar-simple}
\end{eqnarray}
and the covariant derivative $D_- = \partial_- - igB_-(z^-)$ acts on the corresponding $z^-$ variable. Equation~(\ref{eq:scalar-exp-Amin-exp-form-full-S}) provides the path-ordered exponent representation of the scalar propagator~(\ref{eq:scalar-prop}), which we will use to construct a gauge-covariant expansion of the (anti)quark propagator~(\ref{eq:prop-def}).


For the subsequent discussion, it is convenient to introduce a path-ordered exponent with a ``shifted'' phase:
\begin{eqnarray}
&&\mathcal{S}(x^-, y^-;\xi^-)\label{eq:definition-exp-shifted}\\
&&\equiv \mathcal{P}\exp\Big\{ i\int^{x^-}_{y^-} dz^- \Big(e^{i\frac{ P^2_\perp(z^-)}{2p^-}(z^--\xi^-)}iD_- e^{-i\frac{ P^2_\perp(z^-)}{2p^-}(z^--\xi^-)} - \frac{P^2_\perp(z^-)}{2p^-}\Big) \Big\}\,,
\nonumber
\end{eqnarray}
so that definition~(\ref{eq:scalar-simple}) corresponds to a trivial shift: $\mathcal{S}(x^-, y^-) \equiv \mathcal{S}(x^-, y^-;0^-)$.

One can verify that the path-ordered exponents~(\ref{eq:scalar-simple}) and~(\ref{eq:definition-exp-shifted}) satisfy the identity
\begin{eqnarray}
&&e^{-i\frac{ P^2_\perp(x^-)}{2p^-}x^- }\mathcal{S}(x^-, y^-) e^{i\frac{P^2_\perp(y^-)}{2p^-}y^-}\label{eq:parallel-shift-scalar-full}
= e^{-i\frac{ P^2_\perp(x^-)}{2p^-}(x^--\xi^-) }\mathcal{S}(x^-, y^-;\xi^-) e^{i\frac{P^2_\perp(y^-)}{2p^-}(y^--\xi^-)}\,.
\end{eqnarray}
As we will explain below, this identity allows one to shift the position of the exponential factors in the path-ordered exponents and manipulate the form of the gauge factors in the (anti)quark propagator~(\ref{eq:prop-def}), thereby enabling expansion of these factors onto an arbitrary linear piecewise contour.


As shown in Appendix~\ref{qpp-LSZ}, Eq.~(\ref{eq:scalar-exp-Amin-exp-form-full-S}) can be further simplified to
\begin{eqnarray}
&&\lim_{k^2\to0}k^2(k| \frac{1}{P^2 + i\epsilon}|y) = \lim_{k^2\to0} (k_\perp| \Big(  \theta(k^-) e^{i\frac{ k^2_\perp}{2k^-} z^-} e^{-i\frac{P^2_\perp(z^-) }{2k^-} z^-} \Big|^\infty \mathcal{S}(\infty, y^-)
 \nonumber\\
 &&+ \theta(-k^-) e^{i\frac{ k^2_\perp}{2k^-} z^-} e^{-i\frac{P^2_\perp(z^-) }{2k^-} z^-} \Big|^{-\infty} \mathcal{S}(-\infty, y^-) \Big) e^{i\frac{P^2_\perp(y^-)}{2k^-}y^-} |y_\perp) e^{ik^- y^+} \,.
 \label{eq:LSZ-fin}
\end{eqnarray}

Equation~(\ref{eq:LSZ-fin}) is our final result for the path-ordered exponent representation of the scalar propagator~(\ref{eq:scalar-prop}).

\subsection{Structure of the scalar propagator in the background field\label{sec:scl-prop}}
The meaning of Eq.~(\ref{eq:LSZ-fin}) is not immediately transparent. It contains various exponential factors constructed from the transverse covariant derivative $P_k$, as well as a path-ordered exponent involving the $B_-$ field component. Let us clarify the role of each element in this equation.


To this end, we first (temporary) set the transverse background field component to zero, $B_i=0$.\footnote{For brevity, we also fix $k^->0$. The analysis of $k^-<0$ is identical.} Equation~(\ref{eq:LSZ-fin}) then takes the form
\begin{eqnarray}
&&\lim_{k^2\to0}k^2(k| \frac{1}{P^2 + i\epsilon}|y)\Big|_{B_i=0}^{k^->0}\label{eq:prmix-pB0}
= \lim_{k^2\to0} (k_\perp| \mathcal{P}\exp\Big\{ ig\int^{\infty}_{y^-} dz^- e^{i\frac{p^2_\perp}{2p^-}z^-} B_-(z^-) e^{-i\frac{p^2_\perp}{2p^-}z^-} \Big\} e^{i\frac{p^2_\perp}{2k^-}y^-} |y_\perp) e^{ik^- y^+} \,.
\end{eqnarray}
\end{widetext}

We further simplify by neglecting the exponential factors inside the path-ordered exponent:
\begin{eqnarray}
&&\lim_{k^2\to0}k^2(k| \frac{1}{P^2 + i\epsilon}|y)\Big|_{B_i=0;k^->0}\nonumber\\
&&\approx \lim_{k^2\to0} (k_\perp| U(\infty, y^-) e^{i\frac{p^2_\perp}{2k^-}y^-} |y_\perp) e^{ik^- y^+} \,,\label{eq:Uprop-scalar}
\end{eqnarray}
where $U(\infty, y^-)$ is a Wilson line --- a gauge factor along the light-cone direction:
\begin{eqnarray}
&&U(x^-, y^-) = \mathcal{P}\exp\Big\{ ig\int^{x^-}_{y^-} dz^- B_-(z^-)  \Big\}\,.
\label{eq:simple-wl}
\end{eqnarray}

Equation~(\ref{eq:Uprop-scalar}) represents the vacuum scalar propagator
\begin{eqnarray}
&&\lim_{k^2\to0}k^2(k| \frac{1}{p^2 + i\epsilon}|y)\nonumber\\
&&= \lim_{k^2\to0} (k_\perp| e^{i\frac{p^2_\perp}{2k^-}y^-} |y_\perp) e^{ik^- y^+} = e^{iky}\Big|_{k^2=0}\,,
\end{eqnarray}
dressed with a semi-infinite gauge factor encoding the interaction of the scalar quark with the background field.


The transverse position of the gauge factor in Eq.~(\ref{eq:Uprop-scalar}) is not yet fixed. Inserting a complete set of transverse position states, we can rewrite Eq.~(\ref{eq:Uprop-scalar}) as
\begin{widetext}
\begin{eqnarray}
&&\lim_{k^2\to0}k^2(k| \frac{1}{P^2 + i\epsilon}|y)\Big|_{B_i=0;k^->0}\label{eq:Uprop-scalar-tpfix}
\approx \lim_{k^2\to0} \int d^2z_\perp e^{-ik_\perp z_\perp} U(\infty, y^-;z_\perp) (z_\perp| e^{i\frac{p^2_\perp}{2k^-}y^-} |y_\perp) e^{ik^- y^+} \,,
\end{eqnarray}
where 
\begin{eqnarray}
&&U(\infty, y^-;z_\perp) = \mathcal{P}\exp\Big\{ ig\int^{\infty}_{y^-} dz^- B_-(z^-, z_\perp)  \Big\}\,,
\label{eq:simf-lc-int}
\end{eqnarray}
c.f. Eq. (\ref{eq:simple-wl}).

The transverse position of the gauge factor in Eq.~(\ref{eq:Uprop-scalar-tpfix}) is now fixed at $z_\perp$. The shift in the transverse position from $y_\perp$ to $z_\perp$ is governed by the exponential factor
\begin{eqnarray}
&&(x_\perp| e^{i\frac{p^2_\perp}{2k^-}z^-} |y_\perp)\,,
\label{eq:exp-fact-bare}
\end{eqnarray}
originating from the vacuum propagator:
\begin{eqnarray}
&&(x| \frac{1}{p^2 + i\epsilon }|y)\Big|_{x^- > y^-}
= \frac{-i}{2\pi} \int^\infty_0 \frac{dp^-}{2p^-}   e^{-ip^- (x^+ - y^+)} (x_\perp| e^{-i\frac{ p^2_\perp}{2p^-}(x^--y^-) } |y_\perp)\,.
\label{eq:free-prop}
\end{eqnarray}
\end{widetext}
This exponential factor describes the quark's propagation in the transverse direction. Indeed, without it, the propagator~(\ref{eq:free-prop}) would be proportional to a delta function,
\begin{eqnarray}
&&(x_\perp|y_\perp) = \delta^{(2)}(x_\perp - y_\perp)\,,
\end{eqnarray}
and the quark's transverse position would remain unchanged.

\begin{figure}[htb]
 \begin{center}
\includegraphics[width=0.7\linewidth]{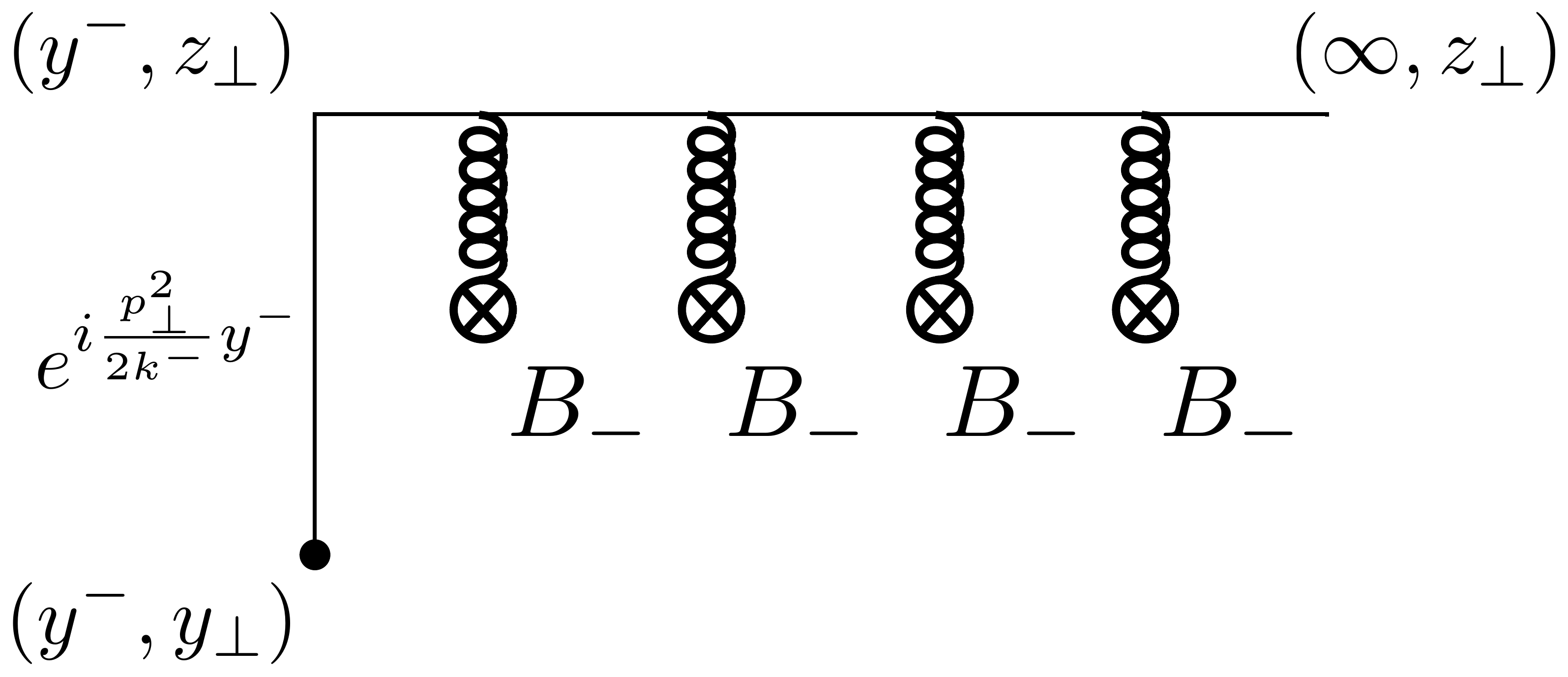}
 \end{center}
\caption{\label{fig:gif4} 
Equation~(\ref{eq:Uprop-scalar-tpfix}) contains a semi-infinite gauge factor~(\ref{eq:simf-lc-int}) along the light-cone direction. The initial shift in the transverse direction from $y_\perp$ to $z_\perp$ is described by an exponential factor.
}
 \end{figure}

With this interpretation of the exponential factor, Eq.~(\ref{eq:Uprop-scalar-tpfix}) admits a simple physical picture. The scalar quark first propagates without any interaction with the background field in the transverse direction, from $y_\perp$ to $z_\perp$. It then propagates in the longitudinal direction to spatial infinity, acquiring the phase $U(\infty, y^-;z_\perp)$, see Fig.~\ref{fig:gif4}.

\begin{figure}[htb]
\begin{center}
\includegraphics[width=0.9\linewidth]{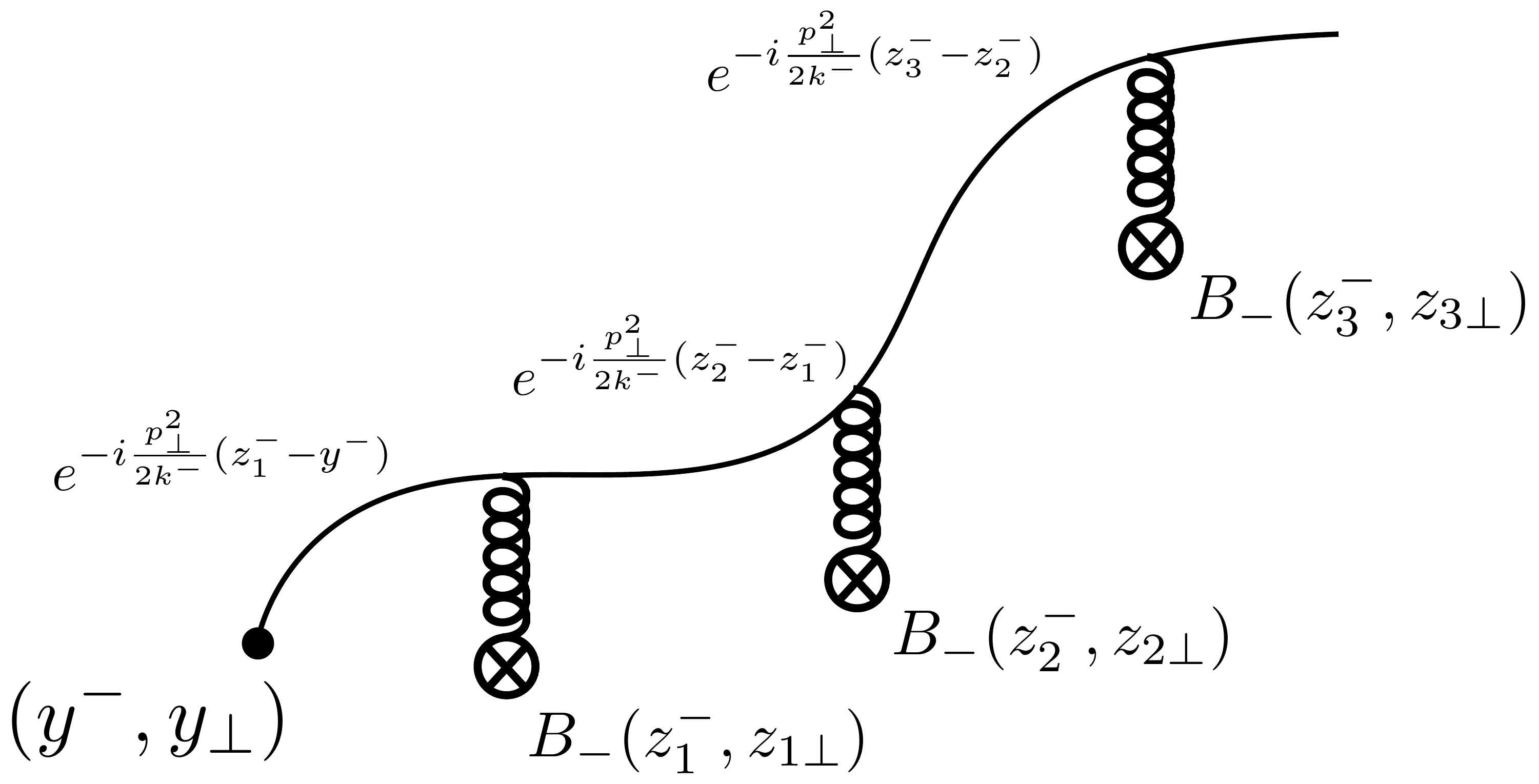}
 \end{center}
\caption{\label{fig:gif5}
In the scalar propagator~(\ref{eq:prmix-pB0}), the exponential factors of the path-ordered exponent describe the quark's shifts in the transverse direction.
}
 \end{figure}

During the quark's longitudinal propagation, its transverse position does not change. From this observation, the role of the exponential factors in the path-ordered exponent of Eq.~(\ref{eq:prmix-pB0}) becomes clear:
\begin{equation}
 \mathcal{P}\exp\Big\{ ig\int^{x^-}_{y^-} dz^- e^{i\frac{p^2_\perp}{2p^-}z^-} B_-(z^-) e^{-i\frac{p^2_\perp}{2p^-}z^-} \Big\}\,.
\label{eq:wl-hebB0}
\end{equation}
These factors account for the quark's shift in the transverse direction during its interaction with the background $B_-$ component, see Fig.~\ref{fig:gif5}. In this regard, Eq.~(\ref{eq:wl-hebB0}) should be understood as a generalization of the light-cone gauge factor~(\ref{eq:simple-wl}) that incorporates the quark's transverse propagation dynamics.

\begin{figure}[htb]
\begin{center}
\includegraphics[width=0.5\linewidth]{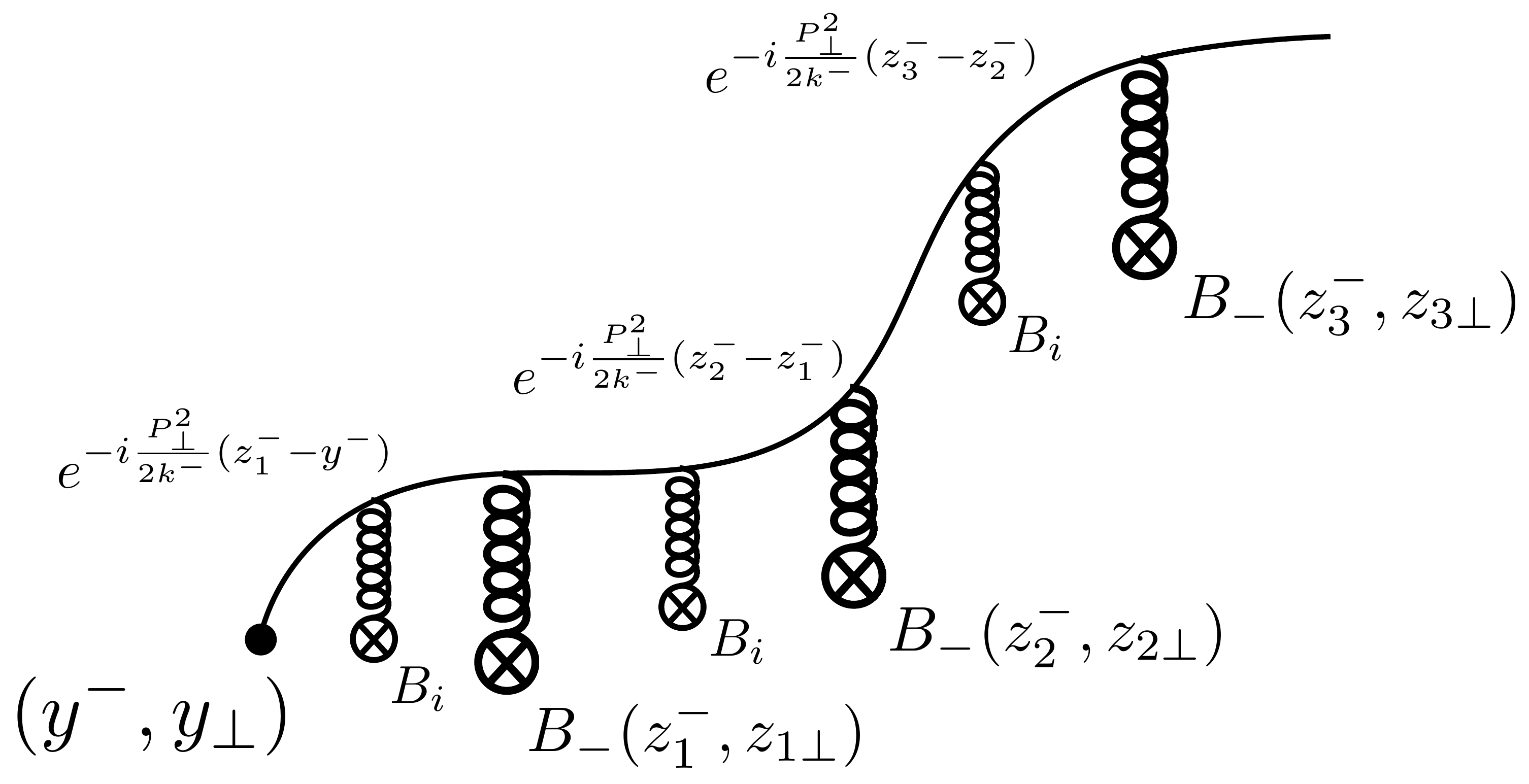}
 \end{center}
\caption{\label{fig:gif6}The path-ordered exponent representation of the scalar propagator~(\ref{eq:LSZ-fin-exp}) describes the quark's interactions with the $B_-$ field component, as well as with the transverse component $B_i$ through the exponential factors~(\ref{eq:exp-fact-cova}), encoding the quark's propagation in the transverse direction.}
 \end{figure}

We are now in a position to understand the full result for the scalar propagator~(\ref{eq:LSZ-fin}). It is a gauge-covariant generalization of Eq.~(\ref{eq:prmix-pB0}) that includes interactions with the transverse field component $B_i$. As we systematically derive, this generalization amounts to replacing the exponential factors~(\ref{eq:exp-fact-bare}) with their covariant counterparts\footnote{Equation~(\ref{eq:LSZ-fin}) was obtained by direct computation in full generality.}
\begin{equation}
(x_\perp| e^{i\frac{P^2_\perp}{2k^-}z^-} |y_\perp)\,.
\label{eq:exp-fact-cova}
\end{equation}
\begin{widetext}
Writing Eq.~(\ref{eq:LSZ-fin}) explicitly, we have
\begin{eqnarray}
&&\lim_{k^2\to0}k^2(k| \frac{1}{P^2 + i\epsilon}|y)\Big|_{k^->0} = \lim_{k^2\to0} (k_\perp| e^{i\frac{ k^2_\perp}{2k^-} z^-} e^{-i\frac{P^2_\perp(z^-) }{2k^-} z^-} \Big|^\infty 
 \label{eq:LSZ-fin-exp}\\
 &&\times \mathcal{P}\exp\Big\{ i\int^{\infty}_{y^-} dz^- \Big(e^{i\frac{ P^2_\perp(z^-)}{2p^-}z^-}iD_- e^{-i\frac{ P^2_\perp(z^-)}{2p^-}z^-} - \frac{P^2_\perp(z^-)}{2p^-}\Big) \Big\} e^{i\frac{P^2_\perp(y^-)}{2k^-}y^-} |y_\perp) e^{ik^- y^+} \,,
 \nonumber
\end{eqnarray}
c.f.\ Eq.~(\ref{eq:prmix-pB0}).

The path-ordered exponent
\begin{eqnarray}
&&\mathcal{S}(x^-, y^-) = \mathcal{P}\exp\Big\{ i\int^{x^-}_{y^-} dz^- \Big(e^{i\frac{ P^2_\perp(z^-)}{2p^-}z^-}iD_- e^{-i\frac{ P^2_\perp(z^-)}{2p^-}z^-} - \frac{P^2_\perp(z^-)}{2p^-}\Big) \Big\}
\end{eqnarray}
appearing in Eqs.~(\ref{eq:LSZ-fin}) and~(\ref{eq:LSZ-fin-exp}) generalizes the expression~(\ref{eq:wl-hebB0}) by incorporating interactions with the transverse background field component $B_i$. These interactions are acquired during the quark's propagation in the transverse direction, see Fig.~\ref{fig:gif6}.
\end{widetext}

In the next section, we explain how the exponential factors~(\ref{eq:exp-fact-cova}) give rise to the transverse gauge links, rendering the propagator~(\ref{eq:LSZ-fin}) fully gauge covariant --- an important aspect of our derivation.


\subsection{Transverse gauge link\label{sec:tr-link}}
The transverse exponential factor (\ref{eq:exp-fact-cova}) should be understood as exponentiation of a two-dimensional scalar propagator:
\begin{eqnarray}
&& (x_\perp|\frac{1}{P^2_\perp}|y_\perp) = \int^\infty_0 dT\: (x_\perp| e^{-T P^2_\perp} |y_\perp)\,.
\label{eq:scal-prop}
\end{eqnarray}
In the propagator~(\ref{eq:LSZ-fin-exp}), the exponentiation parameter $T$ corresponds to the longitudinal coordinate variable.

\begin{figure*}[htb]
\begin{center}
\includegraphics[width=0.8\textwidth]{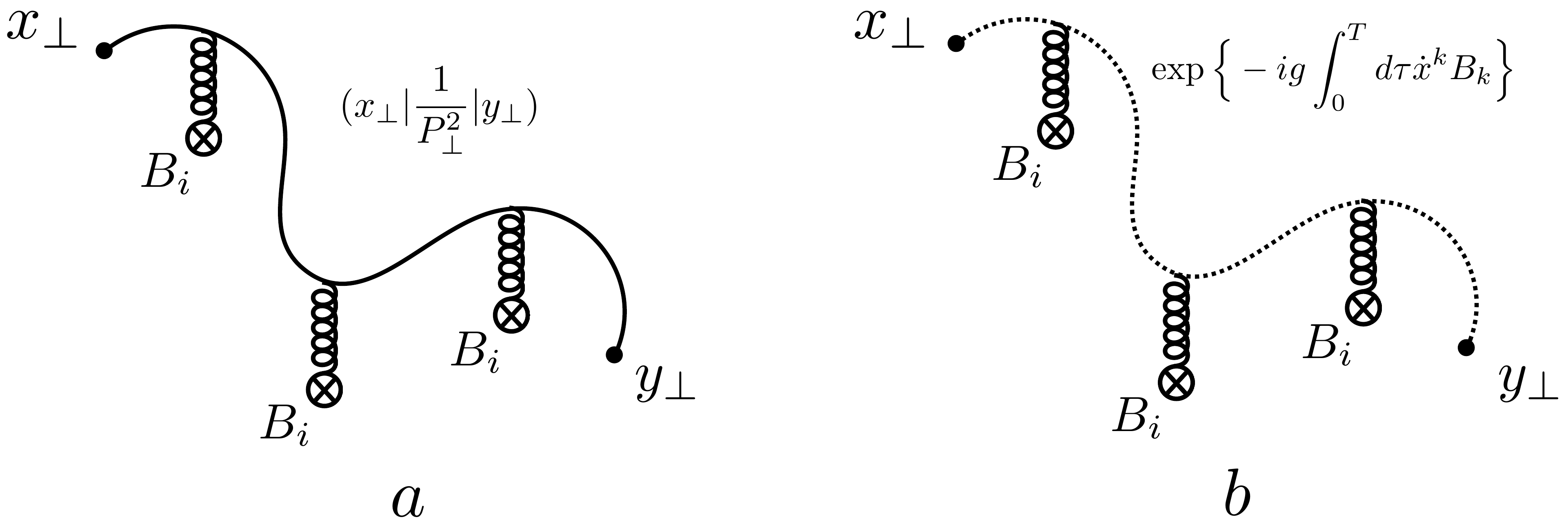}
 \end{center}
\caption{\label{fig:gif7}a) Transverse exponential factor (\ref{eq:exp-fact-cova}) describes quark's propagation between two end points in the transverse plane, see Eq. (\ref{eq:scal-prop-func}); b) As a result of this propagation, quark acquires a transverse scalar phase (\ref{eq:scal-prop-func-phase}).}
 \end{figure*}

The transverse propagator~(\ref{eq:scal-prop}) admits a functional integral representation:
\begin{eqnarray}
&& (x_\perp|\frac{1}{P^2_\perp}|y_\perp) 
= \int^\infty_0 dT  \int^{x(T)=x_\perp}_{x(0)=y_\perp} \mathcal{D} x(\tau)\nonumber\\
&&\times \exp\Big\{-\int^T_0 d\tau \Big(\frac{1}{4}\dot{x}^2 + ig \dot{x}^k B_k\Big)\Big\}
\label{eq:scal-prop-func}
\end{eqnarray}
From this representation, one sees that the transverse exponential factor~(\ref{eq:exp-fact-cova}) describes propagation of a scalar quark in the transverse plane between two endpoints, involving a sum over all possible trajectories, see Fig.~\ref{fig:gif7}a. The interaction with the background field produces a transverse scalar gauge phase
\begin{eqnarray}
&&\exp\Big\{- ig \int^T_0 d\tau \dot{x}^k B_k \Big\}
\label{eq:scal-prop-func-phase}
\end{eqnarray}
acquired by this quark during this propagation, see Fig. \ref{fig:gif7}b.

The gauge phase~(\ref{eq:scal-prop-func-phase}) is different for each trajectory in the transverse plane, making direct use of the functional integral representation~(\ref{eq:scal-prop-func}) quite difficult. The standard remedy is to expand the scalar phases onto a predefined contour, chosen based on the expansion parameter of the problem.

\begin{figure*}[htb]
\begin{center}
\includegraphics[width=0.8\textwidth]{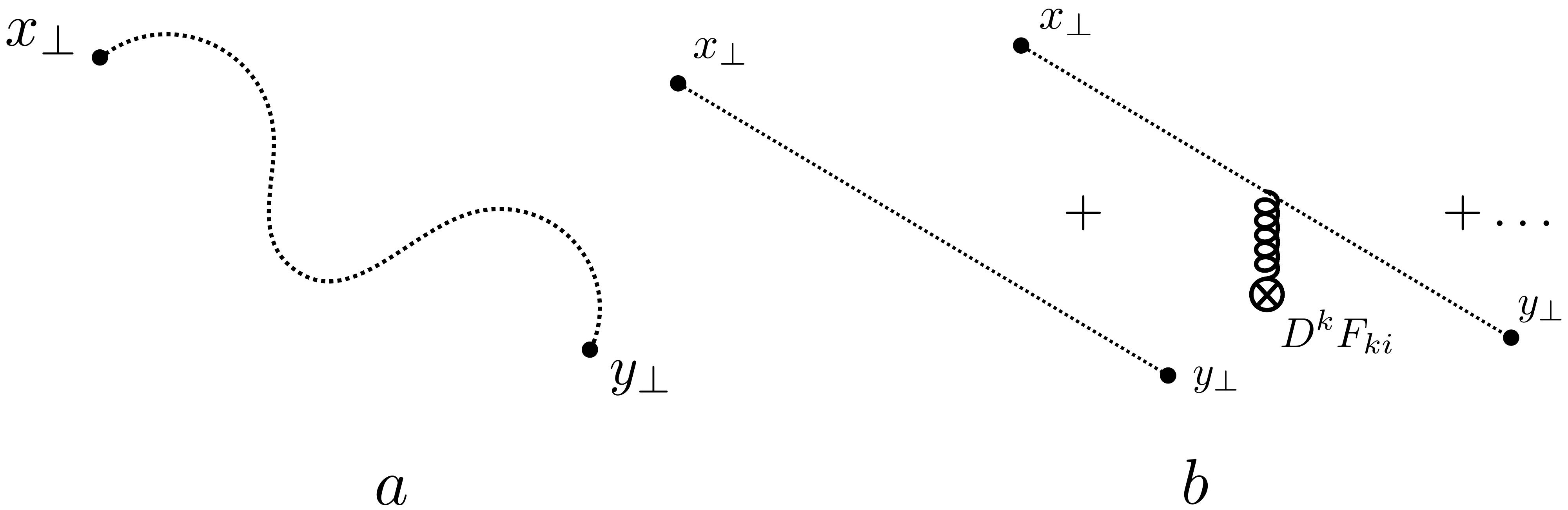}
 \end{center}
\caption{\label{fig:gif8}a) Transverse gauge phase for a scalar quark moving along an arbitrary trajectory in the transverse plane; b) The phase can be expanded onto a straight line connecting end points. The expansion yields an infinite sum of transverse gauge links with insertions of field-strength tensors and their covariant derivatives, see Eq. (\ref{eq:tr-WL-insertion}).}
 \end{figure*}

In our case, the transverse gauge phase~(\ref{eq:scal-prop-func-phase}) can be expanded onto a straight line connecting the endpoints in the transverse plane, see Fig.~\ref{fig:gif8}. Methods for constructing such an expansion are available, see e.g.\ Refs.~\cite{Balitsky:1987bk,Balitsky:2015qba}.

Consider the transverse exponential factor~(\ref{eq:exp-fact-cova}), expanded as
\begin{widetext}
\begin{eqnarray}
&&(x_\perp|e^{is P^2_\perp}|y_\perp) = (x_\perp|e^{is (p^2_\perp + g\{p_k, B_k\} + g^2B_k B_k)}|y) 
\label{eq:trglink-expansion-init}\\
&&= (x_\perp|e^{is p^2_\perp }|y_\perp) + is \int^1_0 du (x_\perp|e^{is \bar{u}p^2_\perp }( g\{p_k, B_k\} + g^2B_k B_k)  e^{isup^2_\perp}|y_\perp) \nonumber\\
&&+ (is)^2 \int^1_0 du \int^u_0 dv (x_\perp|e^{is \bar{u}p^2_\perp }( g\{p_k, B_k\} + g^2B_k B_k) 
e^{is(u-v)p^2_\perp} ( g\{p_m, B_m\} + g^2B_m B_m) e^{isv p^2_\perp}|y_\perp) + \dots\,,
\nonumber
\end{eqnarray}
where $s\equiv z^-/(2k^-)$.

We aim to expand all transverse background fields onto a straight line parametrized as
\begin{eqnarray}
\xi^k_u = u x^k + \bar{u} y^k\,.
\label{eq:lin-param}
\end{eqnarray}
The background fields in Eq. (\ref{eq:trglink-expansion-init}) can be formally expanded onto this path:
\begin{eqnarray}
&&B_k = B_k(\xi_u) + (X - \xi_u)_m\partial^m B_k(\xi_u) + \frac{1}{2}(X-\xi_u)_m(X-\xi_u)_n \partial^m\partial^n B_k(\xi_u) + \dots\,,
\label{eq:str-line-exp}
\end{eqnarray}
where $X$ is the coordinate operator, $X_m|x_\perp) = x_m|x_\perp)$, satisfying $[p_m, X_n] = ig_{mn}$.

For brevity, we consider only the first two terms in Eq.~(\ref{eq:trglink-expansion-init}), containing a single gluon insertion:
\begin{eqnarray}
&&(x_\perp|e^{is p^2_\perp }|y_\perp) + igs \int^1_0 du (x_\perp|e^{is \bar{u}p^2_\perp } \{p_k, B_k\} e^{isup^2_\perp}|y_\perp) \,.
\end{eqnarray}
The methods outlined below, however, apply equally to the higher-order terms in Eq.~(\ref{eq:trglink-expansion-init}) and yield the complete gauge-covariant operator content of the expansion onto the straight line~(\ref{eq:str-line-exp}).


Retaining the leading term of Eq.~(\ref{eq:str-line-exp}), we find
\begin{eqnarray}
&&(x_\perp|e^{is p^2_\perp }|y_\perp) + igs \int^1_0 du (x_\perp|e^{is \bar{u}p^2_\perp } \{p_k, B_k\} e^{isup^2_\perp}|y_\perp)\nonumber\\
&&= (x_\perp|e^{is p^2_\perp }|y_\perp) 
+ 2igs \int^1_0 du (x_\perp|p_k e^{is p^2_\perp }  |y_\perp)B_k(\xi_u) + \dots\,,
\end{eqnarray}
where the ellipsis denotes the higher-order terms of Eq.~(\ref{eq:str-line-exp}). This gives
\begin{equation}
(x_\perp|e^{is p^2_\perp }|y_\perp) + igs \int^1_0 du (x_\perp|e^{is \bar{u}p^2_\perp } \{p_k, B_k\} e^{isup^2_\perp}|y_\perp)
= (x_\perp|e^{is p^2_\perp }|y_\perp) \Big(1 + i g\int^1_0 du (x-y)^k   B_k(\xi_u) \Big)  + \dots\,.
\label{eq:lead-trglink}
\end{equation}
Equation~(\ref{eq:lead-trglink}) corresponds to the leading term of the transverse gauge link along the straight line~(\ref{eq:lin-param}):
\begin{eqnarray}
&&U(x_\perp, y_\perp) = \mathcal{P}\exp\Big(ig\int^1_0 du (x-y)^k B_k(\xi_u)\Big)\,.
\label{eq:transv-WL}
\end{eqnarray}
This identification is confirmed by computing the higher-order terms of Eqs.~(\ref{eq:trglink-expansion-init}) and~(\ref{eq:str-line-exp}). Indeed, one obtains
\begin{eqnarray}
&&(x_\perp|e^{isP^2_\perp}|y_\perp) = (x_\perp|e^{isp^2_\perp}|y_\perp)\Big\{U(x_\perp, y_\perp) + s\int^1_0 du\  u \bar{u}\  (x-y)^k U(x_\perp, \xi_u)D_m F_{m k}(\xi_u)U(\xi_u, y_\perp) \label{eq:tr-WL-insertion}\\
&& 
+ 2is\int^1_0 du\ \bar{u}\int^u_0 dv\ v\  (x-y)^k (x-y)^m U(x_\perp, \xi_u)F_{ks}(\xi_u)U(\xi_u, \xi_v)F_{ms}(\xi_v)U(\xi_v, y_\perp) + \dots \Big\}\,,
\nonumber
\end{eqnarray}
where the ellipsis denotes higher-order terms with insertions of transverse field-strength tensors and their covariant derivatives.

We thus find that the exponential factor~(\ref{eq:exp-fact-cova}), in addition to the ``bare'' factor~(\ref{eq:exp-fact-bare}) describing the quark's propagation in the transverse plane, contains a transverse gauge phase acquired during this propagation. According to Eq.~(\ref{eq:tr-WL-insertion}), this phase can be represented as an infinite sum of transverse gauge links~(\ref{eq:transv-WL}) with insertions of transverse field-strength tensors $F_{ij}$ and their covariant derivatives.


The content of the scalar propagator~(\ref{eq:LSZ-fin}) is now clear: it describes the quark's propagation in both the longitudinal and transverse directions, together with the corresponding gauge phases acquired along the way.


In the subsequent discussion, we will also need the expansion onto the straight line~(\ref{eq:lin-param}) of the following operator combination:
\begin{eqnarray}
&&iD^x_i(x_\perp|e^{is P^2_\perp}|y_\perp) = (x_\perp|P_ie^{is P^2_\perp}|y_\perp) 
\nonumber\\
&&= - \frac{(x-y)_i}{2s} (x_\perp| e^{i s p^2_\perp} |y_\perp) U(x_\perp, y_\perp)+ (x_\perp| e^{i s p^2_\perp} |y_\perp)
\int^1_0 du  u (x-y)^k U(x_\perp, \xi_u)F_{ki}(\xi_u)U(\xi_u, y_\perp) + \dots\,,
\label{eq:tr-WL-insertionPileft}
\end{eqnarray}
where the ellipsis denotes higher-order terms with insertions of transverse field-strength tensors and their covariant derivatives. For our purposes, the explicit form of these terms is not needed, though the methods outlined above can be used to derive them systematically to any given order of the expansion~(\ref{eq:str-line-exp}).


Although the scalar propagator~(\ref{eq:LSZ-fin}) is completely general for the background field~(\ref{eq:bfield-form}), its direct application is not straightforward, as it contains an infinite number of operator contributions involving longitudinal and transverse gauge links. In the next section, we develop an efficient approach for organizing these contributions, allowing us to systematically derive a hierarchy of operators characterizing dijet production in different kinematic limits.

Before developing this technique, however, let us generalize the scalar propagator~(\ref{eq:LSZ-fin}) to the full quark propagator~(\ref{eq:prop-def}) and use it to derive a general formula for the dijet production cross section~(\ref{eq:hadronic-bfm-LO}).



\subsection{Quark propagator in the background field\label{sec:quark-prop}}
To derive the (anti)quark propagator~(\ref{eq:prop-def}) in a form analogous to the scalar result~(\ref{eq:LSZ-fin}), we rewrite it as
\begin{eqnarray}
&&\lim_{k^2\to0}k^2(k| \frac{1}{P^2 + \frac{g}{2}\sigma^{\mu\nu}F_{\mu\nu} + i\epsilon}|y)\label{eq:qprop-init}\\
&&= \lim_{k^2\to0}k^2 \int d^4z  e^{ikz}\frac{1}{2k^-}(z|\frac{1}{p^+ + g( B_- + \frac{1}{4p^-}\sigma^{\mu\nu}F_{\mu\nu}) - \frac{P^2_\perp}{2p^-} + i\epsilon p^-}|y)\,.
\nonumber
\end{eqnarray}

Comparing Eq.~(\ref{eq:qprop-init}) with the corresponding expression for the scalar propagator~(\ref{eq:scal-prop-first}), we see that the desired form can be obtained by the replacement
\begin{eqnarray}
&&B_- \to B_- + \frac{1}{4p^-}\sigma^{\mu\nu}F_{\mu\nu}
\end{eqnarray}
in Eq. (\ref{eq:LSZ-fin}). The result reads
\begin{eqnarray}
&&\lim_{k^2\to0}k^2(k| \frac{1}{P^2 + \frac{g}{2}\sigma^{\mu\nu}F_{\mu\nu}+ i\epsilon}|y) = \lim_{k^2\to0} (k_\perp| \Big(  \theta(k^-) e^{i\frac{ k^2_\perp}{2k^-} z^-} e^{-i\frac{P^2_\perp(z^-) }{2k^-} z^-} \Big|^\infty \mathcal{G}(\infty, y^-)
 \nonumber\\
 &&+ \theta(-k^-) e^{i\frac{ k^2_\perp}{2k^-} z^-} e^{-i\frac{P^2_\perp(z^-) }{2k^-} z^-} \Big|^{-\infty} \mathcal{G}(-\infty, y^-) \Big) e^{i\frac{P^2_\perp(y^-)}{2k^-}y^-} |y_\perp) e^{ik^- y^+} \,,
 \label{eq:LSZ-fin-quark}
\end{eqnarray}
where the path-ordered exponent is 
\begin{eqnarray}
&&\mathcal{G}(x^-, y^-) = \mathcal{G}(x^-, y^-;0^-)\,,
\label{eq:quark-exponent}
\end{eqnarray}
with a ``shifted" path-ordered exponent defined as 
\begin{eqnarray}
&&\mathcal{G}(x^-, y^-;\xi^-)\label{eq:definition-exp-shifted-quark}\\
&&\equiv \mathcal{P}\exp\Big\{ i\int^{x^-}_{y^-} dz^- \Big(e^{i\frac{ P^2_\perp(z^-)}{2p^-}(z^--\xi^-)}\Big(iD_- + \frac{g}{4p^-}\sigma^{\mu\nu}F_{\mu\nu}\Big) e^{-i\frac{ P^2_\perp(z^-)}{2p^-}(z^--\xi^-)} - \frac{P^2_\perp(z^-)}{2p^-}\Big) \Big\}\,,
\nonumber
\end{eqnarray}
c.f. Eq. (\ref{eq:definition-exp-shifted}).

The path-ordered exponent representation~(\ref{eq:LSZ-fin-quark}) of the (anti)quark propagator coincides with that of the scalar propagator~(\ref{eq:LSZ-fin}), up to insertions of
\begin{eqnarray}
&&\frac{g}{4p^-}\sigma^{\mu\nu}F_{\mu\nu}
\label{eq:strength-tensorF-vert}
\end{eqnarray}
into the gauge factors. As discussed in Sec.~\ref{sec:cross-section}, these field-strength tensor insertions describe the interaction of the (anti)quark with the background field via Pauli vertices.


Equation~(\ref{eq:LSZ-fin-quark}) represents the general form of the (anti)quark propagator. Analogously to Eq.~(\ref{eq:parallel-shift-scalar-full}), the path-ordered exponent~(\ref{eq:definition-exp-shifted-quark}) satisfies the identity
\begin{eqnarray}
&&e^{-i\frac{ P^2_\perp(x^-)}{2p^-}x^- }\mathcal{G}(x^-, y^-) e^{i\frac{P^2_\perp(y^-)}{2p^-}y^-} = e^{-i\frac{ P^2_\perp(x^-)}{2p^-}(x^--\xi^-) }\mathcal{G}(x^-, y^-;\xi^-) e^{i\frac{P^2_\perp(y^-)}{2p^-}(y^--\xi^-)}\,.
\label{eq:parallel-shift-scalar-full-quark}
\end{eqnarray}
We will use this identity later to organize the operators in Eq.~(\ref{eq:LSZ-fin-quark}) by expanding them onto an arbitrary linear piecewise contour.


Finally, let us list the remaining propagators appearing in the dijet production cross section~(\ref{eq:hadronic-bfm-LO}):
\begin{eqnarray}
&&\lim_{k^2\to0}(y| \frac{1}{P^2 + \frac{g}{2}\sigma^{\mu\nu}F_{\mu\nu}+ i\epsilon}|k)k^2 = \lim_{k^2\to0} e^{-ik^- y^+} (y_\perp| e^{-i\frac{P^2_\perp(y^-)}{2k^-}y^-} \label{eq:LSZ-fin-quark-second}\\
&&\times \Big(   \mathcal{G}(y^-, -\infty) e^{i\frac{P^2_\perp(z^-) }{2k^-} z^-} e^{-i\frac{ k^2_\perp}{2k^-} z^-} \Big|^{-\infty} \theta(k^-)
 + \mathcal{G}(y^-, \infty) e^{i\frac{P^2_\perp(z^-) }{2k^-} z^-}  e^{-i\frac{ k^2_\perp}{2k^-} z^-} \Big|^\infty \theta(-k^-) \Big) |k_\perp) \,.\nonumber
\end{eqnarray}
We will also need the propagators for the anti-time-ordered products:
\begin{eqnarray}
&&\lim_{k^2\to0}k^2(k| \frac{1}{\tilde{P}^2 + \frac{g}{2}\sigma^{\mu\nu}\tilde{F}_{\mu\nu}- i\epsilon}|x) = \lim_{k^2\to0} (k_\perp| \Big(  \theta(k^-) e^{i\frac{ k^2_\perp}{2k^-} z^-} e^{-i\frac{\tilde{P}^2_\perp(z^-) }{2k^-} z^-} \Big|^{-\infty} \tilde{\mathcal{G}}(-\infty, x^-)
 \nonumber\\
 &&+ \theta(-k^-) e^{i\frac{ k^2_\perp}{2k^-} z^-} e^{-i\frac{\tilde{P}^2_\perp(z^-) }{2k^-} z^-} \Big|^{\infty} \tilde{\mathcal{G}}(\infty, x^-) \Big) e^{i\frac{\tilde{P}^2_\perp(x^-)}{2k^-}x^-} |x_\perp) e^{ik^- x^+}
 \label{eq:proptilde-1}
\end{eqnarray}
and
\begin{eqnarray}
&&\lim_{k^2\to0}(x| \frac{1}{\tilde{P}^2 + \frac{g}{2}\sigma^{\mu\nu}\tilde{F}_{\mu\nu}- i\epsilon}|k)k^2 = \lim_{k^2\to0} e^{-ik^- x^+} (x_\perp| e^{-i\frac{\tilde{P}^2_\perp(x^-)}{2k^-}x^-}\label{eq:proptilde-2}\\
&&\times \Big(   \tilde{\mathcal{G}}(x^-, \infty) e^{i\frac{\tilde{P}^2_\perp(z^-) }{2k^-} z^-} e^{-i\frac{ k^2_\perp}{2k^-} z^-} \Big|^{\infty} \theta(k^-)
+ \tilde{\mathcal{G}}(x^-, -\infty) e^{i\frac{\tilde{P}^2_\perp(z^-) }{2k^-} z^-}  e^{-i\frac{ k^2_\perp}{2k^-} z^-} \Big|^{-\infty }\theta(-k^-) \Big) |k_\perp) \,,
\nonumber
\end{eqnarray}
where $\tilde{\mathcal{G}}(x^-, y^-)$ is defined identically to $\mathcal{G}(x^-, y^-)$, see Eqs.~(\ref{eq:quark-exponent}) and~(\ref{eq:definition-exp-shifted-quark}), with the replacements $B_-\to\tilde{B}_-$ and $F_{\mu\nu}\to\tilde{F}_{\mu\nu}$.
\end{widetext}
The tilde notation in Eqs.~(\ref{eq:proptilde-1}) and~(\ref{eq:proptilde-2}) indicates that the background fields $\tilde{B}_\mu$ belong to the conjugate functional integral in Eq.~(\ref{eq:hadronic-func-split}), giving rise to the anti-time ordering of operators. For brevity, in the following we omit the tilde notation.
This simplification is justified because the operators constructed from the $B_\mu$ and $\tilde{B}_\mu$ fields in our final expressions involve separations that are either space-like or light-like. In both cases, all operators commute with each other, so the distinction between the two types of time ordering is immaterial and the tilde notation can be dropped.


With the path-ordered exponent representations of the (anti)\-quark pro\-pa\-gators,  Eqs.~(\ref{eq:LSZ-fin-quark}), (\ref{eq:LSZ-fin-quark-second}), (\ref{eq:proptilde-1}), and~(\ref{eq:proptilde-2}), we are now in a position to write down a general expression for the dijet production cross section~(\ref{eq:hadronic-bfm-LO}).


\begin{widetext}
\section{General result for the dijet production cross section\label{sec:cross-section-gen-sec}}
Substituting (anti)quark propagators (\ref{eq:LSZ-fin-quark}), (\ref{eq:LSZ-fin-quark-second}), (\ref{eq:proptilde-1}), and (\ref{eq:proptilde-2}) into Eq. (\ref{eq:hadronic-bfm-LO}) we obtain
\begin{eqnarray}
&&d\sigma^{\gamma^\ast p}_{\lambda\lambda'} = \sum_f\sum_{s_1,s_2}\frac{2\pi e^2_f \alpha_{\rm EM}x}{V_4Q^2}   \epsilon^{\lambda\ast}_\mu \epsilon^{\lambda'}_\nu \frac{dk^-_1d^2k_{1\perp}}{(2\pi)^3 2k^-_1} \Big|_{k^2_1=0}\frac{dk^-_2d^2k_{2\perp}}{(2\pi)^3 2k^-_2} \Big|_{k^2_2=0}
\label{eq:hadronic-bfm-general}\\
&&\times \int d^4x e^{iqx}  \langle P, S| \Big\{ \bar{v}_{s_2}(k_2) (-k_{2\perp}| e^{-i\frac{ k^2_{2\perp}}{2k^-_2} z^-} e^{i\frac{P^2_\perp(z^-) }{2k^-_2} z^-} \Big|^{\infty} \mathcal{G}(\infty, x^-)  e^{-i\frac{P^2_\perp(x^-)}{2k^-_2}x^-} |x_\perp) e^{-ik^-_2 x^+} \gamma^\mu 
\nonumber\\
&&\times e^{-ik^-_1 x^+} (x_\perp| e^{-i\frac{P^2_\perp(x^-)}{2k^-_1}x^-}   \mathcal{G}(x^-, \infty) e^{i\frac{P^2_\perp(z^-) }{2k^-_1} z^-} e^{-i\frac{ k^2_{1\perp}}{2k^-_1} z^-} \Big|^{\infty} |k_{1\perp}) u_{s_1}(k_1) - (B=0)\Big\}
\nonumber\\
&&\times \int d^4y e^{-iqy} \Big\{\bar{u}_{s_1}(k_1) (k_{1\perp}|  e^{i\frac{ k^2_{1\perp}}{2k^-_1} z^-} e^{-i\frac{P^2_\perp(z^-) }{2k^-_1} z^-} \Big|^\infty \mathcal{G}(\infty, y^-) e^{i\frac{P^2_\perp(y^-)}{2k^-_1}y^-} |y_\perp) e^{ik^-_1 y^+}\gamma^\nu 
\nonumber\\
&&\times e^{ik^-_2 y^+} (y_\perp| e^{i\frac{P^2_\perp(y^-)}{2k^-_2}y^-} \mathcal{G}(y^-, \infty) e^{-i\frac{P^2_\perp(z^-) }{2k^-_2} z^-}  e^{i\frac{ k^2_{2\perp}}{2k^-_2} z^-} \Big|^\infty |-k_{2\perp}) v_{s_2}(k_2) - (B=0)\Big\} |P, S\rangle \,,
\nonumber
\end{eqnarray}
where the quark and antiquark dijet momenta satisfy $k^-_1>0$ and $k^-_2>0$. Equation~(\ref{eq:hadronic-bfm-general}) represents the most general form of the dijet production cross section in the background field~(\ref{eq:bfield-form}). It describes the splitting of a virtual photon into a quark-antiquark pair that subsequently propagates in the background field $B_\mu$ of the target into the final dijet state, see Fig.~\ref{fig:fig9}a.
\begin{figure}[htb]
\begin{center}
\includegraphics[width=0.8\textwidth]{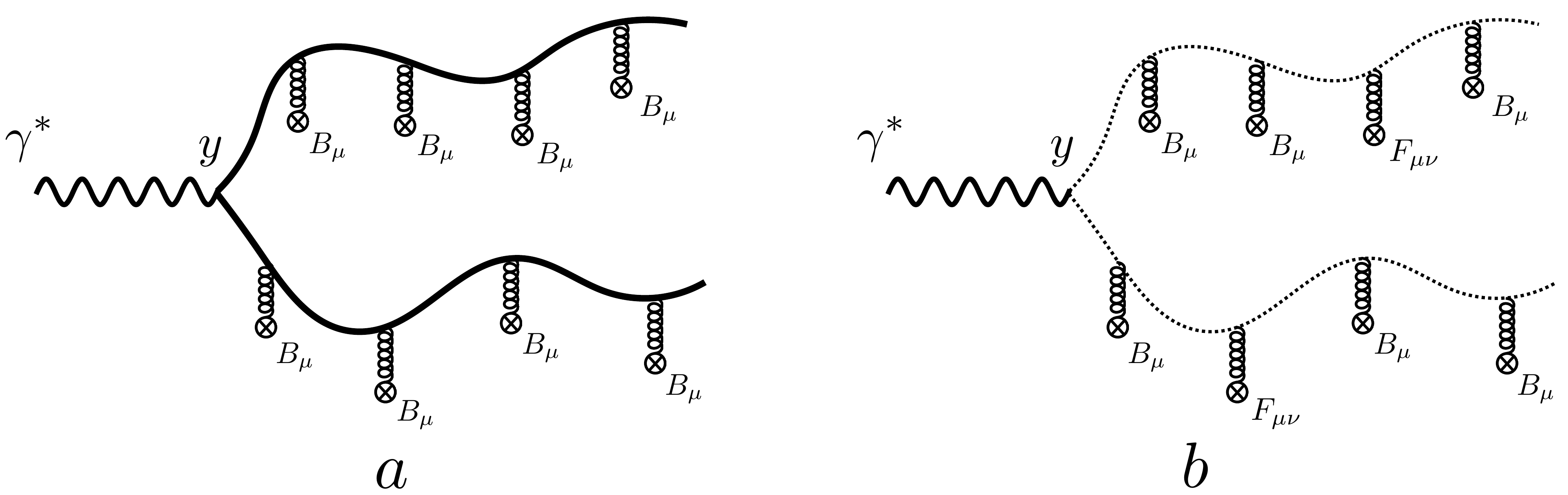}
 \end{center}
 \caption{\label{fig:fig9}a) The virtual photon annihilates into a quark-antiquark pair propagating in the background field of the target; b) the pair acquires a gauge phase constructed from the background fields of the target. The phase runs along an arbitrary trajectory and contains insertions of the background field-strength tensors.}
 \end{figure}

It is convenient to separate from Eq.~(\ref{eq:hadronic-bfm-general}) the virtual photon scattering amplitude on the target:
\begin{eqnarray}
&&iM^\nu_{s_1,s_2}(q, k_1, k_2)\label{eq:amp-gen-def}\\
&&\equiv \int d^4y e^{-iqy} \Big\{\bar{u}_{s_1}(k_1) (k_{1\perp}|  e^{i\frac{ k^2_{1\perp}}{2k^-_1} z^-} e^{-i\frac{P^2_\perp(z^-) }{2k^-_1} z^-} \Big|^\infty \mathcal{G}(\infty, y^-) e^{i\frac{P^2_\perp(y^-)}{2k^-_1}y^-} |y_\perp)\nonumber\\
&&\times e^{ik^-_1 y^+}\gamma^\nu 
e^{ik^-_2 y^+} (y_\perp| e^{i\frac{P^2_\perp(y^-)}{2k^-_2}y^-} \mathcal{G}(y^-, \infty) e^{-i\frac{P^2_\perp(z^-) }{2k^-_2} z^-}  e^{i\frac{ k^2_{2\perp}}{2k^-_2} z^-} \Big|^\infty |-k_{2\perp}) v_{s_2}(k_2) - (B=0)\Big\}\,,
\nonumber
\end{eqnarray}
and similarly for the complex-conjugated amplitude.
\end{widetext}

Equation~(\ref{eq:amp-gen-def}) describes propagation of the quark-antiquark pair in both the longitudinal and transverse directions, and defines the corresponding gauge factors constructed from the $B_\mu$ background fields. In general, these gauge factors run along an arbitrary trajectory of the (anti)quark, see Fig.~\ref{fig:fig9}b. Equation~(\ref{eq:amp-gen-def}) involves a sum over all possible trajectories for the pair propagating in the background field.

\begin{figure*}[htb]
\begin{center}
\includegraphics[width=0.8\textwidth]{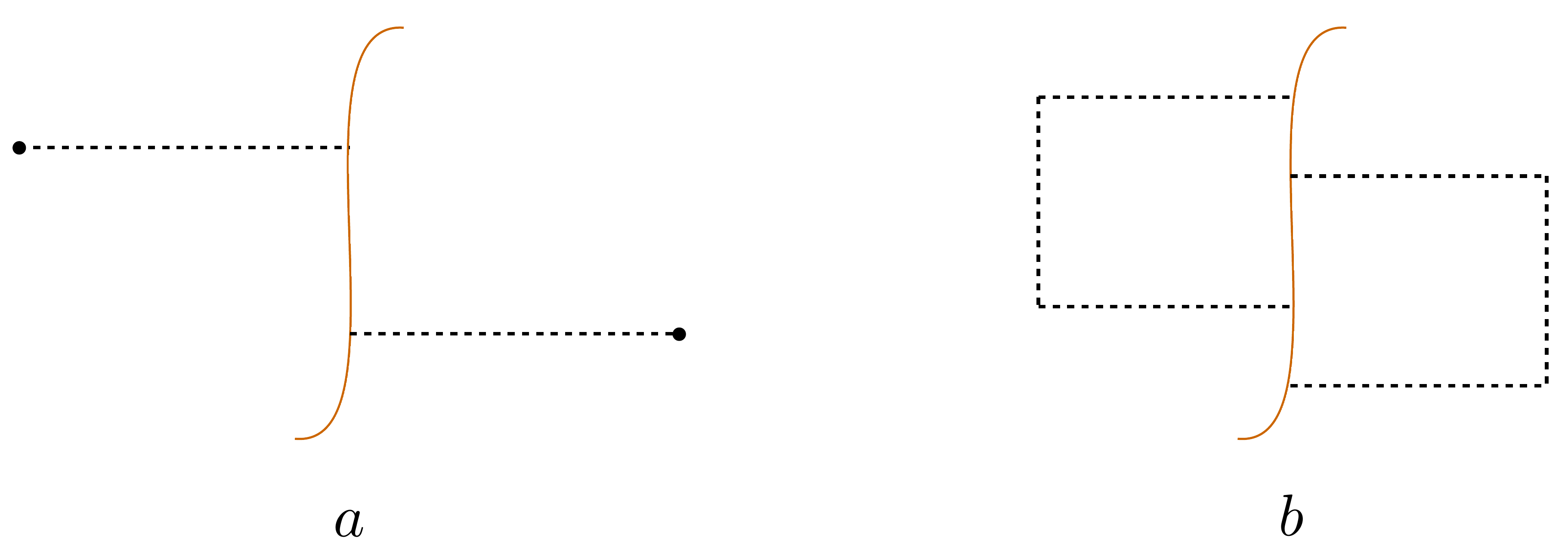}
 \end{center}
 \caption{\label{fig:fig10}a) In the back-to-back limit of dijet production, the operator hierarchy can be derived by expanding the path-ordered exponents onto a contour of TMD operators; b) for dijet production at small $x$, the corresponding hierarchy is obtained by expanding onto a quadrupole contour.}
 \end{figure*}

The operators in Eq.~(\ref{eq:hadronic-bfm-general}) fully encode the interactions between the quantum partons and the background fields of the target. As a result, Eq.~(\ref{eq:hadronic-bfm-general}) can be used to compute dijet production in different kinematic regimes.\footnote{The methods developed in this paper extend beyond dijet production and can be applied to other physical observables.} Such a computation involves deriving the particular hierarchy of operators defining the scattering --- for instance, TMD operators in back-to-back dijet production, or quadrupole operators in the small-$x$ limit. Crucially, in our approach the operators in Eq.~(\ref{eq:hadronic-bfm-general}) are gauge covariant, making the derivation of the corresponding operator hierarchies straightforward. As we explain in the next section, these hierarchies can be obtained by expanding the path-ordered exponents~(\ref{eq:quark-exponent}) in Eq.~(\ref{eq:hadronic-bfm-general}) onto a fixed contour determined by the kinematics of the problem. For instance, in the back-to-back configuration the contour is that of TMD operators, see Fig.~\ref{fig:fig10}a, while in the small-$x$ limit it assumes a quadrupole form, see Fig.~\ref{fig:fig10}b.

The choice of expansion contour is an important element of our analysis, as it must be consistent with the expansion parameter of the problem. For example, expanding the path-ordered exponents in Eq.~(\ref{eq:hadronic-bfm-general}) onto the TMD contour of Fig.~\ref{fig:fig10}a in the back-to-back regime, one finds that the first few terms of the expansion provide the dominant contribution to dijet production, while the higher-order terms --- systematically derivable from Eq.~(\ref{eq:hadronic-bfm-general}) --- are suppressed by a small expansion parameter (twist suppression). In the next section, we present an algorithm for constructing such expansions, allowing one to systematically derive, order by order, a complete set of operators defining the scattering in a given kinematic regime.


Finally, we emphasize that the expansion procedure for the path-ordered exponents, developed in the next section, is completely general and exact. It is a purely formal operation used to derive the operator hierarchies defining the scattering. In this sense, two different expansions of the path-ordered exponents onto two different contours in coordinate space are mathematically equivalent. Our approach therefore enables a systematic re-expansion from one contour choice to another, providing a dictionary between different operator hierarchies in different kinematic regimes. This makes it possible to match operators and compare results obtained in different kinematic approximations. Examples of such matching are given later in this paper.


\section{Expansion of the path-ordered exponents\label{sec:gen-exp}}
In the previous section, we computed the dijet production cross section, Eq.~(\ref{eq:hadronic-bfm-general}), using the general form of the (anti)quark propagators in the background field, Eq.~(\ref{eq:LSZ-fin-quark}). The same approach can be applied to other physical observables computed within QCD factorization. In these propagators, interactions between the quantum partons and the background field are described by the path-ordered exponents~(\ref{eq:definition-exp-shifted-quark}). While the obtained results are completely general, the path-ordered exponents are not straightforward to evaluate.

\begin{figure*}[htb]
\begin{center}
\includegraphics[width=0.9\textwidth]{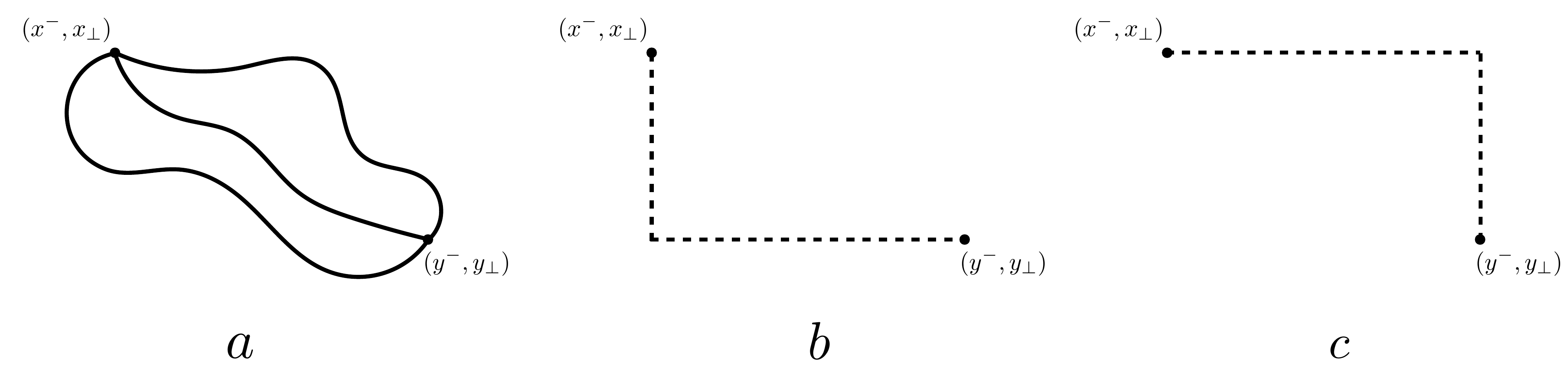}
 \end{center}
 \caption{\label{fig:fig11}a) The path-ordered exponents~(\ref{eq:definition-exp-shifted-quark}) in the (anti)quark propagator~(\ref{eq:LSZ-fin-quark}) involve a sum over all possible quark trajectories, each yielding a distinct gauge factor; b) the path-ordered exponent~(\ref{eq:exp-ex1-init}) can be expanded onto the light-cone direction at the transverse position $y_\perp$; c) alternatively, the same exponent can be expanded onto the light-cone direction at the transverse position $x_\perp$.}
 \end{figure*}

These exponents describe propagation of quantum partons between two points in the background field along an infinite set of arbitrary trajectories, see Fig.~\ref{fig:fig11}a. Each trajectory, determined by the quantum parton dynamics, gives rise to a gauge link connecting the endpoints with an arbitrary number of operator insertions~(\ref{eq:strength-tensorF-vert}). The path of these gauge links is not fixed but is dictated by the trajectory of the quantum parton. The path-ordered exponents~(\ref{eq:definition-exp-shifted-quark}) involve a sum over all possible trajectories, so they entangle the dynamics of the quantum partons with the background field interactions encoded in the gauge factors.

For physical observables, however, we generally aim to derive a factorized form in which the quantum parton dynamics and the background field interactions are separated. In the factorization formula, the former corresponds to a perturbative coefficient, while the latter is given by a (generically non-perturbative) matrix element of background field operators. We therefore need to disentangle these two elements within the path-ordered exponents~(\ref{eq:definition-exp-shifted-quark}).


An efficient way to achieve this is to expand the path-ordered exponents onto a fixed linear piecewise contour, see e.g.\ Fig.~\ref{fig:fig10}. Such an expansion fixes the path of the gauge links independently of the quantum parton dynamics, yielding the desired factorized form. In this section, we develop a general approach for expanding the path-ordered exponents~(\ref{eq:quark-exponent}) onto an arbitrary linear piecewise contour.


In the next two subsections, we consider a path-ordered exponent~(\ref{eq:quark-exponent}) between two points $(x^-,x_\perp)$ and $(y^-,y_\perp)$, represented schematically in Fig.~\ref{fig:fig11}a:
\begin{widetext}
\begin{eqnarray}
&&(x_\perp|\mathcal{G}(x^-, y^-) |y_\perp)
\label{eq:exp-ex1-init}\\
&&= (x_\perp|\mathcal{P}\exp\Big\{ i\int^{x^-}_{y^-} dz^- \Big(e^{i\frac{ P^2_\perp(z^-)}{2p^-}z^-}\Big(iD_- + \frac{g}{4p^-}\sigma^{\mu\nu}F_{\mu\nu}\Big) e^{-i\frac{ P^2_\perp(z^-)}{2p^-}z^-} - \frac{P^2_\perp(z^-)}{2p^-}\Big) \Big\}|y_\perp)\,,
\nonumber
\end{eqnarray}
\end{widetext}
and construct its expansion onto the light-cone direction at a fixed transverse position $y_\perp$, see Fig.~\ref{fig:fig11}b. We begin with the simpler case $B_i=0$ and then generalize to $B_i\neq 0$.


Subsequently, we explain how the transverse position of the path-ordered exponents can be shifted. Combined with the light-cone expansion, this provides a general method for expanding the path-ordered exponents~(\ref{eq:quark-exponent}) onto an arbitrary linear piecewise contour.

\begin{widetext}
\subsection{Expansion onto the light-cone direction. Simple case of $B_i=0$}
We aim to expand the path-ordered exponent in Eq.~(\ref{eq:exp-ex1-init}) onto the light-cone direction at the transverse position $y_\perp$, see Fig.~\ref{fig:fig11}b. Setting the transverse background field component to zero, $B_i=0$, we first expand the path-ordered exponent in Eq.~(\ref{eq:exp-ex1-init}):
\begin{eqnarray}
&&\mathcal{G}(x^-, y^-)\Big|_{B_i=0} 
= \mathcal{P}\exp\Big\{ i\int^{x^-}_{y^-} dz^- e^{i\frac{ p^2_\perp}{2p^-}z^-}\Big(gB_-(z^-) + \frac{g}{4p^-}\sigma^{\mu\nu}F_{\mu\nu}(z^-)\Big) e^{-i\frac{ p^2_\perp}{2p^-}z^-} \Big\}
\nonumber\\
&&= 1 + i\int^{x^-}_{y^-} dz^- e^{i\frac{ p^2_\perp}{2p^-}z^-}\Big(gB_-(z^-) + \frac{g}{4p^-}\sigma^{\mu\nu}F_{\mu\nu}(z^-)\Big) e^{-i\frac{ p^2_\perp}{2p^-}z^-} + \dots\,,
\label{eq:exp-B0-inter-coord-init}
\end{eqnarray}
where the ellipsis denotes higher-order terms.

Inserting a complete set of transverse states for each background field insertion, we rewrite Eq.~(\ref{eq:exp-ex1-init}) as
\begin{eqnarray}
&&(x_\perp|\mathcal{G}(x^-, y^-)|y_\perp)\Big|_{B_i=0} = (x_\perp|y_\perp)
\label{eq:exp-B0-inter-coord}\\
&& + i\int^{x^-}_{y^-} dz^- \int d^2z_\perp (x_\perp|e^{i\frac{ p^2_\perp}{2p^-}z^-}|z_\perp)\Big(gB_-(z^-, z_\perp) + \frac{g}{4p^-}\sigma^{\mu\nu}F_{\mu\nu}(z^-, z_\perp)\Big) \nonumber\\
&&\times (z_\perp|e^{-i\frac{ p^2_\perp}{2p^-}z^-}|y_\perp) + \dots\,.
\nonumber
\end{eqnarray}
The transverse position of the background fields in Eq.~(\ref{eq:exp-B0-inter-coord}) is not fixed at $y_\perp$ --- note the integration over all transverse coordinates $z_\perp$. Shifts in the transverse position away from $y_\perp$ are governed by the exponential factors~(\ref{eq:exp-fact-bare}).


To fix the transverse position of the operators at $y_\perp$, one must commute the exponential factors through the background fields. This commutation amounts to expanding the background fields onto the fixed transverse position $y_\perp$, and can be performed using the operator (Campbell) identity 
 \begin{eqnarray}
&&e^ABe^{-A}= B + [A, B]+\frac{1}{2!}[A, [A, B]] +\frac{1}{3!}[A, [A, [A, B]]] + \dots \,.
\label{eq:commutation-identity}
\end{eqnarray}

Applying Eq.~(\ref{eq:commutation-identity}) to the operator insertions in Eq.~(\ref{eq:exp-B0-inter-coord-init}), we obtain
 \begin{eqnarray}
&&e^{i\frac{ p^2_\perp}{2p^-}z^-}\Big(gB_-(z^-) + \frac{g}{4p^-}\sigma^{\mu\nu}F_{\mu\nu}(z^-)\Big) e^{-i\frac{ p^2_\perp}{2p^-}z^-} = gB_-(z^-) + \frac{g}{4p^-}\sigma^{\mu\nu}F_{\mu\nu}(z^-) 
\nonumber\\
&&+ \frac{iz^-}{2p^-} \mathcal{A}(z^-) +\frac{1}{2!}\Big(\frac{ iz^-}{2p^-}\Big)^2 [p^2_\perp , \mathcal{A}(z^-)] +\frac{1}{3!}\Big(\frac{ iz^-}{2p^-}\Big)^3 [p^2_\perp, [p^2_\perp , \mathcal{A}(z^-)]] + \dots \,,
\label{eq:tr-identity-init}
\end{eqnarray}
where 
\begin{eqnarray}
&&\mathcal{A}(z^-) \equiv [p^2_\perp , gB_-(z^-) + \frac{g}{4p^-}\sigma^{\mu\nu}F_{\mu\nu}(z^-)]\,.
\label{eq:defAins}
\end{eqnarray}

Using the identities\footnote{The transverse momentum here should be understood as an operator $\hat{p}$ in the sense of the Schwinger notation, see the discussion around Eq.~(\ref{eq:sch-eihen}).}
\begin{eqnarray}
&&[p^2_\perp, \mathcal{O}] = i \{p_k, \partial_k \mathcal{O}\},\ \ \ [p_k, \mathcal{O}] = i\partial_k \mathcal{O}
\label{eq:oper-alg}
\end{eqnarray}
for an arbitrary operator $\mathcal{O}$, the operator $\mathcal{A}$ can be rewritten as
\begin{eqnarray}
&&\mathcal{A}(z^-) =  \{p_k, ig\partial_k B_-(z^-) + \frac{ig\sigma^{\mu\nu}}{4p^-} \partial_k F_{\mu\nu}(z^-)\}.
\label{eq:Abi0}
\end{eqnarray}

Substituting Eq. (\ref{eq:tr-identity-init}) into Eq. (\ref{eq:exp-B0-inter-coord-init}), we obtain
\begin{eqnarray}
&&\mathcal{G}(x^-, y^-)\Big|_{B_i=0} = \mathcal{P}\exp\Big\{ i\int^{x^-}_{y^-} dz^- \Big(gB_-(z^-) + \frac{g}{4p^-}\sigma^{\mu\nu}F_{\mu\nu}(z^-) 
\nonumber\\
&&+ \frac{iz^-}{2p^-} \mathcal{A}(z^-) +\frac{1}{2!}\Big(\frac{ iz^-}{2p^-}\Big)^2 [p^2_\perp , \mathcal{A}(z^-)] +\frac{1}{3!}\Big(\frac{ iz^-}{2p^-}\Big)^3 [p^2_\perp, [p^2_\perp , \mathcal{A}(z^-)]] + \dots\Big)  \Big\}\,.
\label{eq:comm-init-bi0-pexp}
\end{eqnarray}
The pattern of this expansion is manifest and can be written down to any required order.

The $B_-$ operators in the path-ordered exponent give rise to the light-cone gauge factors~(\ref{eq:simple-wl}). Expanding in all remaining operators, we obtain
\begin{eqnarray}
&&\mathcal{G}(x^-, y^-)\Big|_{B_i=0} 
= U(x^-, y^-) + i\int^{x^-}_{y^-} dz^- U(x^-, z^-)\mathcal{O}_{\mathcal{A}}(z^-) U(z^-, y^-) 
\label{eq:Bi0exp-general-form}\\
&& + (i)^2\int^{x^-}_{y^-} dz^- \int^{z^-}_{y^-} dz'^- U(x^-, z^-)\mathcal{O}_{\mathcal{A}}(z^-)U(z^-, z'^-) 
\mathcal{O}_{\mathcal{A}}(z'^-)U(z'^-, y^-) + \dots\,,
\nonumber
\end{eqnarray}
where
\begin{eqnarray}
&&\mathcal{O}_{\mathcal{A}}(z^-) = \mathcal{O}_{\mathcal{A}}(z^-;0^-)
\end{eqnarray}
and the ``shifted" operator insertion is 
\begin{eqnarray}
&&\mathcal{O}_{\mathcal{A}}(z^-;\xi^-)\equiv \frac{g}{4p^-}\sigma^{\mu\nu}F_{\mu\nu}(z^-) + \frac{i(z^--\xi^-)}{2p^-}\mathcal{A}(z^-)
+ \frac{1}{2!}\Big(\frac{i(z^--\xi^-)}{2p^-}\Big)^2[ p^2_\perp(z^-), \mathcal{A}(z^-) ]
\nonumber\\
&&+ \frac{1}{3!}\Big(\frac{i(z^--\xi^-)}{2p^-}\Big)^3[p^2_\perp, [ p^2_\perp(z^-), \mathcal{A}(z^-) ]] + \dots \,.
\label{eq:OA-def}
\end{eqnarray}
The general pattern of this series is transparent.

The transverse position of the background field operators in Eq.~(\ref{eq:Bi0exp-general-form}) cannot yet be completely fixed at $y_\perp$, due to the transverse momentum operators $p_k$ contained in the insertions $\mathcal{O}_{\mathcal{A}}$. To fix the position of the background fields, we need to commute not only the exponential factors, as in Eq.~(\ref{eq:tr-identity-init}), but also the remaining transverse momenta $p_k$ through all operators. Commutation through the light-cone gauge factors, for instance, is accomplished using
\begin{eqnarray}
&&p_k U(x^-, y^-) - U(x^-, y^-)p_k 
\label{eq:comm-WL-b0}\\
&&= - g \int^{x^-}_{y^-}dz^- U(x^-, z^-) \partial_k B_-(z^-)U(z^-, y^-)\,,
\nonumber
\end{eqnarray}
see also Eq. (\ref{eq:oper-alg}).

Commuting all transverse momenta to the left, we get
\begin{align}
&\mathcal{G}(x^-, y^-)\Big|_{B_i=0} 
= U(x^-, y^-) + \frac{ig}{4p^-} \int^{x^-}_{y^-} dz^- U(x^-, z^-) \sigma^{\mu\nu}F_{\mu\nu}(z^-) U(z^-, y^-)
\\
&- \frac{g}{2p^-} \int^{x^-}_{y^-} dz^- z^- U(x^-, z^-) \Big( \partial^2_\perp B_-(z^-) + \frac{\sigma^{\mu\nu}}{4p^-} \partial^2_\perp F_{\mu\nu}(z^-) \Big)  U(z^-, y^-)\nonumber\\
&- \frac{igp_k}{p^-} \int^{x^-}_{y^-} dz^- z^- U(x^-, z^-) 
\Big( \partial_k B_-(z^-) + \frac{\sigma^{\mu\nu}}{4p^-} \partial_k F_{\mu\nu}(z^-) \Big)  U(z^-, y^-)\nonumber\\
&- \frac{ig^2}{p^-} \int^{x^-}_{y^-}dz^- \int^{z^-}_{y^-} dz'^- z'^- U(x^-, z^-) \partial_k B_-(z^-)U(z^-, z'^-)
\Big( \partial_k B_-(z'^-) + \frac{\sigma^{\mu\nu}}{4p^-} \partial_k F_{\mu\nu}(z'^-) \Big)  U(z'^-, y^-) +  \dots\quad 
\nonumber
\end{align}

We find that position of all background field operators can be now fixed at $y_\perp$ using 
\begin{eqnarray}
&&\mathcal{O}|y_\perp) = \mathcal{O}(y_\perp)|y_\perp)
\end{eqnarray}
as
\begin{eqnarray}
&&(x_\perp|\mathcal{G}(x^-, y^-)|y_\perp)\Big|_{B_i=0} 
= (x_\perp|U(x^-, y^-;y_\perp) + \frac{ig}{4p^-} \int^{x^-}_{y^-} dz^- U(x^-, z^-;y_\perp) \sigma^{\mu\nu}F_{\mu\nu}(z^-,y_\perp) U(z^-, y^-;y_\perp) 
\nonumber\\
&&- \frac{g}{2p^-} \int^{x^-}_{y^-} dz^- z^- U(x^-, z^-;y_\perp) \Big( \partial^2_\perp B_-(z^-,y_\perp) + \frac{\sigma^{\mu\nu}}{4p^-} \partial^2_\perp F_{\mu\nu}(z^-,y_\perp) \Big) U(z^-, y^-;y_\perp) \nonumber\\
&&- \frac{igp_k}{p^-} \int^{x^-}_{y^-} dz^- z^- 
U(x^-, z^-;y_\perp) \Big( \partial_k B_-(z^-,y_\perp) + \frac{\sigma^{\mu\nu}}{4p^-} \partial_k F_{\mu\nu}(z^-,y_\perp) \Big) U(z^-, y^-;y_\perp) \nonumber\\
&&- \frac{ig^2}{p^-} \int^{x^-}_{y^-}dz^- \int^{z^-}_{y^-} dz'^- z'^- U(x^-, z^-;y_\perp)
\partial_k B_-(z^-,y_\perp)U(z^-, z'^-;y_\perp)\nonumber\\
&&\times\Big( \partial_k B_-(z'^-,y_\perp) + \frac{\sigma^{\mu\nu}}{4p^-} \partial_k F_{\mu\nu}(z'^-,y_\perp) \Big)  U(z'^-, y^-;y_\perp) +  \dots|y_\perp)\,.
\label{eq:expB0-col}
\end{eqnarray}
\end{widetext}
The commutation procedure thus corresponds to expanding the path-ordered exponent in Eq.~(\ref{eq:exp-ex1-init}) onto the light-cone direction at $y_\perp$, see Fig.~\ref{fig:fig11}b --- precisely the result we sought.


To summarize, the expansion of the path-ordered exponents onto the light-cone direction at a fixed transverse position proceeds in two steps: expanding the exponents as in Eq.~(\ref{eq:Bi0exp-general-form}), followed by commuting the transverse momentum operators using Eqs.~(\ref{eq:oper-alg}) and~(\ref{eq:comm-WL-b0}).

The result of the expansion depends on the direction of the commutation. For instance, commuting all $p_k$ operators in Eq.~(\ref{eq:Bi0exp-general-form}) to the right fixes the background field operators at $x_\perp$, corresponding to an expansion onto the light-cone direction at position $x_\perp$, see Fig.~\ref{fig:fig11}c.


The operators in Eq.~(\ref{eq:expB0-col}) are not gauge covariant, since in this subsection we neglected the transverse background field component $B_i$. In the next subsection, we consider the general case $B_i\neq 0$. As we will see, the methods developed here generalize readily, with the main identities replaced by their gauge-covariant counterparts. This yields a fully gauge-covariant procedure for expanding the path-ordered exponents~(\ref{eq:quark-exponent}) onto the light-cone direction at a fixed transverse position.


\begin{widetext}
\subsection{Expansion onto the light-cone direction. General case of $B_i\neq 0$\label{sec:generalBineq0}}
To extend the approach of the previous subsection to the general case $B_i\neq 0$, we first need to derive a formula for commuting the exponential factors through the background field operators in the argument of the path-ordered exponent, cf.\ Eq.~(\ref{eq:tr-identity-init}). After some algebra, we obtain
\begin{eqnarray}
&&e^{i\frac{ P^2_\perp(z^-)}{2p^-}z^-}(iD_- + \frac{g}{4p^-}\sigma^{\mu\nu}F_{\mu\nu}) e^{-i\frac{ P^2_\perp(z^-)}{2p^-}z^-} - \frac{P^2_\perp(z^-)}{2p^-}\label{eq:generalpord-exp}\\
&&= gB_-(z^-) + \frac{g}{4p^-}\sigma^{\mu\nu}F_{\mu\nu}(z^-) 
+ \frac{iz^-}{2p^-}\mathcal{B}(z^-)
+ \frac{1}{2!}\Big(\frac{iz^-}{2p^-}\Big)^2[ P^2_\perp(z^-), \mathcal{B}(z^-) ]\nonumber\\
&&+ \frac{1}{3!} \Big(\frac{iz^-}{2p^-}\Big)^3 [  P^2_\perp(z^-) , [P^2_\perp(z^-) , \mathcal{B}(z^-)]] + \dots\,,
\nonumber
\end{eqnarray}
where 
\begin{eqnarray}
&&\mathcal{B}(z^-)\equiv [P^2_\perp(z^-), gB_-(z^-) + \frac{g}{4p^-}\sigma^{\mu\nu}F_{\mu\nu}(z^-)] - i \partial_- P^2_\perp(z^-)\,
\end{eqnarray}
is the gauge-covariant generalization of Eq.~(\ref{eq:defAins}). Using the operator identities
\begin{eqnarray}
&&[P^2_\perp(z^-) , \mathcal{O}] = i\{P_k(z^-), D_k\mathcal{O}  \},\ \ \ [P_k, \mathcal{O}] = iD_k \mathcal{O} \,,
\label{eq:op-com-gen-two}
\end{eqnarray}
cf. Eq. (\ref{eq:oper-alg}), this expression becomes 
\begin{eqnarray}
&&\mathcal{B}(z^-) = \{P_k(z^-), ig F_{k-}(z^-)  + \frac{ig\sigma^{\mu\nu}}{4p^-}  D_kF_{\mu\nu}(z^-) \}\,,
\label{eq:operB}
\end{eqnarray}
cf. Eq. (\ref{eq:Abi0}). The first term of this operator corresponds to the scalar part of the path-ordered exponent, while the second, $\sim\sigma^{\mu\nu}$, reflecting its spinor degrees of freedom, is the Pauli vertex.


Using Eq.~(\ref{eq:generalpord-exp}), the path-ordered exponent takes the form
\begin{eqnarray}
&&\mathcal{G}(x^-, y^-) 
= \mathcal{P}\exp\Big\{ i\int^{x^-}_{y^-} dz^- \Big(gB_-(z^-) + \frac{g}{4p^-}\sigma^{\mu\nu}F_{\mu\nu}(z^-) 
\\
&&+ \frac{iz^-}{2p^-}\mathcal{B}(z^-)
+ \frac{1}{2!}\Big(\frac{iz^-}{2p^-}\Big)^2[ P^2_\perp(z^-), \mathcal{B}(z^-) ]
+ \frac{1}{3!} \Big(\frac{iz^-}{2p^-}\Big)^3 [  P^2_\perp(z^-) , [P^2_\perp(z^-) , \mathcal{B}(z^-)]] + \dots\Big) \Big\}\,,
\nonumber
\end{eqnarray}
cf. Eq. (\ref{eq:comm-init-bi0-pexp}). 

The $B_-$ operators give rise to Wilson lines. Expanding in all remaining operators, we obtain a general expansion of the path-ordered exponent, cf.\ Eq.~(\ref{eq:Bi0exp-general-form}):
\begin{eqnarray}
&&\mathcal{G}(x^-, y^-) 
=  U(x^-, y^-) + i\int^{x^-}_{y^-} dz^- U(x^-, z^-)\mathcal{O}_{\mathcal{B}}(z^-) U(z^-, y^-)
\label{eq:Bi0exp-general-form-Bneq0}\\
&&+ (i)^2\int^{x^-}_{y^-} dz^- \int^{z^-}_{y^-} dz'^- U(x^-, z^-)\mathcal{O}_{\mathcal{B}}(z^-)U(z^-, z'^-) 
\mathcal{O}_{\mathcal{B}}(z'^-)U(z'^-, y^-) + \dots\,,
\nonumber
\end{eqnarray}
where
\begin{eqnarray}
&&\mathcal{O}_{\mathcal{B}}(z^-) = \mathcal{O}_{\mathcal{B}}(z^-;\xi^-)
\end{eqnarray}
and a ``shifted" operator insertion is 
\begin{eqnarray}
&&\mathcal{O}_{\mathcal{B}}(z^-;\xi^-)\equiv \frac{g}{4p^-}\sigma^{\mu\nu}F_{\mu\nu}(z^-) + \frac{i(z^--\xi^-)}{2p^-}\mathcal{B}(z^-)
+ \frac{1}{2!}\Big(\frac{i(z^--\xi^-)}{2p^-}\Big)^2[ P^2_\perp(z^-), \mathcal{B}(z^-) ]
\nonumber\\
&&+ \frac{1}{3!}\Big(\frac{i(z^--\xi^-)}{2p^-}\Big)^3[P^2_\perp(z^-), [ P^2_\perp(z^-), \mathcal{B}(z^-) ]] + \dots \,.
\label{eq:general-Ob}
\end{eqnarray}
The pattern of this expansion is manifest from Eqs.~(\ref{eq:Bi0exp-general-form-Bneq0}) and~(\ref{eq:general-Ob}), and can be written down to any required order.

As in the case $B_i=0$, the transverse position of the operators cannot yet be completely fixed. To do so, we commute the covariant transverse momenta $P_k$ through all background field operators. Commutation through the light-cone gauge factors is accomplished using
\begin{eqnarray}
&&P_k(x^-) U(x^-, y^-) - U(x^-, y^-)P_k(y^-)\label{eq:com-gen-UP}\\
&&= g \int^{x^-}_{y^-}dz^- U(x^-, z^-) F_{-k}(z^-)U(z^-, y^-)\,,
\nonumber
\end{eqnarray}
cf. Eq. (\ref{eq:comm-WL-b0}).

Commuting the covariant transverse momenta to the left, we obtain
\begin{eqnarray}
&&\mathcal{G}(x^-, y^-) 
= U(x^-, y^-) + \frac{ig}{4p^-}\int^{x^-}_{y^-} dz^- U(x^-, z^-) \sigma^{\mu\nu}F_{\mu\nu}(z^-)U(z^-, y^-)
\label{eq:lc-exp-get}\\
&&- \frac{g}{2p^-}\int^{x^-}_{y^-} dz^-z^-\ U(x^-, z^-)  \Big( D_kF_{k-}(z^-)  + \frac{\sigma^{\mu\nu}}{4p^-}  D^2_\perp F_{\mu\nu}(z^-) \Big) \} U(z^-, y^-)\nonumber\\
&&- \frac{ig}{p^-}\int^{x^-}_{y^-} dz^-z^- P_k(x^-) U(x^-, z^-) \Big( F_{k-}(z^-)
+ \frac{\sigma^{\mu\nu}}{4p^-}  D_kF_{\mu\nu}(z^-) \Big)  U(z^-, y^-)\nonumber\\
&&+ \frac{ig^2}{p^-} \int^{x^-}_{y^-}dz^- \int^{z'^-}_{y^-} dz'^-z'^-  U(x^-, z^-) F_{-k}(z^-)U(z^-, z'^-) \Big( F_{k-}(z'^-)  + \frac{\sigma^{\mu\nu}}{4p^-}  D_kF_{\mu\nu}(z'^-) \Big) U(z'^-, y^-)+
\dots\,.
\nonumber
\end{eqnarray}

We can now fix the transverse position of all operators at $y_\perp$:
\begin{eqnarray}
&&(x_\perp|\mathcal{G}(x^-, y^-) |y_\perp)
= (x_\perp|U(x^-, y^-;y_\perp) + \frac{ig}{4p^-}\int^{x^-}_{y^-} dz^- U(x^-, z^-;y_\perp) \sigma^{\mu\nu}F_{\mu\nu}(z^-,y_\perp)U(z^-, y^-;y_\perp)\label{eq:expBfull-col}\\
&&\times 
- \frac{g}{2p^-}\int^{x^-}_{y^-} dz^-z^-\ U(x^-, z^-;y_\perp)  \Big( D_kF_{k-}(z^-,y_\perp) + \frac{\sigma^{\mu\nu}}{4p^-}  D^2_\perp F_{\mu\nu}(z^-,y_\perp) \Big) \} U(z^-, y^-;y_\perp)  \nonumber\\
&&
- \frac{ig}{p^-}\int^{x^-}_{y^-} dz^-z^- P_k(x^-, y_\perp) U(x^-, z^-;y_\perp) \Big( F_{k-}(z^-,y_\perp)  + \frac{\sigma^{\mu\nu}}{4p^-}  D_kF_{\mu\nu}(z^-,y_\perp) \Big)  U(z^-, y^-;y_\perp)\nonumber\\
&& 
+ \frac{ig^2}{p^-} \int^{x^-}_{y^-}dz^- \int^{z'^-}_{y^-} dz'^-z'^-U(x^-, z^-;y_\perp) F_{-k}(z^-,y_\perp) U(z^-, z'^-;y_\perp) \Big( F_{k-}(z'^-,y_\perp)  
+ \frac{\sigma^{\mu\nu}}{4p^-}  D_kF_{\mu\nu}(z'^-;y_\perp) \Big)\nonumber\\
&&\times U(z'^-, y^-;y_\perp)+ \dots|y_\perp)\,.\nonumber
\end{eqnarray}
\end{widetext}

Equation~(\ref{eq:expBfull-col}) is the gauge-covariant generalization of Eq.~(\ref{eq:expB0-col}) and corresponds to the expansion of the path-ordered exponent onto the light-cone direction at $y_\perp$, see Fig.~\ref{fig:fig11}b.


To summarize, the expansion of the path-ordered exponents onto the light-cone direction at a fixed transverse position is performed by applying Eq.~(\ref{eq:Bi0exp-general-form-Bneq0}), followed by commuting the covariant transverse momenta using Eqs.~(\ref{eq:op-com-gen-two}) and~(\ref{eq:com-gen-UP}).


A key feature of our approach is that it allows one to unambiguously determine the gauge-covariant operator content to any required order of the expansion --- one of the main advantages of the formalism. This is essential for deriving the hierarchy of operators defining a given scattering reaction.


The methods of this subsection provide a complete procedure for expanding the path-ordered exponents~(\ref{eq:quark-exponent}) onto the light-cone direction at a fixed transverse position. However, physical observables often require expansion onto contours with transverse separation between different light-cone segments, see e.g.\ Fig.~\ref{fig:fig10}. To expand onto such contours, in addition to the light-cone expansion developed here, we need a technique for shifting the path-ordered exponents~(\ref{eq:quark-exponent}) in the transverse direction. As discussed in Sec.~\ref{sec:scl-prop}, a transverse shift is governed by the exponential factor~(\ref{eq:exp-fact-cova}). A parallel shift can therefore be achieved by commuting this factor with the path-ordered exponents. We develop this operation in the next subsection.


\subsection{Transverse shift of the path-ordered exponents\label{sec:pshift}}
Expressions for physical observables in our formalism typically contain combinations of the path-ordered exponents~(\ref{eq:quark-exponent}) and exponential factors~(\ref{eq:exp-fact-cova}) corresponding to the quantum parton propagators in the background field. Together, these elements describe the interaction of the quantum partons with the background field of the target in full generality. For instance, in the dijet production formula~(\ref{eq:hadronic-bfm-general}), the (anti)quark propagators are given by, see Eqs.~(\ref{eq:LSZ-fin-quark}), (\ref{eq:LSZ-fin-quark-second}), (\ref{eq:proptilde-1}), and~(\ref{eq:proptilde-2}),
\begin{equation}
(k_{1\perp}|  e^{i\frac{ k^2_{1\perp}}{2k^-_1} z^-} e^{-i\frac{P^2_\perp(z^-) }{2k^-_1} z^-} \Big|^\infty \mathcal{G}(\infty, y^-) e^{i\frac{P^2_\perp(y^-)}{2k^-_1}y^-} |y_\perp)\,
\label{eq:prop-ex-tr-shift}
\end{equation}
describing the parton's propagation in the background field from a point $(y^-, y_\perp)$ to the future infinity.


Using the methods of the previous subsection, one can expand the path-ordered exponent $\mathcal{G}(\infty, y^-)$ in this propagator onto the light-cone direction. From Eq.~(\ref{eq:Bi0exp-general-form-Bneq0}), the general form of this expansion reads
\begin{widetext}
\begin{eqnarray}
&&(k_{1\perp}|  e^{i\frac{ k^2_{1\perp}}{2k^-_1} z^-} e^{-i\frac{P^2_\perp(z^-) }{2k^-_1} z^-} \Big|^\infty \mathcal{G}(\infty, y^-) e^{i\frac{P^2_\perp(y^-)}{2k^-_1}y^-} |y_\perp)\label{eq:staple-expan-prop}\\
&&= \int d^2z_\perp (k_{1\perp}|  e^{i\frac{ k^2_{1\perp}}{2k^-_1} z^-} e^{-i\frac{P^2_\perp(z^-) }{2k^-_1} z^-} \Big|^\infty 
\Big( U(\infty, y^-) + i\int^{\infty}_{y^-} dz^- U(\infty, z^-)\mathcal{O}_{\mathcal{B}}(z^-) U(z^-, y^-)\nonumber\\
&&+ (i)^2\int^{\infty}_{y^-} dz^- \int^{z^-}_{y^-} dz'^- U(\infty, z^-)\mathcal{O}_{\mathcal{B}}(z^-)
 U(z^-, z'^-) 
\mathcal{O}_{\mathcal{B}}(z'^-)U(z'^-, y^-) + \dots \Big)|z_\perp) (z_\perp| e^{i\frac{P^2_\perp(y^-)}{2k^-_1}y^-} |y_\perp)\,,
\nonumber
\end{eqnarray}
where the ellipsis denotes higher-order insertions of $\mathcal{O}_{\mathcal{B}}$ into the light-cone gauge factors; the operator insertions $\mathcal{O}_{\mathcal{B}}$ are defined in Eq.~(\ref{eq:general-Ob}).

As explained in the previous subsection, to fix the transverse position of the background field operators we commute all covariant transverse derivatives $P_k$ of the operator insertions $\mathcal{O}_{\mathcal{B}}$ to the left. The result for the first few terms follows from Eq.~(\ref{eq:lc-exp-get}):
\begin{eqnarray}
&&(k_{1\perp}|  e^{i\frac{ k^2_{1\perp}}{2k^-_1} z^-} e^{-i\frac{P^2_\perp(z^-) }{2k^-_1} z^-} \Big|^\infty \mathcal{G}(\infty, y^-) e^{i\frac{P^2_\perp(y^-)}{2k^-_1}y^-} |y_\perp)= \int d^2z_\perp (k_{1\perp}|  e^{i\frac{ k^2_{1\perp}}{2k^-_1} z^-} e^{-i\frac{P^2_\perp(z^-) }{2k^-_1} z^-} \Big|^\infty |z_\perp)
\label{eq:col-exp-shift-prelim}\\
&&\times \Big(U(\infty, y^-;z_\perp) + \frac{ig}{4k^-_1}\int^{\infty}_{y^-} dz^- U(x^-, z^-;z_\perp)
\sigma^{\mu\nu}F_{\mu\nu}(z^-,z_\perp)U(z^-, y^-;z_\perp) + \dots \Big)(z_\perp|e^{i\frac{P^2_\perp(y^-)}{2k^-_1}y^-} |y_\perp)\,.
\nonumber
\end{eqnarray}
\end{widetext}

As discussed in Sec.~\ref{sec:tr-link}, the exponential factors~(\ref{eq:exp-fact-cova}) in Eq.~(\ref{eq:col-exp-shift-prelim}) contain transverse gauge contributions, extractable by expanding the fields onto a straight line in the transverse plane, see Eq.~(\ref{eq:tr-WL-insertion}). For the exponential factor at the future infinity, this simplifies considerably: the gluon fields there are defined by a pure gauge, so only the leading term of the expansion~(\ref{eq:tr-WL-insertion}) survives. We therefore write
\begin{eqnarray}
&&(k_{1\perp}|  e^{i\frac{ k^2_{1\perp}}{2k^-_1} z^-} e^{-i\frac{P^2_\perp(z^-) }{2k^-_1} z^-} \Big|^\infty |z_\perp)= \int d^2x_\perp (k_{1\perp}|x_\perp)  \nonumber\\
&&\times e^{i\frac{ k^2_{1\perp}}{2k^-_1} z^-} (x_\perp|e^{-i\frac{p^2_\perp }{2k^-_1} z^-} \Big|^\infty |z_\perp) U(x_\perp, z_\perp; \infty)
\label{eq:inf-link-full}
\end{eqnarray}
The transverse gauge links at the future infinity, $U(x_\perp, z_\perp; \infty)$ in Eq.~(\ref{eq:inf-link-full}), ultimately connect the light-cone gauge factors in the dijet production cross section, ensuring gauge invariance of the resulting operators.


To simplify the analysis, we impose the supplementary boundary condition 
$$\lim_{x^-\to\infty}B_i(x) = 0,$$ allowing us to drop the transverse gauge links at the future infinity. Equation~(\ref{eq:inf-link-full}) then reduces to
\begin{eqnarray}
&&(k_{1\perp}|  e^{i\frac{ k^2_{1\perp}}{2k^-_1} z^-} e^{-i\frac{P^2_\perp(z^-) }{2k^-_1} z^-} \Big|^\infty |z_\perp)\Big|_{B_i(\infty)=0}\nonumber\\
&&= e^{-ik_{1\perp}z_\perp}
\label{eq:inf-link-simp}
\end{eqnarray}
and Eq.~(\ref{eq:col-exp-shift-prelim}) becomes
\begin{widetext}
\begin{eqnarray}
&&(k_{1\perp}|  e^{i\frac{ k^2_{1\perp}}{2k^-_1} z^-} e^{-i\frac{P^2_\perp(z^-) }{2k^-_1} z^-} \Big|^\infty \mathcal{G}(\infty, y^-) e^{i\frac{P^2_\perp(y^-)}{2k^-_1}y^-} |y_\perp) = \int d^2z_\perp e^{-ik_{1\perp}z_\perp}\Big(U(\infty, y^-;z_\perp) 
\nonumber\\
&&+ \frac{ig}{4k^-_1}\int^{\infty}_{y^-} dz^- U(x^-, z^-;z_\perp) \sigma^{\mu\nu}F_{\mu\nu}(z^-,z_\perp)U(z^-, y^-;z_\perp) + \dots \Big)(z_\perp|e^{i\frac{P^2_\perp(y^-)}{2k^-_1}y^-} |y_\perp)\,.
\label{eq:col-exp-shift-notrinf}
\end{eqnarray}
\end{widetext}

\begin{figure}[htb]
\begin{center}
\includegraphics[width=0.9\linewidth]{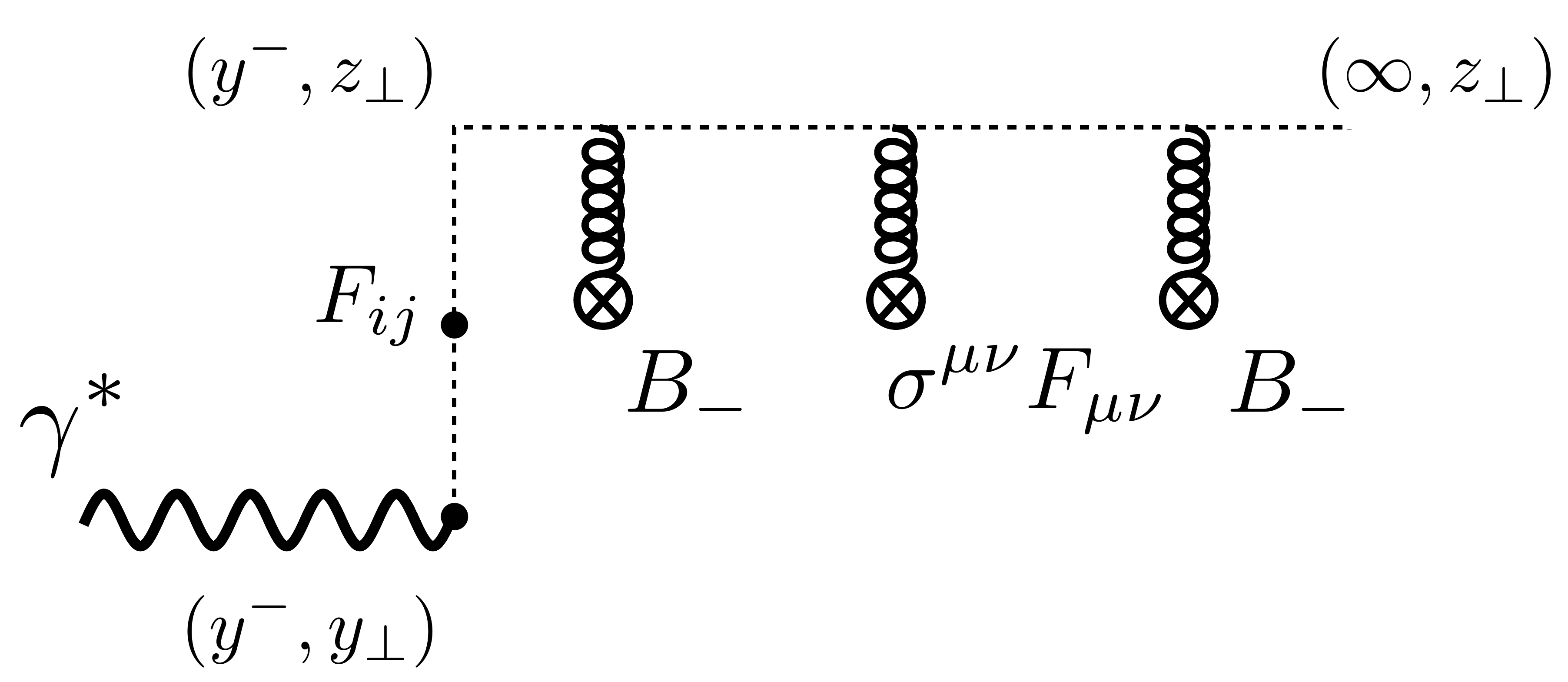}
 \end{center}
 \caption{\label{fig:fig12}Expansion contour defining the operators in Eq.~(\ref{eq:col-exp-shift-notrinf}) for the (anti)quark propagator in the background field. The transverse segment corresponds to an infinite sum of transverse gauge links with various transverse field-strength tensor insertions, see Eq.~(\ref{eq:tr-WL-insertion}). Note the $\sigma^{\mu\nu}F_{\mu\nu}$ operator insertion in the light-cone segment of the contour.}
 \end{figure}

Consider now the remaining exponential factor~(\ref{eq:exp-fact-cova}) in Eq.~(\ref{eq:col-exp-shift-notrinf}). This factor can be represented as an infinite sum of transverse gauge links with various field-strength tensor insertions, see Eq.~(\ref{eq:tr-WL-insertion}). The content of Eq.~(\ref{eq:col-exp-shift-notrinf}) is then clear: it corresponds to expanding the background fields in the propagator~(\ref{eq:prop-ex-tr-shift}) onto the linear piecewise contour shown in Fig.~\ref{fig:fig12}. This contour consists of a transverse segment connecting $y_\perp$ and $z_\perp$ at the longitudinal position $y^-$, followed by a light-cone segment from $(y^-, z_\perp)$ to the future infinity.

One drawback of this contour choice is that the exponential factor~(\ref{eq:exp-fact-cova}) in Eq.~(\ref{eq:col-exp-shift-notrinf}) is located at a finite longitudinal position $y^-$, where it involves an infinite number of transverse gauge factors with various field-strength tensor insertions, see Eq.~(\ref{eq:tr-WL-insertion}). This complicates the resulting operator hierarchy. As noted above, the exponential factor~(\ref{eq:exp-fact-cova}) reduces to a simple transverse gauge link only at spatial infinity, where the background gluon field is a pure gauge.



For this reason, when computing physical observables it is advantageous to move the exponential factors~(\ref{eq:exp-fact-cova}) to spatial infinity. As we will see, this operation corresponds to shifting the light-cone gauge factors in the transverse direction.

To move the exponential factor in Eq.~(\ref{eq:prop-ex-tr-shift}) to spatial infinity, we commute it through the path-ordered exponent using the relation~(\ref{eq:parallel-shift-scalar-full-quark}):\footnote{Here we drop the transverse Wilson lines at the future infinity, see Eq.~(\ref{eq:inf-link-simp}).}
\begin{eqnarray}
&&(k_{1\perp}|  e^{i\frac{ k^2_{1\perp}}{2k^-_1} z^-} e^{-i\frac{P^2_\perp(z^-) }{2k^-_1} z^-} \Big|^\infty \mathcal{G}(\infty, y^-) e^{i\frac{P^2_\perp(y^-)}{2k^-_1}y^-} |y_\perp)\nonumber\\
&&= e^{i\frac{ k^2_{1\perp}}{2k^-_1} y^- } (k_{1\perp}| \mathcal{G}(\infty, y^-;y^-)|y_\perp) \,,
\label{eq:prop-ex-tr-shift-commute}
\end{eqnarray}
where we used
\begin{align}
& \mathcal{G}(\infty, y^-) e^{i\frac{P^2_\perp(y^-)}{2k^-_1}y^-} = e^{i\frac{ P^2_\perp(z^-)}{2k^-_1} y^- }\Big|^\infty \mathcal{G}(\infty, y^-;y^-)
\label{eq:paralshift}\\
&= e^{i\frac{ P^2_\perp(z^-)}{2k^-_1} y^- }\Big|^\infty \mathcal{P}\exp\Big\{ i\int^{\infty}_{y^-} dz^- \Big(e^{i\frac{ P^2_\perp(z^-)}{2k^-_1}(z^--y^-)}\nonumber\\
&\times \Big(iD_- + \frac{g}{4k^-_1}\sigma^{\mu\nu}F_{\mu\nu}\Big) e^{-i\frac{ P^2_\perp(z^-)}{2k^-_1}(z^--y^-)}
- \frac{P^2_\perp(z^-)}{2k^-_1}\Big) \Big\}\,.
\nonumber
\end{align}

The ``shifted'' path-ordered exponent $\mathcal{G}(\infty, y^-;y^-)$ in Eq.~(\ref{eq:prop-ex-tr-shift-commute}) can be further expanded onto the light-cone direction using the methods developed above. From Eq.~(\ref{eq:Bi0exp-general-form-Bneq0}), the general form of this expansion is
\begin{widetext}
\begin{eqnarray}
&&\mathcal{G}(x^-, y^-, \xi^-) 
= U(x^-, y^-) + i\int^{x^-}_{y^-} dz^- U(x^-, z^-)\mathcal{O}_{\mathcal{B}}(z^-;\xi^-) U(z^-, y^-)
\label{eq:result-exp-shift-exp}\\
&&+ (i)^2\int^{x^-}_{y^-} dz^- \int^{z^-}_{y^-} dz'^- U(x^-, z^-)\mathcal{O}_{\mathcal{B}}(z^-;\xi^-)U(z^-, z'^-) 
\mathcal{O}_{\mathcal{B}}(z'^-;\xi^-)U(z'^-, y^-) + \dots\,,
\nonumber
\end{eqnarray}
where the ``shifted'' operator insertions into the light-cone gauge factors are defined in Eq.~(\ref{eq:general-Ob}).

Using this expansion, the propagator~(\ref{eq:prop-ex-tr-shift}) after commuting the exponential factor takes the form
\begin{eqnarray}
&&(k_{1\perp}|  e^{i\frac{ k^2_{1\perp}}{2k^-_1} z^-} e^{-i\frac{P^2_\perp(z^-) }{2k^-_1} z^-} \Big|^\infty \mathcal{G}(\infty, y^-) e^{i\frac{P^2_\perp(y^-)}{2k^-_1}y^-} |y_\perp) = e^{i\frac{ k^2_{1\perp}}{2k^-_1} y^- } (k_{1\perp}| \Big(U(\infty, y^-)\label{eq:light-cone-general}\\
&&+ i\int^{\infty}_{y^-} dz^- U(\infty, z^-)\mathcal{O}_{\mathcal{B}}(z^-;y^-) U(z^-, y^-)\nonumber\\
&&+ (i)^2\int^{\infty}_{y^-} dz^- \int^{z^-}_{y^-} dz'^- U(\infty, z^-)\mathcal{O}_{\mathcal{B}}(z^-;y^-)U(z^-, z'^-) 
\mathcal{O}_{\mathcal{B}}(z'^-;y^-)
 U(z'^-, y^-) + \dots\Big) |y_\perp) \,.
 \nonumber
\end{eqnarray}
The commutation of the exponential factor thus amounts to the replacement
\begin{eqnarray}
&&\mathcal{O}_{\mathcal{B}}(z^-)\to\mathcal{O}_{\mathcal{B}}(z^-;y^-)\,,
\end{eqnarray}
cf. Eqs. (\ref{eq:staple-expan-prop}) and (\ref{eq:light-cone-general}).

To fix the transverse position of operators in Eq.~(\ref{eq:light-cone-general}), we commute all covariant transverse momenta to the left using Eqs.~(\ref{eq:op-com-gen-two}) and~(\ref{eq:com-gen-UP}), as in the derivation of Eq.~(\ref{eq:lc-exp-get}). The result is
\begin{eqnarray}
&&\mathcal{G}(\infty, y^-; y^-) 
= U(\infty, y^-) + \frac{ig}{4k^-_1}\int^{\infty}_{y^-} dz^- U(\infty, z^-) \sigma^{\mu\nu}F_{\mu\nu}(z^-)U(z^-, y^-)
\nonumber\\
&& + \frac{ig}{k^-_1}\int^{\infty}_{y^-} dz^- (z^--y^-)
P_k(\infty) U(\infty, z^-) F_{-k}(z^-) U(z^-, y^-) + \dots\,,
\label{eq:col-exp-expl}
\end{eqnarray}
cf.\ Eq.~(\ref{eq:lc-exp-get}), where we wrote only the terms linear in the field-strength tensor. Using our approach, the gauge-covariant operators in Eq.~(\ref{eq:col-exp-expl}) can be reconstructed to any required order. As a result, one can unambiguously reconstruct the covariant structure of the hierarchy of operators defining interactions of the quantum partons and the background fields.

Substituting Eq.~(\ref{eq:col-exp-expl}) into Eq.~(\ref{eq:prop-ex-tr-shift-commute}), we obtain
\begin{eqnarray}
&&(k_{1\perp}|  e^{i\frac{ k^2_{1\perp}}{2k^-_1} z^-} e^{-i\frac{P^2_\perp(z^-) }{2k^-_1} z^-} \Big|^\infty \mathcal{G}(\infty, y^-) e^{i\frac{P^2_\perp(y^-)}{2k^-_1}y^-} |y_\perp)\label{eq:prop-ex-tr-shift-commute-colexp}\\
&&= e^{i\frac{ k^2_{1\perp}}{2k^-_1} y^- } (k_{1\perp}|U(\infty, y^-) + \frac{ig}{4k^-_1}\int^{\infty}_{y^-} dz^-
 U(\infty, z^-) \sigma^{\mu\nu}F_{\mu\nu}(z^-)U(z^-, y^-)\nonumber\\
&&+ \frac{igk_{1k}}{k^-_1}\int^{\infty}_{y^-} dz^- (z^--y^-)
 U(\infty, z^-) F_{-k}(z^-) U(z^-, y^-) + \dots|y_\perp) \,.
\nonumber
\end{eqnarray}

Fixing the transverse position of all operators in Eq.~(\ref{eq:prop-ex-tr-shift-commute-colexp}), we arrive at
\begin{eqnarray}
&&(k_{1\perp}|  e^{i\frac{ k^2_{1\perp}}{2k^-_1} z^-} e^{-i\frac{P^2_\perp(z^-) }{2k^-_1} z^-} \Big|^\infty \mathcal{G}(\infty, y^-) e^{i\frac{P^2_\perp(y^-)}{2k^-_1}y^-} |y_\perp)\label{eq:prop-ex-tr-shift-commute-colexp-fixtrpos}\\
&&= e^{i\frac{ k^2_{1\perp}}{2k^-_1} y^- } e^{-ik_{1\perp}y_\perp} \Big( U(\infty, y^-; y_\perp) + \frac{ig}{4k^-_1}\int^{\infty}_{y^-} dz^-
U(\infty, z^-; y_\perp) \sigma^{\mu\nu}F_{\mu\nu}(z^-, y_\perp)U(z^-, y^-;y_\perp)\nonumber\\
&&+ \frac{igk_{1k}}{k^-_1}\int^{\infty}_{y^-} dz^- (z^--y^-)
 U(\infty, z^-;y_\perp) F_{-k}(z^-, y_\perp) 
 U(z^-, y^-;y_\perp) + \dots\Big) \,.\nonumber
\end{eqnarray}
\end{widetext}

\begin{figure}[htb]
\begin{center}
\includegraphics[width=0.9\linewidth]{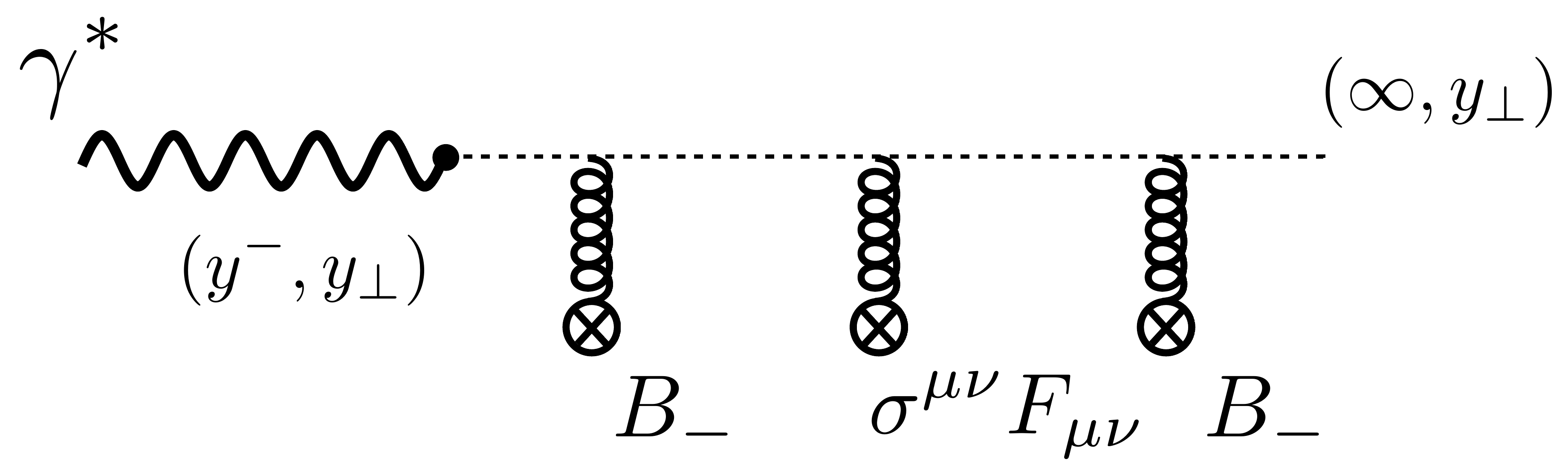}
 \end{center}
 \caption{\label{fig:fig13}The light-cone expansion contour defining the operators in Eq.~(\ref{eq:prop-ex-tr-shift-commute-colexp-fixtrpos}) for the (anti)quark propagator in the background field. The transverse gauge link at the future infinity is not shown explicitly, as it is suppressed by the supplementary boundary condition $\lim_{x^-\to\infty}B_i(x)=0$.}
 \end{figure}

Equation~(\ref{eq:prop-ex-tr-shift-commute-colexp-fixtrpos}) corresponds to expanding the path-ordered exponent in the (anti)quark propagator~(\ref{eq:prop-ex-tr-shift}) onto the light-cone direction shown in Fig.~\ref{fig:fig13}. Comparing with the contour in Fig.~\ref{fig:fig12} for operators in Eq.~(\ref{eq:col-exp-shift-notrinf}), we see that applying the commutation relation~(\ref{eq:parallel-shift-scalar-full-quark}) for the exponential factors~(\ref{eq:exp-fact-cova}) effectively shifts the transverse position of the path-ordered exponents. Once the exponential factors are at spatial infinity --- either the future ($z^-=\infty$) or past ($z^-=-\infty$) --- they reduce to simple transverse gauge links.

The transverse shift introduced in this subsection completes our approach for expanding the path-ordered exponents~(\ref{eq:quark-exponent}) and their ``shifted'' versions~(\ref{eq:definition-exp-shifted-quark}). Using this approach, the path-ordered exponents in the (anti)quark propagators can be expanded onto an arbitrary linear piecewise contour, see e.g.\ Fig.~\ref{fig:fig10}, by a suitable sequence of commutations of the exponential factors~(\ref{eq:exp-fact-cova}) with the path-ordered exponents~(\ref{eq:quark-exponent}), the latter being expanded onto the light-cone direction using the methods of the previous subsections. The construction is fully gauge covariant, enabling unambiguous derivation of the operator hierarchy for physical observables to any required order. This constitutes the main advantage of our approach.


Once the expansion of the path-ordered exponents is performed, the expressions for physical observables assume a factorized form, from which one can compute the perturbative coefficients and derive the corresponding factorization formulas.


The ability to expand onto an arbitrary linear piecewise contour makes our approach applicable to physical observables in different kinematic regimes. In particular, this is instrumental for understanding interplay effects between large and small $x$.


In our framework, the choice of expansion contour is not fixed: one can derive an expansion onto an arbitrary combination of transverse and light-cone segments. The construction is mathematically exact. However, the contour must be chosen consistently with the expansion parameter of the problem, giving rise to a power counting that organizes the terms of the formal expansion.


\section{Power counting of operators in the back-to-back limit of the dijet production\label{sec:power-cop}}
In our approach, we first derive a general expression for a physical observable in terms of (anti)quark propagators in the background field, constructed from the path-ordered exponents~(\ref{eq:definition-exp-shifted-quark}) and exponential factors~(\ref{eq:exp-fact-cova}), see e.g.\ Eq.~(\ref{eq:hadronic-bfm-general}) for the dijet production cross section. While completely general and exact, these expressions do not have a factorized form, since the path-ordered exponents~(\ref{eq:definition-exp-shifted-quark}) and exponential factors~(\ref{eq:exp-fact-cova}) entangle the dynamics of the quantum partons with the interactions with the background fields.


To disentangle these effects and derive a factorization formula for the physical observable, one needs to expand the path-ordered exponents onto a linear piecewise contour of appropriate choice. This expansion can be carried out using the methods developed in the previous sections.


The result takes the form of an infinite series of operators along the light-cone direction connected by transverse gauge links, see e.g.\ Fig.~\ref{fig:fig10}. These operators constitute the hierarchy defining the physical observable. They contain various insertions of gluon field-strength tensors and their covariant derivatives into the light-cone gauge factors. Using our approach, the gauge-covariant form of these operators can be systematically derived.


One may ask whether such formal expansions are useful for describing physical observables, since it is clearly impossible to reconstruct all terms of the infinite series --- even though our approach can derive all terms up to any fixed order. This is where the choice of expansion contour for the path-ordered exponents becomes important. In high-energy scattering, one can typically identify an expansion parameter defined by the kinematics of the problem. The choice of the expansion contour must be consistent with this parameter, so that higher-order terms of the expansion are suppressed by higher powers of the expansion parameter relative to the leading-order terms. When this condition is met, the expansion series converges and only a finite number of terms need to be retained in practice.


\subsection{Power counting in the back-to-back limit}
To illustrate out formalism, let us consider the back-to-back limit of dijet production. This limit is characterized by a wide separation between the transverse momenta of the quantum partons and those of the background fields. In this regime, a typical transverse momentum $\bar{P}_\perp$ of the quantum (anti)quark is large, while the imbalance between the quark and antiquark transverse momenta, $\Delta_\perp$, is small. The ratio $\Delta_\perp/\bar{P}_\perp\ll1$ serves as the expansion parameter of the problem, defined by the back-to-back kinematics. In this regime, interactions with the background fields do not significantly alter the transverse momenta of the quark-antiquark pair; the transverse momenta carried by the background fields, $\sim \Delta_\perp$, are suppressed relative to those of the quark-antiquark pair, $\sim \bar{P}_\perp$.

This suppression gives rise to a power counting for the operators representing the background field content. For instance, $\bar{P}_i F_{\mu\nu} \gg \Delta_i F_{\mu\nu} \sim \partial_i F_{\mu\nu}$, so the $\partial_i F_{\mu\nu}$ contribution is power-suppressed relative to $\bar{P}_i F_{\mu\nu}$.


Consequently, different operators contribute at different levels of importance, and computing physical observables requires isolating the dominant operator contributions according to the power counting. This is achieved by an appropriate choice of the expansion contour for the path-ordered exponents in the general expression for the physical observable, consistent with the expansion parameter of the problem, as discussed above.


Once the expansion contour is identified and the leading-power contribution to the physical observable is separated, the corresponding perturbative coefficients can be readily computed. Some care is needed, however, in estimating the size of different terms, as a certain interplay exists between the power counting of operators and the perturbative coefficients. For instance, in the back-to-back limit a perturbative coefficient for the $F_{\mu\nu}$ contribution could involve a combination $(q_i-q'_i) F_{\mu\nu}$, where $q_i$ and $q'_i$ are transverse momenta before and after the interaction with the background field. While each term individually is enhanced, $q_i F_{\mu\nu} \sim q'_i F_{\mu\nu} \sim \bar{P}_i F_{\mu\nu} \gg \partial_i F_{\mu\nu}$, their difference is suppressed: $(q_i-q'_i) F_{\mu\nu}\sim \Delta_i F_{\mu\nu}\sim \partial_i F_{\mu\nu}$. For this reason, the combination $(q_i-q'_i) F_{\mu\nu}$ should be attributed to the subleading-power corrections. This interplay must be taken into account in the power counting.


Let us now consider the power counting in more detail by computing the leading-power contribution to dijet production in the back-to-back limit. The appropriate expansion contour for the path-ordered exponents is suggested by our previous analysis. The power counting in the back-to-back limit identifies the leading-power contribution as the one with the fewest insertions of field-strength tensors and their transverse derivatives, corresponding to the twist counting of operators --- each field-strength tensor insertion is power-suppressed as $F_{-i}\sim \Delta_i B_-$.


Inspecting the expansion of the (anti)quark propagator onto the light-cone contour in Fig.~\ref{fig:fig13}, see Eq.~(\ref{eq:light-cone-general}), one notes that each successive term of the expansion involves an additional field-strength tensor insertion compared to the leading terms. The higher-order terms are therefore suppressed by higher powers of $\Delta_\perp/\bar{P}_\perp \ll 1$, confirming that the expansion~(\ref{eq:light-cone-general}) onto the contour in Fig.~\ref{fig:fig13} is consistent with the expansion parameter.


We thus conclude that the expansion contour in Fig.~\ref{fig:fig13} is suitable for the power counting of dijet production in the back-to-back limit. Substituting the expansion of the propagators onto the light-cone contour, Eq.~(\ref{eq:prop-ex-tr-shift-commute-colexp-fixtrpos}), into the dijet production cross section~(\ref{eq:hadronic-bfm-general}), we obtain
\begin{widetext}
\begin{eqnarray}
&&d\sigma^{\gamma^\ast p}_{\lambda\lambda'}\Big|^{\rm back-to-back} = \sum_f\sum_{s_1,s_2}\frac{2\pi e^2_f \alpha_{\rm EM}x}{V_4Q^2}   \epsilon^{\lambda\ast}_\mu \epsilon^{\lambda'}_\nu \frac{dk^-_1d^2k_{1\perp}}{(2\pi)^3 2k^-_1} \Big|_{k^2_1=0}\frac{dk^-_2d^2k_{2\perp}}{(2\pi)^3 2k^-_2} \Big|_{k^2_2=0} \int d^4x e^{i(q-k_1-k_2)x}
\label{eq:djet-btb-col-lead-init}\\
&&\times \langle P, S| \bar{v}_{s_2}(k_2)  \Big\{ -\Big(\frac{\gamma^\mu \sigma^{-k}}{2k^-_1} + \frac{\sigma^{-k}\gamma^\mu}{2k^-_2} \Big) ig \int_{x^-}^{\infty} dz^- U(\infty, z^-; x_\perp) F_{-k}(z^-, x_\perp)U(z^-, \infty; x_\perp) \nonumber\\
&& 
-\Big( \frac{k_{1k}\gamma^\mu}{k^-_1} - \frac{k_{2k}\gamma^\mu}{k^-_2} \Big) ig \int_{x^-}^{\infty} dz^-(z^--x^-) U(\infty, z^-; x_\perp) F_{-k}(z^-, x_\perp) U(z^-, \infty; x_\perp) \Big\} u_{s_1}(k_1)\nonumber\\
&& 
\times \int d^4y e^{-i(q-k_1-k_2)y} \bar{u}_{s_1}(k_1) \Big\{ \Big(\frac{\sigma^{-m}\gamma^\nu}{2k^-_1}+\frac{\gamma^\nu\sigma^{-m}}{2k^-_2} \Big) ig \int^{\infty}_{y^-} dz'^- U(\infty, z'^-; y_\perp) F_{-m}(z'^-, y_\perp) U(z'^-, \infty;y_\perp)\nonumber\\
&&
+ \Big( \frac{k_{1m}\gamma^\nu}{k^-_1} - \frac{k_{2m}\gamma^\nu}{k^-_2}\Big) ig \int_{y^-}^{\infty} dz'^-(z'^- - y^-) U(\infty, z'^-;y_\perp) F_{-m}(z'^-, y_\perp) U(z'^-, \infty;y_\perp) \Big\} v_{s_2}(k_2) |P, S\rangle + \dots\,,
\nonumber
\end{eqnarray}
where the ellipsis denotes power-suppressed higher-order terms. Within our approach, these terms can be straightforwardly reconstructed; the form of some of them can be read off from Eq.~(\ref{eq:expBfull-col}).


Integrating over $x^-$ and $y^-$, and invoking the translational invariance of the operators, we obtain
\begin{eqnarray}
&&d\sigma^{\gamma^\ast p}_{\lambda\lambda'}\Big|^{\rm back-to-back} = \sum_f\sum_{s_1,s_2}\frac{2\pi e^2_f \alpha_{\rm EM}x}{Q^2}   \epsilon^{\lambda\ast}_\mu \epsilon^{\lambda'}_\nu \frac{dk^-_1d^2k_{1\perp}}{(2\pi)^3 2k^-_1} \Big|_{k^2_1=0}\frac{dk^-_2d^2k_{2\perp}}
 {(2\pi)^3 2k^-_2} \Big|_{k^2_2=0} 2\pi\delta(q^--k^-_1-k^-_2)
 \nonumber\\
 &&\times  \bar{v}_{s_2}(k_2)
(ig)^2 \Big\{ \Big(\frac{\gamma^\mu \sigma^{-k}}{2k^-_1} + \frac{\sigma^{-k}\gamma^\mu}{2k^-_2} \Big) \frac{i}{q^+-k^+_1-k^+_2} + \Big( \frac{k_{1k}\gamma^\mu}{k^-_1} - \frac{k_{2k}\gamma^\mu}{k^-_2} \Big)\frac{1}{(q^+-k^+_1-k^+_2)^2} \Big\} t^c  u_{s_1}(k_1)
\nonumber\\
&&\times \bar{u}_{s_1}(k_1)
\Big\{ \Big(\frac{\sigma^{-m}\gamma^\nu}{2k^-_1}
+\frac{\gamma^\nu\sigma^{-m}}{2k^-_2} \Big) \frac{i}{q^+-k^+_1-k^+_2}- \Big( \frac{k_{1m}\gamma^\nu}{k^-_1} - \frac{k_{2m}\gamma^\nu}{k^-_2}\Big)\frac{1}{(q^+-k^+_1-k^+_2)^2}  \Big\} t^d v_{s_2}(k_2)
\nonumber\\
&&\times 
\int d^2x_\perp e^{i(k_{1\perp}+k_{2\perp}-q_\perp)x_\perp} \langle P, S|
\int_{-\infty}^{\infty} dz^-
e^{-i(k^+_1+k^+_2-q^+)z^-}  F^a_{-k}(z^- , x_\perp) U^{ac}(z^-, \infty; x_\perp) 
\nonumber\\
&&\times U^{db}(\infty, 0^-;0_\perp) F^b_{-m}(0^-, 0_\perp) |P, S\rangle + \dots\,,
\label{eq:btb-ltwist-pr}
\end{eqnarray}
where all operators were written in the adjoint representation using
\begin{eqnarray}
&&U(\infty, z^-) t^aU(z^-, \infty) = U^{ac}(z^-, \infty) t^c\,.
\end{eqnarray}

From Eq.~(\ref{eq:btb-ltwist-pr}), we find that the dijet production cross section in the back-to-back configuration at leading power is determined by a matrix element of the gluon TMD operator. For a longitudinally polarized virtual photon, see Eq.~(\ref{eq:gauge-vec}), we have
\begin{eqnarray}
&&d\sigma^{\gamma^\ast p}_{00}\Big|^{\rm back-to-back} = \sum_f \frac{2\pi e^2_f g^2 \alpha_{\rm EM}x}{2Q^2}  \frac{dk^-_1d^2k_{1\perp}}{(2\pi)^3 2k^-_1} \Big|_{k^2_1=0}\frac{dk^-_2d^2k_{2\perp}}{(2\pi)^3 2k^-_2} \Big|_{k^2_2=0} 2\pi\delta(q^--k^-_1-k^-_2)
\label{eq:btb-ltwist}\\
&&\times \Big(\frac{Q}{q^-}\Big)^2\Big( \frac{k_{1k}}{k^-_1} - \frac{k_{2k} }{k^-_2} \Big) \Big( \frac{k_{1m}}{k^-_1} - \frac{k_{2m} }{k^-_2}\Big) \frac{8 k^-_1 k^-_2}{(q^+-k^+_1-k^+_2)^4} \int d^2x_\perp e^{i(k_{1\perp}+k_{2\perp}-q_\perp)x_\perp} \int_{-\infty}^{\infty} dz^- e^{-i(k^+_1+k^+_2-q^+)z^-}\nonumber\\
&&\times  
  \langle P, S|
    F^a_{-k}(z^- , x_\perp) U^{ac}(z^-, \infty; x_\perp) U^{cd}(x_\perp, 0;\infty)U^{db}(\infty, 0^-;0_\perp) F^b_{-m}(0^-, 0_\perp) |P, S\rangle+ \dots\,,
\nonumber
\end{eqnarray}
where gauge invariance of the TMD operator was restored by inserting a transverse gauge link at future infinity, see the discussion in Sec.~\ref{sec:pshift}.
\end{widetext}

\subsection{Leading power cross section in the back-to-back limit}
While the operator in Eq.~(\ref{eq:btb-ltwist}) provides the leading-power contribution, the corresponding perturbative coefficient contains factors of subleading power. To see this explicitly, it is convenient to introduce the variables
\begin{equation}
z = \frac{k^-_1}{q^-};\ \ \ \Delta_k = k_{1k} + k_{2k};\ \ \ \bar{P}_k = zk_{1k} - \bar{z}k_{2k}\,,
\label{eq:dijet-var}
\end{equation}
where $\bar{z}\equiv 1-z$. In the back-to-back kinematics, $\Delta_\perp/\bar{P}_\perp \ll 1$, the perturbative coefficient in Eq.~(\ref{eq:btb-ltwist}) contains terms suppressed by the expansion parameter $\Delta_\perp/\bar{P}_\perp$, e.g.
\begin{eqnarray}
&& \frac{k_{1k}}{k^-_1} - \frac{k_{2k} }{k^-_2} = \frac{ \bar{P}_k }{z\bar{z}q^-} - \frac{( z - \bar{z})\Delta_k }{z\bar{z}q^-}\,.
\label{eq:sup-pert-coef}
\end{eqnarray}

To isolate the leading-power contribution to dijet production, these suppressed terms must be removed. From Eq.~(\ref{eq:btb-ltwist}), at leading power we obtain
\begin{widetext}
\begin{eqnarray}
&&d\sigma^{\gamma^\ast p}_{00}\Big|^{\rm back-to-back}_{\rm lead.\ power} = \sum_f \frac{2\pi e^2_f g^2 \alpha_{\rm EM}x}{Q^2} \Big(\frac{Q}{q^-}\Big)^2\frac{dk^-_1d^2k_{1\perp}}{(2\pi)^3 2k^-_1} \Big|_{k^2_1=0}\frac{dk^-_2d^2k_{2\perp}}{(2\pi)^3 2k^-_2} \Big|_{k^2_2=0}\label{eq:lead-order-col-btb}\\
&&\times 2\pi\delta(q^--k^-_1-k^-_2)
2\bar{P}_k \bar{P}_m \frac{32 z^3 \bar{z}^3 (q^-)^4}{(\bar{P}^2_\perp + z\bar{z}Q^2)^4}  \int d^2x_\perp e^{i(k_{1\perp}+k_{2\perp}-q_\perp)x_\perp} \int_{-\infty}^{\infty} dz^- e^{-i(k^+_1+k^+_2-q^+)z^-}
\nonumber\\
&&\times  \langle P, S|
    F^a_{-k}(z^- , x_\perp) U^{ac}(z^-, \infty; x_\perp) U^{cd}(x_\perp, 0_\perp;\infty)U^{db}(\infty, 0^-;0_\perp) F^b_{-m}(0^-, 0_\perp) |P, S\rangle\,.
\nonumber
\end{eqnarray}

Similarly, for a transversely polarized virtual photon, from Eq.~(\ref{eq:btb-ltwist-pr}) we find
\begin{eqnarray}
&&d\sigma^{\gamma^\ast p}_{0\lambda'=\pm1}\Big|^{\rm back-to-back}_{\rm lead.\ power} = \sum_f \frac{2\pi e^2_f g^2 \alpha_{\rm EM}x}{Q^2} \frac{Q}{q^-} \epsilon^{\lambda'=\pm1}_l \frac{dk^-_1d^2k_{1\perp}}{(2\pi)^3 2k^-_1} \Big|_{k^2_1=0}\frac{dk^-_2d^2k_{2\perp}}{(2\pi)^3 2k^-_2} \Big|_{k^2_2=0}\label{eq:interf-first}\\
&&\times 2\pi\delta(q^--k^-_1-k^-_2)
\bar{P}^k \Big( g^{ml} 
+ \frac{2\bar{P}^m\bar{P}^l}{\bar{P}^2_{\perp} + z\bar{z}Q^2 } \Big) \frac{16z^2\bar{z}^2( 1 - 2 z)(q^-)^3}{(\bar{P}^2_{\perp} + z\bar{z}Q^2)^3} \int d^2x_\perp e^{i(k_{1\perp}+k_{2\perp}-q_\perp)x_\perp}\nonumber\\
&&\times \int_{-\infty}^{\infty} dz^- e^{-i(k^+_1+k^+_2-q^+)z^-}
\langle P, S| F^a_{-k}(z^- , x_\perp) U^{ac}(z^-, \infty; x_\perp) U^{cd}(x_\perp, 0_\perp;\infty) U^{db}(\infty, 0^-;0_\perp)\nonumber\\
&&\times F^b_{-m}(0^-, 0_\perp) |P, S\rangle\,,
\nonumber
\end{eqnarray}
and
\begin{eqnarray}
&&d\sigma^{\gamma^\ast p}_{\lambda\pm10}\Big|^{\rm back-to-back}_{\rm lead.\ power} = \sum_f \frac{2\pi e^2_f g^2 \alpha_{\rm EM}x}{Q^2}   \epsilon^{\lambda=\pm1\ast}_s \frac{Q}{q^-} \frac{dk^-_1d^2k_{1\perp}}{(2\pi)^3 2k^-_1} \Big|_{k^2_1=0}\frac{dk^-_2d^2k_{2\perp}}{(2\pi)^3 2k^-_2} \Big|_{k^2_2=0}\label{eq:interf-second}\\
&&\times 2\pi\delta(q^--k^-_1-k^-_2)
\Big( g^{sk} + \frac{2\bar{P}^s \bar{P}^k}{\bar{P}^2_\perp + z\bar{z}Q^2} \Big) \bar{P}^m \frac{16z^2\bar{z}^2 ( 1 - 2z ) (q^-)^3}{(\bar{P}^2_\perp + z\bar{z}Q^2)^3} \int d^2x_\perp e^{i(k_{1\perp}+k_{2\perp}-q_\perp)x_\perp}\nonumber\\
&&\times \int_{-\infty}^{\infty} dz^- e^{-i(k^+_1+k^+_2-q^+)z^-}
\langle P, S| F^a_{-k}(z^- , x_\perp) U^{ac}(z^-, \infty; x_\perp) U^{cd}(x_\perp, 0_\perp;\infty) U^{db}(\infty, 0^-;0_\perp)\nonumber\\
&&\times F^b_{-m}(0^-, 0_\perp) |P, S\rangle\,,
\nonumber
\end{eqnarray}
as well as
\begin{eqnarray}
&&d\sigma^{\gamma^\ast p}_{\lambda=\pm1\lambda'=\pm1}\Big|^{\rm back-to-back}_{\rm lead.\ power} =   \sum_f \frac{2\pi e^2_f g^2 \alpha_{\rm EM}x}{Q^2} \frac{dk^-_1d^2k_{1\perp}}{(2\pi)^3 2k^-_1} \Big|_{k^2_1=0}\frac{dk^-_2d^2k_{2\perp}}{(2\pi)^3 2k^-_2} \Big|_{k^2_2=0} 
\label{eq:lead-order-col-btb-TT}\\
&&\times 2\pi\delta(q^--k^-_1-k^-_2)
\frac{4z^2\bar{z}^2 (q^-)^2}{(\bar{P}^2_\perp + z\bar{z}Q^2)^2} \Big[ -\delta_{\lambda\lambda'} \Big( \frac{z}{\bar{z}} + \frac{\bar{z}}{z}\Big) ( g_{sl} + i\lambda \epsilon_{sl} ) +  2 \delta_{\lambda(-\lambda')} \Big( g_{1s}g_{1l} \nonumber\\
&&-  g_{2s}g_{2 l} + i \lambda ( g_{1s}g_{2 l} +  g_{2s} g_{1l}) \Big) \Big]
\Big(  g^{ks}g^{ml} + \frac{2g^{ks}\bar{P}^m \bar{P}^l }{\bar{P}^2_\perp + z\bar{z}Q^2} + \frac{2g^{ml} \bar{P}^k \bar{P}^s }{\bar{P}^2_\perp + z\bar{z}Q^2} + \frac{4 \bar{P}^k \bar{P}^s \bar{P}^m \bar{P}^l }{(\bar{P}^2_\perp + z\bar{z}Q^2)^2}\Big)\nonumber\\
&&\times \int d^2x_\perp e^{i(k_{1\perp}+k_{2\perp}-q_\perp)x_\perp} \int dz^- e^{-i(k^+_1+k^+_2-q^+)z^-}\nonumber\\
&&\times 
\langle P, S| F^a_{-k}(z^- , x_\perp) U^{ac}(z^-, \infty; x_\perp) U^{cd}(x_\perp, 0_\perp;\infty)U^{db}(\infty, 0^-;0_\perp) F^b_{-m}(0^-, 0_\perp) |P, S\rangle\,.
\nonumber
\end{eqnarray}

In an experimental setting, the quark jet is not distinguished from the antiquark jet. One can therefore symmetrize the cross sections under the interchange $k_1\leftrightarrow k_2$ (equivalently $z\leftrightarrow \bar{z}$). The symmetrized cross sections~(\ref{eq:interf-first}) and~(\ref{eq:interf-second}) then vanish due to the $(1-2z)$ factor, which is antisymmetric under $z\leftrightarrow \bar{z}$. At leading power in the back-to-back limit, it therefore suffices to consider only the cross sections~(\ref{eq:lead-order-col-btb}) and~(\ref{eq:lead-order-col-btb-TT}).


The dijet production cross section in the back-to-back limit is determined by a matrix element of the gluon TMD operator. This matrix element can be parametrized in terms of the leading-power gluon TMD distributions, see Refs.~\cite{Mulders:2000sh,Boer:2016xqr}:
\begin{eqnarray}
&& \int_{-\infty}^{\infty} \frac{dz^-}{2\pi} e^{-i x_{\rm djet} P^+ z^-} \int \frac{d^2x_\perp}{(2\pi)^2} e^{i \Delta_\perp x_\perp} \langle P, S|
    F^a_{-k}(z^- , x_\perp) U^{ac}(z^-, \infty; x_\perp) U^{cd}(x_\perp, 0_\perp;\infty)\nonumber\\
&&\times U^{db}(\infty, 0^-;0_\perp) F^b_{-m}(0^-, 0_\perp) |P, S\rangle\label{eq:TMD-gluon-param}\\
&&= \frac{x_{\rm djet}P^+}{2}\Big[ - g^{km} f_1(x_{\rm djet}, \Delta^2_\perp) + \frac{\Delta^{km}_\perp}{M^2} h^\perp_1(x_{\rm djet}, \Delta^2_\perp) + i\epsilon^{km} S_L g_1(x_{\rm djet}, \Delta^2_\perp)
\nonumber\\
&&+ \frac{\epsilon^{\{k}_{\ \ \ s}\Delta^{m\}s}_\perp S_L}{2M^2} h^\perp_{1L}(x_{\rm djet}, \Delta^2_\perp) -\frac{g^{km} \epsilon^{sl}S^s_{T}\Delta^l_\perp}{M} f^\perp_{1T}(x_{\rm djet}, \Delta^2_\perp) + \frac{i\epsilon^{km} (\Delta_\perp \cdot S_T)}{M} g_{1T}(x_{\rm djet}, \Delta^2_\perp)
\nonumber\\
&&+ \frac{\Delta^s_\perp\epsilon^{s\{k} S^{m\}}_T + S^s_T \epsilon^{s\{k} \Delta^{m\}}_\perp}{4M}h_1(x_{\rm djet}, \Delta^2_\perp) + \frac{\epsilon^{\{k}_{\ \ \ s} \Delta^{m\}s l}_\perp S^l_T}{2M^3} h^\perp_{1T}(x_{\rm djet}, \Delta^2_\perp)\Big]\,,
\nonumber
\end{eqnarray}
where
\begin{eqnarray}
&&x_{\rm djet} \equiv \frac{\bar{P}^2_{\perp} + z\bar{z}Q^2}{2z\bar{z}(P\cdot q)},
\end{eqnarray}
see the corresponding exponential factor in Eqs. (\ref{eq:lead-order-col-btb}) and (\ref{eq:lead-order-col-btb-TT}). In Eq. (\ref{eq:TMD-gluon-param}) we use the standard notations: $a^{\{k} b^{m\}} \equiv a^k b^m + a^m b^k$, $\epsilon^{12} = - \epsilon^{21} = 1$, $g^{11} = g^{22} = -1$,
\begin{eqnarray}
&&\Delta^{km}_\perp \equiv \Delta^k_\perp \Delta^m_\perp + \frac{1}{2}\Delta^2_\perp g^{km}\,,
\end{eqnarray}
and 
\begin{eqnarray}
&&\Delta^{kms}_\perp \equiv \Delta^k_\perp \Delta^m_\perp \Delta^s_\perp + \frac{1}{4} \Delta^2_\perp \Big(g^{km} \Delta^s_\perp + g^{ks} \Delta^m_\perp + g^{ms} \Delta^k_\perp\Big)\,.
\end{eqnarray}

Substituting the parametrization~(\ref{eq:TMD-gluon-param}) into Eq.~(\ref{eq:lead-order-col-btb}), we obtain
\begin{eqnarray}
&&d\sigma^{\gamma^\ast p}_{00}\Big|^{\rm back-to-back}_{\rm lead.\ power} = \sum_f (2\pi)^4 e^2_f g^2 \alpha_{\rm EM}x \frac{dk^-_1d^2k_{1\perp}}{(2\pi)^3 2k^-_1} \Big|_{k^2_1=0}\frac{dk^-_2d^2k_{2\perp}}{(2\pi)^3 2k^-_2} \Big|_{k^2_2=0}\label{eq:00-btb-param}\\
&&\times 2\pi \delta(q^--k^-_1-k^-_2) q^-
\frac{16 z^2 \bar{z}^2 \bar{P}^2_\perp }{(\bar{P}^2_\perp + z\bar{z}Q^2)^3} \Big( f_1(x_{\rm djet}, \Delta^2_\perp) + \frac{\Delta^2_\perp}{2M^2} \cos (2 (\phi_{\bar{P}} - \phi_{\Delta})) h^\perp_1(x_{\rm djet}, \Delta^2_\perp)\nonumber\\
&&+ \frac{ \Delta^2_\perp S_L }{2M^2}  \sin(2(\phi_{\bar{P}} - \phi_{\Delta}))
  h^\perp_{1L}(x_{\rm djet}, \Delta^2_\perp)  + \frac{ | \Delta_\perp| |S_\perp| }{M} \sin(\phi_\Delta - \phi_S )f^\perp_{1T}(x_{\rm djet}, \Delta^2_\perp)
\nonumber\\
&&+ \frac{  |\Delta_\perp| |S_\perp| }{2M} \sin(2\phi_{\bar{P}} - \phi_\Delta - \phi_S) h_1(x_{\rm djet}, \Delta^2_\perp) + \frac{ |\Delta_\perp| |S_\perp| \Delta^2_\perp }{4M^3} \sin(2\phi_{\bar{P}} - 3\phi_\Delta + \phi_S) h^\perp_{1T}(x_{\rm djet}, \Delta^2_\perp)\Big)\,,
\nonumber
\end{eqnarray}
where $\phi_{\bar{P}}$, $\phi_{\Delta}$, and $\phi_S$ denote the azimuthal angles of the $\bar{P}_\perp$, $\Delta_\perp$, and $S_T$ vectors in the transverse plane, such that $\epsilon^{km} \bar{P}_k \Delta_m = |\bar{P}_\perp||\Delta_\perp|\sin( \phi_{\Delta} - \phi_{\bar{P}})$, etc.

Similarly, from Eq.~(\ref{eq:lead-order-col-btb-TT}) at $\lambda = \lambda'=\pm 1$ we obtain\footnote{For brevity, we do not present the $\delta_{\lambda(-\lambda')}$ contribution to the dijet production cross section. This term contains various angular modulations whose explicit form is not essential for our discussion. A thorough phenomenological analysis of dijet production goes beyond the scope of this paper.}
\begin{eqnarray}
&&d\sigma^{\gamma^\ast p}_{\lambda=\lambda'=\pm1}\Big|^{\rm back-to-back}_{\rm lead.\ power} =   \sum_f (2\pi)^4 e^2_f g^2 \alpha_{\rm EM}x \frac{dk^-_1d^2k_{1\perp}}{(2\pi)^3 2k^-_1} \Big|_{k^2_1=0}\frac{dk^-_2d^2k_{2\perp}}{(2\pi)^3 2k^-_2} \Big|_{k^2_2=0}\label{eq:11-btb-param}\\
&&\times 2\pi\delta(q^--k^-_1-k^-_2) q^-
 \frac{ 4 z\bar{z} (z^2 + \bar{z}^2) }{(\bar{P}^2_\perp + z\bar{z}Q^2)^3} \Big( \frac{ \bar{P}^4_\perp + z^2\bar{z}^2Q^4 }{2z\bar{z}Q^2} f_1(x_{\rm djet}, \Delta^2_\perp) - \frac{ \Delta^2_\perp \bar{P}^2_\perp}{2M^2} \cos (2 (\phi_{\bar{P}} - \phi_{\Delta})) h^\perp_1(x_{\rm djet}, \Delta^2_\perp)
 \nonumber\\
&& - \frac{ \bar{P}^4_\perp - z^2\bar{z}^2Q^4  }{2z\bar{z} Q^2 }  \lambda S_L g_1(x_{\rm djet}, \Delta^2_\perp)
- \frac{ \Delta^2_\perp P^2_\perp S_L }{2M^2} \sin (2(\phi_{\bar{P}} - \phi_{\Delta}))  h^\perp_{1L}(x_{\rm djet}, \Delta^2_\perp) + \frac{\bar{P}^4_\perp  + z^2\bar{z}^2Q^4 }{ 2z\bar{z}Q^2 }
\nonumber\\
&&\times \frac{ | \Delta_\perp| |S_\perp| }{M}  \sin(\phi_\Delta - \phi_S ) f^\perp_{1T}(x_{\rm djet}, \Delta^2_\perp)  - \frac{ \bar{P}^4_\perp - z^2\bar{z}^2Q^4 }{2z\bar{z}Q^2} \frac{ \lambda |\Delta_\perp| |S_\perp| }{M} \cos(\phi_\Delta - \phi_S) g_{1T}(x_{\rm djet}, \Delta^2_\perp)  
\nonumber\\
&&- \frac{  |\Delta_\perp| |S_\perp| P^2_\perp }{2M} \sin( 2\phi_{\bar{P}} - \phi_\Delta - \phi_S) h_1(x_{\rm djet}, \Delta^2_\perp) 
- \frac{ |\Delta_\perp| |S_\perp| \bar{P}^2_\perp \Delta^2_\perp }{4M^3} \sin(2\phi_{\bar{P}} - 3\phi_\Delta + \phi_S)   h^\perp_{1T}(x_{\rm djet}, \Delta^2_\perp)\Big) \,.
\nonumber
\end{eqnarray}
The cross sections~(\ref{eq:00-btb-param}) and~(\ref{eq:11-btb-param}) are sensitive not only to the unpolarized distribution $f_1(x, \Delta^2_\perp)$ and the linearly polarized distribution $h^\perp_1(x, \Delta^2_\perp)$, but also to the $T$-even distributions $g_1$ and $g_{1T}$, as well as the $T$-odd distributions $h^\perp_{1L}(x, \Delta^2_\perp)$, $f^\perp_{1T}(x, \Delta^2_\perp)$, $h_1(x, \Delta^2_\perp)$, and $h^\perp_{1T}(x, \Delta^2_\perp)$, providing access to the spin dependence of the target. The dijet production cross section in the back-to-back limit thus offers an efficient channel for mapping the gluon content of the dense QCD medium of the target.
\end{widetext}

\subsection{Subleading power corrections in the back-to-back limit}
One can, of course, compute power corrections to the above results. This involves evaluating the $\Delta_\perp/\bar{P}_\perp$-suppressed terms of the perturbative coefficient in Eq.~(\ref{eq:btb-ltwist}), as well as deriving the contribution of power-suppressed operators in the expansion of the path-ordered exponents, corresponding to the higher-order terms of the expansion onto the light-cone contour~(\ref{eq:prop-ex-tr-shift-commute-colexp}). Since this expansion is consistent with the expansion parameter, only the first few higher-order terms are needed. Substituting them from Eq.~(\ref{eq:prop-ex-tr-shift-commute-colexp-fixtrpos}) into the cross section~(\ref{eq:hadronic-bfm-general}), one can identify all operators appearing at subleading power. These operators belong to the hierarchy defining dijet production in the back-to-back limit.


For example, one such subleading contribution takes the following form, cf.\ Eq.~(\ref{eq:djet-btb-col-lead-init}):
\begin{widetext}
\begin{eqnarray}
&&d\sigma^{\gamma^\ast p}_{00}\Big|^{\rm back-to-back}_{\rm sublead.;(DF)F} = \sum_f\sum_{s_1,s_2}\frac{2\pi e^2_f \alpha_{\rm EM}x}{V_4(q^-)^2} \frac{dk^-_1d^2k_{1\perp}}{(2\pi)^3 2k^-_1} \Big|_{k^2_1=0}\frac{dk^-_2d^2k_{2\perp}}{(2\pi)^3 2k^-_2} \Big|_{k^2_2=0} \int d^4x e^{i(q-k_1-k_2)x}\langle P, S| \bar{v}_{s_2}(k_2)
\label{eq:DF-contribution}\\
&&\times \Big\{ i\gamma^- \Big( \frac{1}{2k^-_1} + \frac{1}{2k^-_2} \Big) ig \int_{x^-}^{\infty} dz^-(z^--x^-) U(\infty, z^-; x_\perp) D_{k} F_{-k}(z^-, x_\perp) U(z^-, \infty; x_\perp) \Big\} u_{s_1}(k_1) \int d^4y e^{-i(q-k_1-k_2)y}
\nonumber\\
&&\times   \bar{u}_{s_1}(k_1) \Big\{ \gamma^- \Big( \frac{k_{1m}}{k^-_1} - \frac{k_{2m}}{k^-_2}\Big) ig \int_{y^-}^{\infty} dz'^-(z'^- - y^-) U(\infty, z'^-;y_\perp) F_{-m}(z'^-, y_\perp) 
U(z'^-, \infty;y_\perp) \Big\} v_{s_2}(k_2) +\bar{v}_{s_2}(k_2)\nonumber\\
&&\times \Big\{ - \gamma^- \Big( \frac{k_{1k}}{k^-_1} - \frac{k_{2k} }{k^-_2} \Big) ig \int_{x^-}^{\infty} dz^-(z^--x^-) U(\infty, z^-; x_\perp) F_{-k}(z^-, x_\perp)
 U(z^-, \infty; x_\perp) \Big\} u_{s_1}(k_1) \int d^4y e^{-i(q-k_1-k_2)y}\nonumber\\
 &&\times  \bar{u}_{s_1}(k_1) \Big\{ i\gamma^- \Big( \frac{1}{2k^-_1} + \frac{1}{2k^-_2}\Big) ig \int_{y^-}^{\infty} dz'^-(z'^- - y^-) 
U(\infty, z'^-;y_\perp) D_{m} F_{-m}(z'^-, y_\perp) U(z'^-, \infty;y_\perp) \Big\} v_{s_2}(k_2) 
 |P, S\rangle\,.
\nonumber
\end{eqnarray}
Following the steps of the previous computation, one can simplify Eq.~(\ref{eq:DF-contribution}) and derive a result analogous to Eq.~(\ref{eq:lead-order-col-btb}). The subleading-power contribution is determined by the operator\footnote{Here we do not explicitly fix the transverse position of the background field operators.}
\begin{eqnarray}
&&(-k_2|ig\int_{-\infty}^{\infty} dz^- e^{i(q^+-k^+_1-k^+_2)z^-}U(\infty, z^-) D_{k} F_{-k}(z^-) U(z^-, \infty)|k_1)
\label{eq:oper-sublead}\\
&&\times(k_1|ig\int_{-\infty}^{\infty} dz'^- e^{i(q^+-k^+_1-k^+_2)z'^-}U(\infty, z'^-) F_{-m}(z'^-) U(z'^-, \infty)|-k_2)\,.
\nonumber
\end{eqnarray}

Using an identity
\begin{eqnarray}
&& igU(\infty, z^-) D_k F_{-k}(z^-) U(z^-, \infty)= P_kU(\infty, z^-) F_{-k}(z^-) U(z^-, \infty) - U(\infty, z^-) F_{-k}(z^-) U(z^-, \infty) P_k 
\\
&&+ g\int_{z^-}^{\infty} dx^- \Big( U(\infty, z^-) F_{-k}(z^-) U(z^-,x^-)F_{-k}(x^-)U(x^-, \infty)
- U(\infty, x^-)F_{-k}(x^-)U(x, z^-) F_{-k}(z^-) U(z^-, \infty)\Big)\,,\nonumber
\end{eqnarray}
one can rewrite the operator (\ref{eq:oper-sublead}) as
\begin{eqnarray}
&&(-k_2|ig\int_{-\infty}^{\infty} dz^- e^{i(q^+-k^+_1-k^+_2)z^-}U(\infty, z^-) D_{k} F_{-k}(z^-) U(z^-, \infty)|k_1)\label{eq:rel-op} (k_1|ig\int_{-\infty}^{\infty} dz'^- e^{i(q^+-k^+_1-k^+_2)z'^-}
U(\infty, z'^-)
\nonumber\\
&&\times F_{-m}(z'^-) U(z'^-, \infty)|-k_2) = (-k_2|\int_{-\infty}^{\infty} dz^- e^{i(q^+-k^+_1-k^+_2)z^-} \Big\{ -\Delta_k U(\infty, z^-) F_{-k}(z^-) U(z^-, \infty)
 \nonumber\\
&&+ g\int_{z^-}^{\infty} dx^- \Big( U(\infty, z^-) F_{-k}(z^-) U(z^-,x^-)F_{-k}(x^-)U(x^-, \infty) - U(\infty, x^-)F_{-k}(x^-)U(x, z^-) F_{-k}(z^-) U(z^-, \infty)\Big)\Big\} |k_1)
\nonumber\\
&&\times  (k_1|ig\int_{-\infty}^{\infty} dz'^- e^{i(q^+-k^+_1-k^+_2)z'^-}U(\infty, z'^-) F_{-m}(z'^-) U(z'^-, \infty)|-k_2)\,.
\end{eqnarray}
\end{widetext}
This relation illustrates two points. First, it reflects the ambiguity in the choice of operator basis that typically arises in computations at subleading power. Second, it shows that operator~(\ref{eq:oper-sublead}) in Eq.~(\ref{eq:DF-contribution}) is suppressed by an extra power of $\Delta_\perp/\bar{P}_\perp$ in the back-to-back limit, as seen from the first term in Eq.~(\ref{eq:rel-op}). This means that to compute the cross section at subleading power, one must combine this operator contribution, Eq.~(\ref{eq:DF-contribution}), with the $\Delta_\perp/\bar{P}_\perp$-suppressed corrections to the perturbative coefficients in Eq.~(\ref{eq:btb-ltwist}). Similarly, Eq.~(\ref{eq:rel-op}) indicates that TMD operators with three field-strength tensor insertions should also be attributed to subleading-power corrections of order $\Delta_\perp/\bar{P}_\perp$.

To summarize, computing the subleading-power contribution to the dijet production cross section requires a careful inspection of the first few orders of the expansion of the path-ordered exponents in the cross section~(\ref{eq:hadronic-bfm-general}) onto the light-cone contour in Fig.~\ref{fig:fig13}, including the perturbative coefficients, with the expansion parameter $\Delta_\perp/\bar{P}_\perp\ll 1$ applied throughout. Since this analysis is not the main focus of this paper, we defer it to a separate publication~\cite{ColDijet}.

\section{The high-energy power counting\label{sec:djprod-hepc}}
In the previous sections, we developed an approach for expanding the path-ordered exponents~(\ref{eq:definition-exp-shifted-quark}) appearing in expressions for physical observables --- see e.g.\ Eq.~(\ref{eq:hadronic-bfm-general}) for dijet production --- onto a given linear piecewise contour. As discussed above, the choice of expansion contour should be consistent with the expansion parameter of the problem. For the dijet production in the back-to-back limit, we expanded the path-ordered exponents onto the light-cone contour in Fig.~\ref{fig:fig13}. From the leading-order terms of this expansion, we derived the leading-power contribution to dijet production in the back-to-back limit, determined by a matrix element of the gluon TMD operator, see Eq.~(\ref{eq:lead-order-col-btb}). We also outlined a procedure for computing subleading-power corrections. The goal of this section is to carry out an analogous analysis for dijet production in the high-energy limit of small $x$.


In the small-$x$ limit, parton emission is characterized by a wide separation in longitudinal momenta, while the transverse momenta remain comparable. Within the background field approach to QCD factorization, this implies that the typical transverse momenta of the quantum partons and the background fields are of the same order, $P_\perp\sim \Delta_\perp$, rendering the back-to-back power counting developed in the previous section inapplicable.




To compute physical observables at small $x$, we therefore need to introduce an alternative power counting. In our approach, the power counting of various contributions arising from the expansion of the path-ordered exponents at small $x$ is defined by the high-energy power counting of the background fields~(\ref{eq:pw-couting-sx}). Applying this power counting to each term of the expansion, we can estimate its size as a power of the boost parameter $\lambda$, which serves as the expansion parameter.


The leading-power contribution to a physical observable is of order $\sim \lambda^0$, commonly referred to as the eikonal order in the small-$x$ literature. As discussed above, the expansion contour for the path-ordered exponents should be consistent with the expansion parameter, so that the leading terms of the expansion correspond to the eikonal contribution $\sim \lambda^0$.


By retaining a finite number of higher-order terms, one can also fully reconstruct corrections to the eikonal order, i.e.\ the subleading-power corrections of order $\sim \lambda^{-1}$, commonly referred to as sub-eikonal corrections.


To illustrate how the high-energy power counting~(\ref{eq:pw-couting-sx}) is used to compute physical observables at small $x$, let us apply it to the expansion of the path-ordered exponents onto the light-cone direction. The general form of this expansion is given in Eq.~(\ref{eq:Bi0exp-general-form-Bneq0}). Each successive term of the expansion corresponds to a higher number of operator insertions
\begin{eqnarray}
&&\int dz^- \mathcal{O}_{\mathcal{B}}(z^-)
\label{eq:oper-ins}
\end{eqnarray}
into the light-cone gauge link. The light-cone gauge factors themselves are irrelevant for the power counting, since according to Eq.~(\ref{eq:pw-couting-sx})
\begin{eqnarray}
&& \mathcal{P}\exp\Big\{ ig\int dz^- B_-(z^-)  \Big\} \sim \mathcal{P}\exp\Big\{ ig\int dz^- \tilde{B}_-(z^-)  \Big\}\nonumber\\
&&\sim \lambda^0\,,
\end{eqnarray}
so the gauge links do not affect the power counting of the operators.

It therefore suffices to consider the power counting of the operator insertions~(\ref{eq:oper-ins}) without the gauge link factors. The general form of the operator insertion~(\ref{eq:oper-ins}), see Eq.~(\ref{eq:general-Ob}), reads
\begin{widetext}
\begin{eqnarray}
&&\int dz^- \mathcal{O}_{\mathcal{B}}(z^-)\sim \frac{g}{4p^-}\int dz^-\sigma^{\mu\nu}F_{\mu\nu}(z^-) + \int dz^- \frac{iz^-}{2p^-}\mathcal{B}(z^-)
+ \frac{1}{2!}\int dz^- \Big(\frac{iz^-}{2p^-}\Big)^2
\nonumber\\
&&\times [ P^2_\perp(z^-), \mathcal{B}(z^-) ] + \frac{1}{3!} \int dz^- \Big(\frac{iz^-}{2p^-}\Big)^3[P^2_\perp, [ P^2_\perp(z^-), \mathcal{B}(z^-) ]] \dots \,,
\label{eq:exp-lcone-gen}
\end{eqnarray}
\end{widetext}
where operator $\mathcal{B}$ is given in Eq. (\ref{eq:operB}).

From the high-energy power counting~(\ref{eq:pw-couting-sx}), one observes that each transverse covariant momentum $P_k$ contributes a factor $\sim \lambda^0$ and thus does not affect the power counting of terms in Eq.~(\ref{eq:exp-lcone-gen}). We can therefore neglect the transverse momentum $P_k$ when estimating the size of individual terms.


Consider first the first term in Eq.~(\ref{eq:exp-lcone-gen}). Its size can be estimated using the high-energy power counting~(\ref{eq:pw-couting-sx}) for the background gluon field-strength tensors:
\begin{align}
&F_{-i}(x^+, x^-, x_\perp) \sim \lambda \tilde{F}_{-i}(\lambda^{-1}x^+, \lambda x^-, x_\perp)\,,
\label{eq:st-param1}
\\
&F_{-+}(x^+, x^-, x_\perp) \sim \tilde{F}_{-+}(\lambda^{-1}x^+, \lambda x^-, x_\perp)\,,
\label{eq:st-param2}
\\
&F_{ij}(x^+, x^-, x_\perp) \sim \tilde{F}_{ij}(\lambda^{-1}x^+, \lambda x^-, x_\perp)\,,
\label{eq:st-param3}
\\
&F_{+i}(x^+, x^-, x_\perp) \sim \lambda^{-1} \tilde{F}_{+i}(\lambda^{-1} x^+, \lambda x^-, x_\perp)\,.
\label{eq:st-param4}
\end{align}
The operators in the first term of Eq.~(\ref{eq:exp-lcone-gen}) can be estimated as\footnote{In our computation, the background field is taken in the gauge $B_+=0$, for which the field-strength tensors $F_{-+}$ and $F_{+i}$ vanish.}
\begin{equation}
\int dz^- \sigma^{-k} F_{-k}(z^-) \sim \int dz^- \sigma^{-k} \tilde{F}_{-k}( z^-) \sim \lambda^0
\label{eq:powc-sigmafik}
\end{equation}
and
\begin{equation}
\int dz^- \sigma^{km} F_{km}(z^-) \sim \frac{1}{\lambda}\int dz^- \sigma^{km} \tilde{F}_{km}(z^-)
\sim \lambda^{-1}\,.
\label{eq:powc-sigmafik-tr}
\end{equation}

From Eq.~(\ref{eq:powc-sigmafik-tr}), we conclude that each insertion of the fully transverse field-strength tensor $\sigma^{km} F_{km}$ is suppressed by a power $\sim \lambda^{-1}$. When computing physical observables at the eikonal order, this contribution should therefore be neglected; at the sub-eikonal order, it suffices to include this insertion only once.


At the same time, from Eq.~(\ref{eq:powc-sigmafik}) the $\sigma^{-k} F_{-k}$ contribution is unsuppressed, $\sim \lambda^0$. This operator should therefore be retained in each $\mathcal{O}_B$ insertion of the light-cone expansion~(\ref{eq:Bi0exp-general-form-Bneq0}) already at the eikonal order. This could potentially pose a difficulty, since computing physical observables at the eikonal order would then require an infinite number of such insertions. Fortunately, multiple insertions of~(\ref{eq:powc-sigmafik}) are typically trivial due to the algebra of $\gamma$-matrices: $\sigma^{-k}\sigma^{-m} = 0$, etc. It is therefore generally sufficient to include only a finite number of these insertions in the light-cone expansion~(\ref{eq:Bi0exp-general-form-Bneq0}).


Now consider the remaining terms of Eq.~(\ref{eq:exp-lcone-gen}). In terms of the power counting, these terms have the following pattern:\footnote{As noted above, the power counting of the transverse derivatives $P_k$ is trivial, $\sim \lambda^0$, and can be neglected.}
\begin{align}
&\int dz^- \frac{iz^-}{2p^-}\mathcal{B}(z^-)
+ \frac{1}{2!}\int dz^- \Big(\frac{iz^-}{2p^-}\Big)^2[ P^2_\perp(z^-), \mathcal{B}(z^-) ] \nonumber\\
&+ \frac{1}{3!}\int dz^- \Big(\frac{iz^-}{2p^-}\Big)^3[P^2_\perp(z^-), [ P^2_\perp(z^-), \mathcal{B}(z^-) ]] + \dots 
\nonumber\\
&\sim \sum^\infty_{n=1} \int dz^- (z^-)^n \mathcal{B}(z^-)\,.
\label{eq:stermsBetc}
\end{align}

Using Eq. (\ref{eq:operB}) we can further estimate the power counting  of these terms as 
\begin{align}
&\sum^\infty_{n=1} \int dz^- (z^-)^n \mathcal{B}(z^-) \sim \sum^\infty_{n=1} \lambda^{-n} \int d z^- ( z^-)^n \label{eq:sec-terms_bsh}\\
&\times \Big(  \tilde{F}_{k-}( z^-) + \sigma^{-m}  D_k\tilde{F}_{-m}( z^-) + \lambda^{-1} \sigma^{ms} D_k\tilde{F}_{ms}( z^-) \Big)\,.
\nonumber
\end{align}
Each term in this series is suppressed by $\lambda^{-n}$, originating from the factor $(z^-)^n$ in Eq.~(\ref{eq:stermsBetc}). This suppression is a consequence of the Lorentz boost of the background field encoded in the power counting~(\ref{eq:pw-couting-sx}).


These terms first contribute at the sub-eikonal order, where only the $n=1$ term is relevant:
\begin{eqnarray}
&&\int dz^- \frac{iz^-}{2p^-}\mathcal{B}(z^-)
+ \frac{1}{2!}\int dz^- \Big(\frac{iz^-}{2p^-}\Big)^2[ P^2_\perp(z^-), \mathcal{B}(z^-) ]\nonumber\\
&&+ \frac{1}{3!}\int dz^- \Big(\frac{iz^-}{2p^-}\Big)^3[P^2_\perp(z^-), [ P^2_\perp(z^-), \mathcal{B}(z^-) ]] + \dots 
\nonumber\\
&&\sim \lambda^{-1} \int d z^- z^- \Big(  \tilde{F}_{k-}( z^-) + \sigma^{-m}  D_k\tilde{F}_{-m}( z^-) \Big)\nonumber\\
&&\sim \lambda^{-1}\,.
\end{eqnarray}
At the sub-eikonal order in the light-cone expansion of Eq.~(\ref{eq:Bi0exp-general-form-Bneq0}), this term should therefore be included only once, making its treatment straightforward. At the eikonal order, these terms do not contribute.

We can thus estimate the power counting of the operator insertion~(\ref{eq:oper-ins}) at the eikonal order as
\begin{eqnarray}
&&\int dz^- \mathcal{O}_{\mathcal{B}}(z^-)\sim \frac{g}{2p^-}\int dz^-\sigma^{-k}F_{-k}(z^-)\nonumber\\
&&\sim \lambda^0 \,,
\end{eqnarray}

The expansion of the path-ordered exponents onto the light-cone direction~(\ref{eq:Bi0exp-general-form-Bneq0}) at the eikonal order within the power counting~(\ref{eq:pw-couting-sx}) therefore takes a simple form:
\begin{widetext}
\begin{eqnarray}
&&\mathcal{G}(x^-, y^-) \Big|_{\rm eik.} = \mathcal{P}\exp\Big\{ ig\int^{x^-}_{y^-} dz^- \Big(B_-(z^-) + \frac{1}{2p^-}\sigma^{-k}F_{-k}(z^-) \Big) \Big\}
\nonumber\\
&&= U(x^-, y^-) + \frac{ig}{2p^-}\int^{x^-}_{y^-} dz^- U(x^-, z^-)\sigma^{-k}F_{-k}(z^-) U(z^-, y^-) \,.
\label{eq:staple-expan-prop-eik-expon}
\end{eqnarray}
\end{widetext}
Our approach also allows one to derive the sub-eikonal corrections to this result in a systematic fashion.


The power counting procedure outlined above can be applied to computing physical observables once the path-ordered exponents have been expanded onto a contour of appropriate choice. In the next section, we carry this out explicitly by applying the high-energy power counting~(\ref{eq:pw-couting-sx}) to compute the dijet production cross section~(\ref{eq:hadronic-bfm-general}) at the eikonal order.


\section{Dijet production at the eikonal order\label{sec:djpro-eik}}
Let us apply the formalism developed in the previous sections to compute the dijet production cross section~(\ref{eq:hadronic-bfm-general}) in the high-energy limit of small $x$ at the eikonal order. To this end, we apply the high-energy power counting~(\ref{eq:pw-couting-sx}) to the expansion of the (anti)quark propagators in the cross section~(\ref{eq:hadronic-bfm-general}):
\begin{equation}
(k_{1\perp}|  e^{i\frac{ k^2_{1\perp}}{2k^-_1} z^-} e^{-i\frac{P^2_\perp(z^-) }{2k^-_1} z^-} \Big|^\infty \mathcal{G}(\infty, y^-) e^{i\frac{P^2_\perp(y^-)}{2k^-_1}y^-} |y_\perp)\,.
\label{eq:prop-ex-tr-eik}
\end{equation}

As a first step, we expand the path-ordered exponent $\mathcal{G}(\infty, y^-)$ in Eq.~(\ref{eq:prop-ex-tr-eik}) onto the light-cone direction, corresponding to expanding the background fields onto the contour in Fig.~\ref{fig:fig12}.\footnote{The background fields of the path-ordered exponent in Eq.~(\ref{eq:prop-ex-tr-eik}) are expanded onto the light-cone direction at $z_\perp$ in Fig.~\ref{fig:fig12}.} Applying the power counting procedure of the previous section, at the eikonal order we obtain
\begin{widetext}
\begin{eqnarray}
&&(k_{1\perp}|  e^{i\frac{ k^2_{1\perp}}{2k^-_1} z^-} e^{-i\frac{P^2_\perp(z^-) }{2k^-_1} z^-} \Big|^\infty \mathcal{G}(\infty, y^-) e^{i\frac{P^2_\perp(y^-)}{2k^-_1}y^-} |y_\perp)\Big|_{\rm eik.}\label{eq:prop-eik-general} \\
&&= \int d^2z_\perp (k_{1\perp}|  e^{i\frac{ k^2_{1\perp}}{2k^-_1} z^-} e^{-i\frac{P^2_\perp(z^-) }{2k^-_1} z^-} \Big|^\infty |z_\perp)
\Big(U(\infty, y^-;z_\perp) + \frac{ig}{2k^-_1}\int^{\infty}_{y^-} dz^- U(\infty, z^-;z_\perp)\nonumber\\
&&\times \sigma^{-k}F_{-k}(z^-,z_\perp) U(z^-, y^-;z_\perp)\Big) (z_\perp| e^{i\frac{P^2_\perp(y^-)}{2k^-_1}y^-} |y_\perp) \,.
\nonumber
\end{eqnarray}
\end{widetext}

While this result has a simple form, it cannot be directly used to compute the dijet production cross section~(\ref{eq:hadronic-bfm-general}). The expression involves an exponential factor~(\ref{eq:exp-fact-cova}) at a finite position $y^-$, implying an infinite sum of transverse gauge links with various field-strength tensor insertions, see Eq.~(\ref{eq:tr-WL-insertion}). To handle this, we need to move the exponential factor to spatial infinity, where the gluon field is a pure gauge and the factor reduces to a simple transverse gauge link, see Eq.~(\ref{eq:tr-WL-insertion}).


\subsection{Exponential factor at the future infinity}
The first option is to move the exponential factor~(\ref{eq:exp-fact-cova}) in the (anti)quark propagator~(\ref{eq:prop-ex-tr-eik}) to future infinity. This was carried out in Sec.~\ref{sec:pshift} and corresponds to expanding all background fields in Eq.~(\ref{eq:prop-ex-tr-eik}) onto the light-cone contour in Fig.~\ref{fig:fig13}. The general form of this expansion is given in Eq.~(\ref{eq:light-cone-general}), with a ``shifted'' operator insertion
\begin{widetext}
\begin{eqnarray}
&&\int dz^- \mathcal{O}_{\mathcal{B}}(z^-;y^-)\label{eq:exp-lcone-gen-eik}\\
&&\sim \frac{g}{4p^-}\int dz^-\sigma^{\mu\nu}F_{\mu\nu}(z^-) + \int dz^- \frac{i(z^--y^-)}{2p^-}\mathcal{B}(z^-)
+ \frac{1}{2!}\int dz^- \Big(\frac{i(z^--y^-)}{2p^-}\Big)^2
\nonumber\\
&&\times [ P^2_\perp(z^-), \mathcal{B}(z^-) ] + \frac{1}{3!} \int dz^- \Big(\frac{i(z^--y^-)}{2p^-}\Big)^3[P^2_\perp, [ P^2_\perp(z^-), \mathcal{B}(z^-) ]] \dots \,,
\nonumber
\end{eqnarray}
cf. Eq. (\ref{eq:exp-lcone-gen}).

Applying the high-energy power counting to this expansion, we find that the power counting of the first term $\sim \sigma^{\mu\nu} F_{\mu\nu}$ is unchanged compared to the analysis of the previous section, see Eqs.~(\ref{eq:powc-sigmafik}) and~(\ref{eq:powc-sigmafik-tr}). However, the power counting of the remaining terms changes significantly. These terms can be estimated as
\begin{eqnarray}
&&\int dz^- \frac{i(z^--y^-)}{2p^-}\mathcal{B}(z^-)
+ \frac{1}{2!}\int dz^- \Big(\frac{i(z^--y^-)}{2p^-}\Big)^2[ P^2_\perp(z^-), \mathcal{B}(z^-) ] 
\label{eq:stermsBetc-eik}\\
&&+ \frac{1}{3!}\int dz^- \Big(\frac{i(z^--y^-)}{2p^-}\Big)^3[P^2_\perp(z^-), [ P^2_\perp(z^-), \mathcal{B}(z^-) ]] + \dots \sim \sum^\infty_{n=1} \int dz^- (z^--y^-)^n \mathcal{B}(z^-)\,,
\nonumber
\end{eqnarray}
cf. Eq. (\ref{eq:stermsBetc}), leading to 
\begin{eqnarray}
&&\sum^\infty_{n=1} \int dz^- (z^--y^-)^n \mathcal{B}(z^-) 
\label{eq:sec-terms_bsh-eik}\\
&&\sim \sum^\infty_{n=1}  \int d z^- ( \lambda^{-1}z^- - y^-)^n \Big(  \tilde{F}_{k-}( z^-) + \sigma^{-m}  D_k\tilde{F}_{-m}( z^-) + \lambda^{-1} \sigma^{ms} D_k\tilde{F}_{ms}( z^-) \Big)\,,
\nonumber
\end{eqnarray}
cf. Eq. (\ref{eq:sec-terms_bsh}).

Due to the longitudinal shift $y^-$, an infinite number of terms~(\ref{eq:stermsBetc-eik}) now contribute already at the eikonal order:
\begin{eqnarray}
&&\int dz^- \frac{i(z^--y^-)}{2p^-}\mathcal{B}(z^-)
+ \frac{1}{2!}\int dz^- \Big(\frac{i(z^--y^-)}{2p^-}\Big)^2[ P^2_\perp(z^-), \mathcal{B}(z^-) ] 
\label{eq:stermsBetc-eik-lead}\\
&&+ \frac{1}{3!}\int dz^- \Big(\frac{i(z^--y^-)}{2p^-}\Big)^3[P^2_\perp(z^-), [ P^2_\perp(z^-), \mathcal{B}(z^-) ]] + \dots \nonumber\\
&&\sim \sum^\infty_{n=1}  \int d z^- ( y^-)^n \Big(  \tilde{F}_{k-}( z^-) + \sigma^{-m}  D_k\tilde{F}_{-m}( z^-) \Big)\sim \lambda^0\,,
\nonumber
\end{eqnarray}
\end{widetext}
making the direct use of the light-cone expansion formula~(\ref{eq:light-cone-general}) for computing the dijet production problematic. The longitudinal shift $y^-$ arises from expanding the propagator~(\ref{eq:prop-ex-tr-eik}) onto the light-cone contour in Fig.~\ref{fig:fig13}. Nevertheless, the expansion formula~(\ref{eq:light-cone-general}) can still be utilized through the following observation, which we find to be specific to computing physical observables at small $x$ within the high-energy power counting~(\ref{eq:pw-couting-sx}).


Since the dijet production cross section~(\ref{eq:hadronic-bfm-general}) involves integration of the (anti)quark propagators~(\ref{eq:prop-ex-tr-eik}) over the longitudinal position $y^-$ of the virtual photon annihilation into the quark-antiquark pair, let us apply the light-cone expansion~(\ref{eq:light-cone-general}) only for $y^->0$ (and $x^->0$ for the complex conjugate amplitude).


In this case, the insertions~(\ref{eq:exp-lcone-gen-eik}) of the expansion formula~(\ref{eq:light-cone-general}) in the dijet production cross section~(\ref{eq:hadronic-bfm-general}) contain an integration of the form
\begin{eqnarray}
&&\int^{\infty}_0 dy^- e^{-i(q^+ - \frac{ k^2_{1\perp}}{2k^-_1}-\frac{ k^2_{2\perp}}{2k^-_2})y^-}  \int^\infty_{y^-} dz^- (y^-)^nO(z^-)\nonumber\\
&&= \int^\infty_{0^-} dz^- O(z^-) \int^{z^-}_0 dy^- (y^-)^n e^{-i(q^+ - \frac{ k^2_{1\perp}}{2k^-_1}-\frac{ k^2_{2\perp}}{2k^-_2})y^-}
\nonumber\\
&& \sim \int^\infty_{0} dz^- O(z^-) (z^-)^{n+1} \sim (\lambda^{-1})^{n+1}\,,
\label{eq:int-eik-sup}
\end{eqnarray}
where $O(z^-)$ denotes the various operators on the right-hand side of Eq.~(\ref{eq:exp-lcone-gen-eik}). The integration over $y^-$ thus yields a suppression of at least $\sim \lambda^{-1}$ for $n=0$, corresponding to the first terms of Eq.~(\ref{eq:exp-lcone-gen-eik}).


The choice of the lower integration limit $y^-=0$ is crucial: only this choice yields the necessary suppression. While this choice might be interpreted as reflecting the shock-wave localization of the background field, we find such an interpretation misleading, since the high-energy power counting~(\ref{eq:pw-couting-sx}) does not assume localization of the field around any fixed longitudinal position. In this sense, the power counting~(\ref{eq:pw-couting-sx}) provides a Lorentz-covariant description of the background field.


From Eq.~(\ref{eq:int-eik-sup}), we conclude that at the eikonal order the contribution of the operator insertions~(\ref{eq:exp-lcone-gen-eik}) in the propagator expansion~(\ref{eq:light-cone-general}) can be neglected for $y^->0$. We can therefore make the following replacement in the cross section~(\ref{eq:hadronic-bfm-general}) at the eikonal order:
\begin{eqnarray}
&&(k_{1\perp}|  e^{i\frac{ k^2_{1\perp}}{2k^-_1} z^-} e^{-i\frac{P^2_\perp(z^-) }{2k^-_1} z^-} \Big|^\infty \mathcal{G}(\infty, y^-) e^{i\frac{P^2_\perp(y^-)}{2k^-_1}y^-} |y_\perp)\Big|^{y^->0}_{\rm eik.}\nonumber\\
&&\to e^{i\frac{ k^2_{1\perp}}{2k^-_1} y^- }e^{-ik_{1\perp}y_\perp} U(\infty, y^-; y_\perp) \,,
 \label{eq:light-cone-general-eik-ypos}
\end{eqnarray}
and similarly for the other propagators.

This replacement does not hold for $y^- < 0$. Indeed, in that region the integration over the operator insertions~(\ref{eq:exp-lcone-gen-eik}) takes the form
\begin{eqnarray}
&&\int^0_{-\infty} dy^- e^{-i(q^+ - \frac{ k^2_{1\perp}}{2k^-_1}-\frac{ k^2_{2\perp}}{2k^-_2})y^-}  \int^\infty_{y^-} dz^- (y^-)^nO(z^-)\nonumber\\
&&= \int^\infty_{-\infty} dz^- O(z^-) \int^{z^-}_{-\infty} dy^- (y^-)^n e^{-i(q^+ - \frac{ k^2_{1\perp}}{2k^-_1}-\frac{ k^2_{2\perp}}{2k^-_2})y^-}
\nonumber\\
&&\sim \int^\infty_{-\infty} dz^- O(z^-) \sim \lambda^0\,,
\label{eq:int-eik-sup-2}
\end{eqnarray}
where in the last estimate we retain only the lowest power of $z^-$. The integration over $y^- < 0$ does not lead to any suppression of the operator insertions~(\ref{eq:exp-lcone-gen-eik}), rendering the replacement~(\ref{eq:light-cone-general-eik-ypos}) invalid in that region. While we use the replacement~(\ref{eq:light-cone-general-eik-ypos}) for $y^->0$\footnote{And $x^->0$ for the complex conjugate amplitude.} in computing the dijet production cross section~(\ref{eq:hadronic-bfm-general}) at the eikonal order, the $y^-<0$ region must be handled differently.

\subsection{Exponential factor at the past infinity}
In the previous section, we derived Eq.~(\ref{eq:light-cone-general-eik-ypos}) by commuting the exponential factor~(\ref{eq:exp-fact-cova}) in the quark propagator~(\ref{eq:prop-ex-tr-eik}) to future infinity, i.e.\ to the left. However, we found that the replacement~(\ref{eq:light-cone-general-eik-ypos}) is not applicable in the integration region $y^-<0$ of the dijet production cross section~(\ref{eq:hadronic-bfm-general}).


Alternatively, one can commute the exponential factor in the propagator~(\ref{eq:prop-ex-tr-eik}) to past infinity. As we will show, this yields the necessary suppression of the operator insertions~(\ref{eq:exp-lcone-gen-eik}) in the region $y^- < 0$ and produces a replacement analogous to Eq.~(\ref{eq:light-cone-general-eik-ypos}).


The commutation to past infinity proceeds in three steps. First, we rewrite the path-ordered exponents in the propagator as
\begin{widetext}
\begin{eqnarray}
&&(k_{1\perp}|  e^{i\frac{ k^2_{1\perp}}{2k^-_1} z^-} e^{-i\frac{P^2_\perp(z^-) }{2k^-_1} z^-} \Big|^\infty \mathcal{G}(\infty, y^-) e^{i\frac{P^2_\perp(y^-)}{2k^-_1}y^-} |y_\perp)
\label{eq:prop-ex-tr-eik-step1}\\
&&= \int d^2z_\perp (k_{1\perp}|  e^{i\frac{ k^2_{1\perp}}{2k^-_1} z^-} e^{-i\frac{P^2_\perp(z^-) }{2k^-_1} z^-} \Big|^\infty \mathcal{G}(\infty, -\infty) |z_\perp)(z_\perp| \mathcal{G}(-\infty, y^-)e^{i\frac{P^2_\perp(y^-)}{2k^-_1}y^-} |y_\perp)\,,
\nonumber
\end{eqnarray}
corresponding to the contour deformation shown in Fig.~\ref{fig:fig14}a.
\begin{figure}[htb]
\begin{center}
\includegraphics[width=0.9\textwidth]{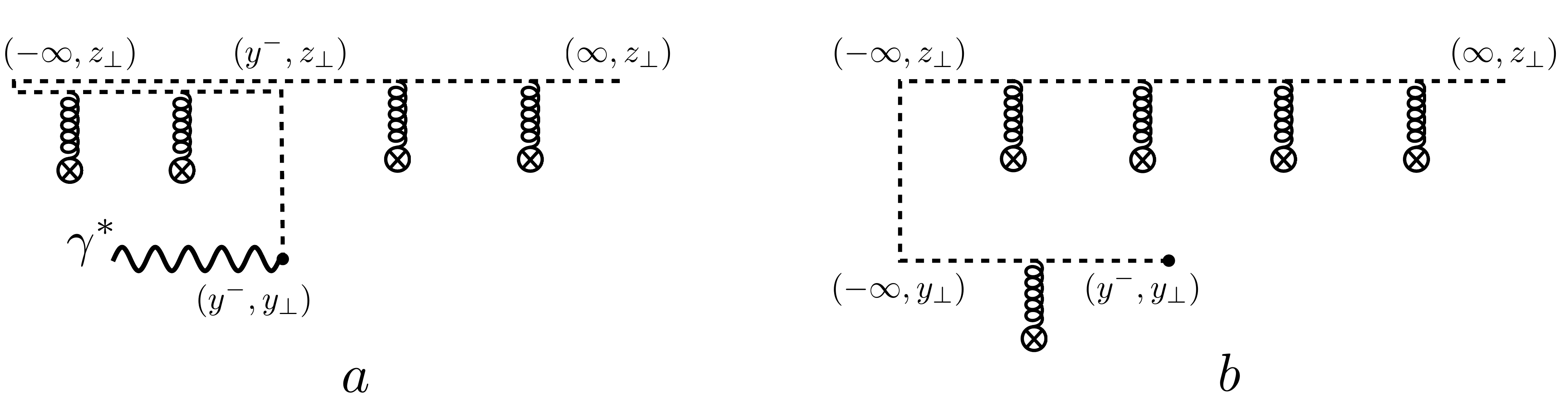}
 \end{center}
\caption{\label{fig:fig14}a) Deformation of the expansion contour in Eq. (\ref{eq:prop-ex-tr-eik-step1}); b) The expansion contour can be further deformed, which yields the (anti)quark propagator (\ref{eq:prop-ex-tr-eik-ymneg}) for the integration region $y^-<0$.}
 \end{figure}

Second, we commute the exponential factor~(\ref{eq:exp-fact-cova}) through the semi-infinite path-ordered exponent:
\begin{eqnarray}
&&(k_{1\perp}|  e^{i\frac{ k^2_{1\perp}}{2k^-_1} z^-} e^{-i\frac{P^2_\perp(z^-) }{2k^-_1} z^-} \Big|^\infty \mathcal{G}(\infty, y^-) e^{i\frac{P^2_\perp(y^-)}{2k^-_1}y^-} |y_\perp)
\label{eq:prop-ex-tr-eik-step2}\\
&&= \int d^2z_\perp (k_{1\perp}|  e^{i\frac{ k^2_{1\perp}}{2k^-_1} z^-} e^{-i\frac{P^2_\perp(z^-) }{2k^-_1} z^-} \Big|^\infty \mathcal{G}(\infty, -\infty) |z_\perp)(z_\perp| e^{i\frac{ P^2_\perp(-\infty)}{2k^-_1}y^- }\mathcal{G}(-\infty, y^-;y^-) |y_\perp)\,.
\nonumber
\end{eqnarray}

Third, we expand each path-ordered exponent onto the light-cone direction, corresponding to expanding the background fields onto the contour in Fig.~\ref{fig:fig14}b. The exponential factor at past infinity can now be replaced by a simple transverse gauge link, see Eq.~(\ref{eq:tr-WL-insertion}).


The expansion of the infinite path-ordered exponent onto the light-cone direction at the eikonal order reads
\begin{eqnarray}
&&\mathcal{G}(\infty, -\infty) \Big|_{\rm eik.}= U(\infty, -\infty) + \frac{ig}{2p^-}\int^{\infty}_{-\infty} dz^- U(\infty, z^-)\sigma^{-k}F_{-k}(z^-) U(z^-, -\infty) \,,\nonumber
\end{eqnarray}
see Eq. (\ref{eq:staple-expan-prop-eik-expon}).

The expansion of the ``shifted'' path-ordered exponent is
\begin{eqnarray}
&&\mathcal{G}(-\infty, y^-, y^-) 
= U(-\infty, y^-) + i\int^{-\infty}_{y^-} dz^- U(-\infty, z^-)\mathcal{O}_{\mathcal{B}}(z^-;y^-) U(z^-, y^-)
\label{eq:lexp-eik-bt}\\
&&+ (i)^2\int^{-\infty}_{y^-} dz^- \int^{z^-}_{y^-} dz'^- U(-\infty, z^-)\mathcal{O}_{\mathcal{B}}(z^-;y^-)U(z^-, z'^-) 
\mathcal{O}_{\mathcal{B}}(z'^-;y^-)U(z'^-, y^-) + \dots\,,
\nonumber
\end{eqnarray}
 see Eq. (\ref{eq:result-exp-shift-exp}).

Substituting this expansion into the dijet production cross section~(\ref{eq:hadronic-bfm-general}), the operator insertions~(\ref{eq:exp-lcone-gen-eik}) in Eq.~(\ref{eq:lexp-eik-bt}) give rise to integrals of the form
\begin{eqnarray}
&&\int^0_{-\infty} dy^- e^{-i(q^+ - \frac{ p^2_{1\perp}}{2k^-_1}-\frac{ p^2_{2\perp}}{2k^-_2})y^-}  \int^{-\infty}_{y^-} dz^- (y^-)^nO(z^-)\nonumber\\
&&= \int^{-\infty}_{0} dz^- O(z^-) \int^0_{z^-} dy^- (y^-)^n e^{-i(q^+ - \frac{ p^2_{1\perp}}{2k^-_1}-\frac{ p^2_{2\perp}}{2k^-_2})y^-}
\sim \int^{-\infty}_{0} dz^- O(z^-) (z^-)^{n+1} \sim (\lambda^{-1})^{n+1}\,,
\label{eq:int-eik-sup2}
\end{eqnarray}
yielding a suppression of at least $\sim \lambda^{-1}$. At the eikonal order, the operator insertions~(\ref{eq:exp-lcone-gen-eik}) can therefore be neglected, and only the first term of expansion~(\ref{eq:lexp-eik-bt}) need be retained. This makes the light-cone expansion~(\ref{eq:lexp-eik-bt}) suitable for computing the dijet production. At the eikonal order, for $y^-<0$ we use the replacement
\begin{eqnarray}
&&\mathcal{G}(-\infty, y^-, y^-) \Big|^{y^-<0}_{\rm eik}
\to U(-\infty, y^-)\,.
\end{eqnarray}

Combining all terms, the (anti)quark propagator in the background field at the eikonal order takes the following form in the region $y^-<0$ of the dijet production cross section~(\ref{eq:hadronic-bfm-general}):\footnote{At future infinity, we use the boundary condition $\lim_{x^-\to\infty}B_i(x)=0$, see Eq.~(\ref{eq:inf-link-simp}).}
\begin{eqnarray}
&&(k_{1\perp}|  e^{i\frac{ k^2_{1\perp}}{2k^-_1} z^-} e^{-i\frac{P^2_\perp(z^-) }{2k^-_1} z^-} \Big|^\infty \mathcal{G}(\infty, y^-) e^{i\frac{P^2_\perp(y^-)}{2k^-_1}y^-} |y_\perp)\Big|^{y^-<0}_{\rm eik.}\label{eq:prop-ex-tr-eik-ymneg}\\
&&\to \int d^2z_\perp e^{-ik_{1\perp}z_\perp} \Big(U(\infty, -\infty;z_\perp)
+ \frac{ig}{2p^-}\int^{\infty}_{-\infty} dz^- U(\infty, z^-;z_\perp)\sigma^{-k}F_{-k}(z^-,z_\perp) U(z^-, -\infty;z_\perp)\Big)\nonumber\\
&&\times  (z_\perp| e^{i\frac{ P^2_\perp(-\infty)}{2k^-_1}y^- } |y_\perp) U(-\infty, y^-;y_\perp)\,,
\nonumber
\end{eqnarray}
corresponding to expanding the background fields of the (anti)quark propagator~(\ref{eq:prop-ex-tr-eik}) onto the contour in Fig.~\ref{fig:fig14}b.

\subsection{Quark propagator in the background field at the eikonal order}
Combining the results of the previous two sections, we obtain the following expression for the (anti)quark propagator in the background field at the eikonal order:
\begin{eqnarray}
&&(k_{1\perp}|  e^{i\frac{ k^2_{1\perp}}{2k^-_1} z^-} e^{-i\frac{P^2_\perp(z^-) }{2k^-_1} z^-} \Big|^\infty \mathcal{G}(\infty, y^-) e^{i\frac{P^2_\perp(y^-)}{2k^-_1}y^-} |y_\perp)\Big|_{\rm eik.}\label{eq:pr-eik}\\
&&\to \theta(y^-) e^{i\frac{ k^2_{1\perp}}{2k^-_1} y^- }e^{-ik_{1\perp}y_\perp} U(\infty, y^-; y_\perp)
+\theta(-y^-) \int d^2z_\perp e^{-ik_{1\perp}z_\perp} \Big(U(\infty, -\infty;z_\perp) \nonumber\\
&&+ \frac{ig}{2p^-}\int^{\infty}_{-\infty} dz^- U(\infty, z^-;z_\perp)\sigma^{-k}F_{-k}(z^-,z_\perp) U(z^-, -\infty;z_\perp)\Big)
(z_\perp| e^{i\frac{ P^2_\perp(-\infty)}{2k^-_1}y^- } |y_\perp) U(-\infty, y^-;y_\perp)\,.
\nonumber
\end{eqnarray}
Analogous expressions for the other propagators in the dijet production cross section~(\ref{eq:hadronic-bfm-general}) are derived in the same manner.

\subsection{Computation of the dijet production at the eikonal order}
Substituting Eq.~(\ref{eq:pr-eik}) and analogous results for the remaining propagators into Eq.~(\ref{eq:hadronic-bfm-general}), we obtain the dijet production cross section at the eikonal order within the high-energy power counting~(\ref{eq:pw-couting-sx}):
\begin{eqnarray}
&&d\sigma^{\gamma^\ast p}_{\lambda\lambda'}\Big|_{\rm eik.} = \sum_f\sum_{s_1,s_2}\frac{2\pi e^2_f \alpha_{\rm EM}x}{V_4Q^2}   \epsilon^{\lambda\ast}_\mu \epsilon^{\lambda'}_\nu \frac{dk^-_1d^2k_{1\perp}}{(2\pi)^3 2k^-_1} \Big|_{k^2_1=0}\frac{dk^-_2d^2k_{2\perp}}{(2\pi)^3 2k^-_2} \Big|_{k^2_2=0}
 \nonumber\\
 &&\times \langle P, S| M^{\dag\mu}_{s_1,s_2}(q, k_1, k_2) \Big|_{\rm eik.}M^\nu_{s_1,s_2}(q, k_1, k_2)\Big|_{\rm eik.}
|P, S\rangle \,,
\label{eq:hadronic-bfm-eik}
\end{eqnarray}
where the virtual photon scattering amplitude at the eikonal order is
\begin{eqnarray}
&&iM^\nu_{s_1,s_2}(q, k_1, k_2) \Big|_{\rm eik.}
= \int d^4y e^{-iqy} \theta(-y^-) \int d^2z_\perp e^{-ik_{1\perp}z_\perp} \int d^2z'_\perp e^{-ik_{2\perp}z'_\perp}\label{eq:amp-gen-eik}\\
&&\times \Big\{\bar{u}_{s_1}(k_1)  \Big(U(\infty, -\infty;z_\perp)
+ \frac{ig}{2k^-_1}\int^{\infty}_{-\infty} dz^- U(\infty, z^-;z_\perp) \sigma^{-m}F_{-m}(z^-,z_\perp)  U(z^-, -\infty;z_\perp)\Big) \nonumber\\
&&\times (z_\perp| e^{i\frac{ P^2_\perp(-\infty)}{2k^-_1}y^- } |y_\perp) U(-\infty, y^-;y_\perp)
e^{ik^-_1 y^+}\gamma^\nu e^{ik^-_2 y^+} U(y^-, -\infty;y_\perp) (y_\perp|e^{i\frac{P^2_\perp(-\infty)}{2k^-_2}y^-} |z'_\perp)\nonumber\\
&&\times \Big(U(-\infty, \infty;z'_\perp) 
- \frac{ig}{2k^-_2}\int^{-\infty}_{\infty} dz'^- U(-\infty, z'^-;z'_\perp) \sigma^{-n}F_{-n}(z'^-, z'_\perp) U(z'^-, \infty;z'_\perp)\Big) v_{s_2}(k_2) - (B=0)\Big\}\,,
\nonumber
\end{eqnarray}
and the complex conjugated amplitude is 
\begin{eqnarray}
&&-iM^{\dag\mu}_{s_1,s_2}(q, k_1, k_2)\Big|_{\rm eik.} = \int d^4x e^{iqx} \theta(-x^-) \int d^2\omega_\perp e^{ik_{1\perp}\omega_\perp} \int d^2\omega'_\perp e^{ik_{2\perp}\omega'_\perp}\label{eq:ampcc-gen-eik}\\
&&\times \Big\{ \bar{v}_{s_2}(k_2)  \Big(U(\infty, -\infty;\omega'_\perp)
- \frac{ig}{2k^-_2}\int^{\infty}_{-\infty} d\omega'^- U(\infty, \omega'^-;\omega'_\perp) \sigma^{-k}F_{-k}(\omega'^-,\omega'_\perp) U(\omega'^-, -\infty;\omega'_\perp)\Big) \nonumber\\
&&\times (\omega'_\perp|e^{-i\frac{ P^2_\perp(-\infty)}{2k^-_2}x^- } |x_\perp) U(-\infty, x^-;x_\perp) e^{-ik^-_2 x^+} \gamma^\mu e^{-ik^-_1 x^+} U(x^-, -\infty;x_\perp) (x_\perp| e^{-i\frac{P^2_\perp(-\infty)}{2k^-_1}x^-}|\omega_\perp)
\nonumber\\
&&\times \Big(U(-\infty, \infty;\omega_\perp) + \frac{ig}{2k^-_1}\int^{-\infty}_{\infty} d\omega^- U(-\infty, \omega^-;\omega_\perp)
\sigma^{-k}F_{-k}(\omega^-,\omega_\perp) U(\omega^-, \infty;\omega_\perp)\Big) u_{s_1}(k_1) - (B=0)\Big\}\,.
\nonumber
\end{eqnarray}
\end{widetext}
Equations~(\ref{eq:hadronic-bfm-eik}), (\ref{eq:amp-gen-eik}), and~(\ref{eq:ampcc-gen-eik}) provide the complete dijet production cross section at the eikonal order within the high-energy power counting~(\ref{eq:pw-couting-sx}). The operators in Eq.~(\ref{eq:amp-gen-eik}) are arranged along the expansion contour shown in Fig.~\ref{fig:fig15}a, and similarly for Eq.~(\ref{eq:ampcc-gen-eik}). The semi-infinite gauge factors in Eqs.~(\ref{eq:amp-gen-eik}) and~(\ref{eq:ampcc-gen-eik}) cancel, reducing the operator contour to that in Fig.~\ref{fig:fig15}b. Gauge invariance of the operators in Eq.~(\ref{eq:hadronic-bfm-eik}) can be restored by adding transverse gauge links at future infinity.
\begin{figure*}[htb]
\begin{center}
\includegraphics[width=0.9\textwidth]{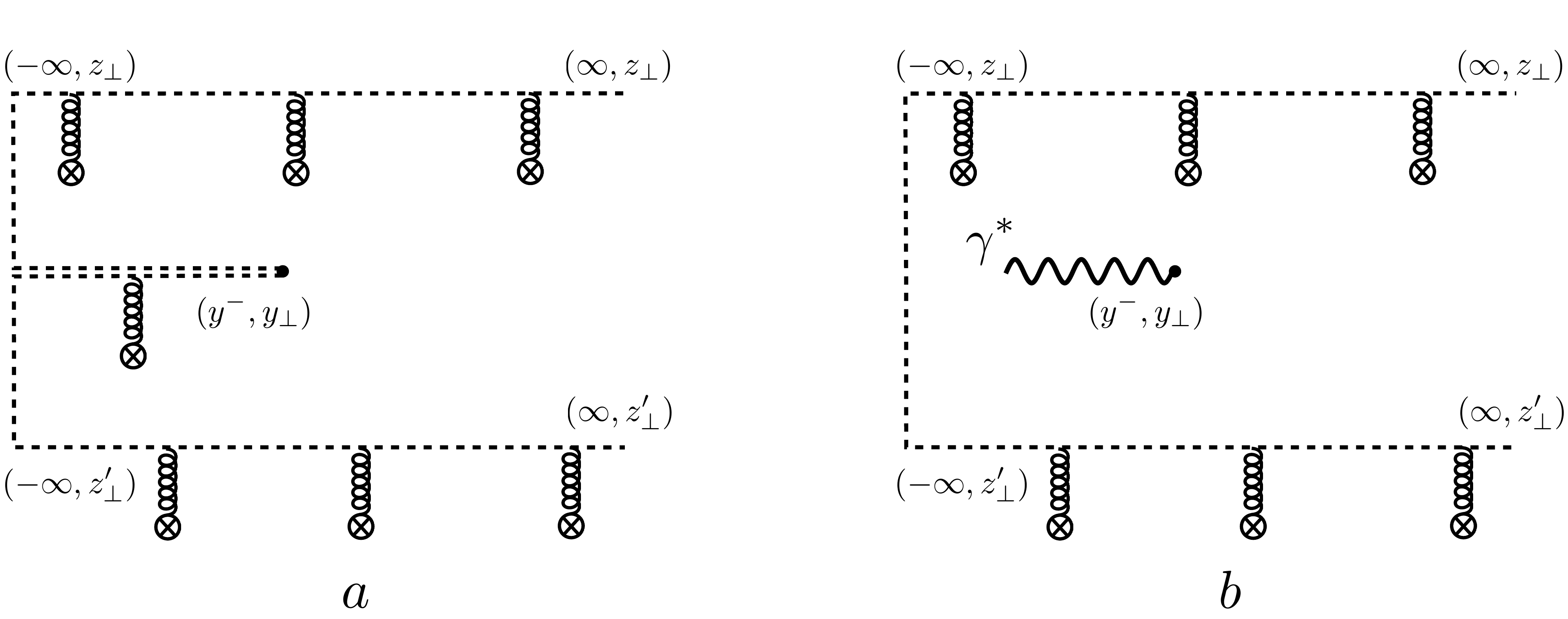}
 \end{center}
\caption{\label{fig:fig15}a) The contour for operators in Eq. (\ref{eq:amp-gen-eik}); b) The semi-infinite gauge factors cancel each other, such that the contour reduces to the staple-like shape. The contour does not depend on the coordinate $(y^-, y_\perp)$ of the virtual photon annihilation.}
 \end{figure*}

Our approach also allows one to derive the contribution of the transverse gauge links at past infinity, corresponding to the exponential factors~(\ref{eq:exp-fact-cova}) in Eqs.~(\ref{eq:amp-gen-eik}) and~(\ref{eq:ampcc-gen-eik}), see Sec.~\ref{sec:tr-link}. Within the high-energy power counting~(\ref{eq:pw-couting-sx}), this contribution is non-trivial and parametrically of the eikonal order, $\sim \lambda^0$. In particular, we will show that the transverse gauge links play an essential role in matching our eikonal result for the dijet production cross section to the back-to-back limit.


The contribution of the transverse gauge links in Eqs.~(\ref{eq:amp-gen-eik}) and~(\ref{eq:ampcc-gen-eik}) can be written explicitly using the expansion of the exponential factors~(\ref{eq:tr-WL-insertion}), assuming that the gauge field at past infinity is a pure gauge so that the corresponding field-strength tensors vanish. The amplitude~(\ref{eq:amp-gen-eik}) in the reference frame~(\ref{eq:ref-fr}) then reads
\begin{widetext}
\begin{eqnarray}
&&iM^\nu_{s_1,s_2}(q, k_1, k_2) \Big|_{\rm eik.}
= 2\pi\delta(q^- - k^-_1 - k^-_2) 2z\bar{z}q^- \frac{-i}{2\pi} \int d^2z_\perp e^{-ik_{1\perp}z_\perp} \int d^2z'_\perp e^{-ik_{2\perp}z'_\perp}\label{eq:amp-gen-eik-form1}\\
&&K_0(\sqrt{z\bar{z}Q^2}| z_\perp - z'_\perp |)
\Big\{\bar{u}_{s_1}(k_1)  \Big(U(\infty, -\infty;z_\perp) + \frac{ig}{2k^-_1}\int^{\infty}_{-\infty} dz^- U(\infty, z^-;z_\perp) \sigma^{-m}F_{-m}(z^-,z_\perp)\nonumber\\
&&\times U(z^-, -\infty;z_\perp)\Big) U(z_\perp, z'_\perp;-\infty) \gamma^\nu \Big(U(-\infty, \infty;z'_\perp) 
- \frac{ig}{2k^-_2}\int^{-\infty}_{\infty} dz'^- U(-\infty, z'^-;z'_\perp)\nonumber\\
&&\times \sigma^{-n}F_{-n}(z'^-, z'_\perp) U(z'^-, \infty;z'_\perp)\Big) v_{s_2}(k_2) - (B=0)\Big\}\,,
\nonumber
\end{eqnarray}
\end{widetext}
and similarly for the complex conjugate amplitude~(\ref{eq:ampcc-gen-eik}). Here we used the notation~(\ref{eq:dijet-var}) and
\begin{equation}
(z_\perp| \frac{1}{ p^2_\perp + z\bar{z}Q^2 } |z'_\perp) = \frac{1}{2\pi} K_0(\sqrt{z\bar{z}Q^2}| z_\perp - z'_\perp |)\,.
\label{eq:fourier1}
\end{equation}
This form of the amplitude can be used for computing the dijet production cross section (\ref{eq:hadronic-bfm-eik}).

The background field operators in Eq.~(\ref{eq:amp-gen-eik-form1}) contain infinite light-cone Wilson lines with Pauli vertex insertions $\sim \sigma^{-k}F_{-k}$. These structures also appeared in the dijet production amplitude in the back-to-back limit, see Eq.~(\ref{eq:djet-btb-col-lead-init}), and originate from the form of the (anti)quark propagator in the background field, Eq.~(\ref{eq:prop-def}), and its path-ordered exponent representation~(\ref{eq:LSZ-fin-quark}).


As in the back-to-back limit, the contribution of the transverse background field component in Eq.~(\ref{eq:amp-gen-eik-form1}) is non-trivial within the high-energy power counting~(\ref{eq:pw-couting-sx}). The transverse field component contributes to the transverse gauge links and, parametrically, this contribution is of the eikonal order. Similarly, the transverse field component contributes to the Pauli vertex insertions $\sim \sigma^{-k}F_{-k}$ through
\begin{eqnarray}
F_{-m} = &&- \partial_m B_- + \partial_- B_m - ig [B_-, B_m]\nonumber\\
&&\sim \lambda^0 + \lambda^0 + \lambda^0\,.
\label{eq-str-ten-est}
\end{eqnarray}
All three terms of the field-strength tensor~(\ref{eq-str-ten-est}) contribute at the eikonal order. This indicates that within the high-energy power counting~(\ref{eq:pw-couting-sx}), the transverse field component must be taken into account already at the eikonal order, alongside the standard contribution of the longitudinal component. This generally makes predictions within the high-energy power counting~(\ref{eq:pw-couting-sx}) distinct from those computed in the CGC framework~(\ref{eq:cgc-frm}). In the following, we illustrate the importance of the transverse field component; in particular, we demonstrate its essential role in matching our eikonal results to the dijet production in the back-to-back limit~(\ref{eq:btb-ltwist-pr}).


An important clarification is in order. While the transverse background field component appears in various operators of Eqs.~(\ref{eq:amp-gen-eik}) and~(\ref{eq:ampcc-gen-eik}), there is no physical distinction between these contributions. The operators in Eqs.~(\ref{eq:amp-gen-eik}) and~(\ref{eq:ampcc-gen-eik}) were derived through the formal expansion of Sec.~\ref{sec:gen-exp}. Although the transverse field component may appear in different operators within this expansion, it physically corresponds to the same interaction of the (anti)quark with the transverse background field at a finite position in coordinate space, see Fig.~\ref{fig:gif6}. Once the transverse field component is included in the analysis, it must therefore be consistently accounted for in all operators derived through our formal manipulations.

For instance, while the form of the amplitude in Eqs.~(\ref{eq:amp-gen-eik}) and~(\ref{eq:amp-gen-eik-form1}) is convenient for matching our eikonal result to the dijet production in the back-to-back limit, for comparison with CGC results it is advantageous to rewrite the amplitude in a form where the transverse field component is attributed entirely to the transverse gauge links.


\begin{widetext}
To perform this rearrangement, we rewrite Eq. (\ref{eq:amp-gen-eik}) as 
\begin{eqnarray}
&&iM^\nu_{s_1,s_2}(q, k_1, k_2) \Big|_{\rm eik.}
= \int d^4y e^{-iqy} \theta(-y^-) \int d^2z_\perp e^{-ik_{1\perp}z_\perp} \int d^2z'_\perp e^{-ik_{2\perp}z'_\perp}\\
&&\times \Big\{\bar{u}_{s_1}(k_1) (z_\perp| \Big(U(\infty, -\infty)
+ \frac{i\sigma^{-m}}{2k^-_1} P_m(\infty) U(\infty, -\infty) - \frac{i\sigma^{-m}}{2k^-_1}U(\infty, -\infty)P_m(-\infty) \Big)\nonumber\\
&&\times e^{i\frac{ P^2_\perp(-\infty)}{2k^-_1}y^- } |y_\perp) e^{ik^-_1 y^+}\gamma^\nu e^{ik^-_2 y^+} (y_\perp|e^{i\frac{P^2_\perp(-\infty)}{2k^-_2}y^-}
\Big(U(-\infty, \infty) - \frac{i\sigma^{-n}}{2k^-_2} P_n(-\infty) U(-\infty, \infty) \nonumber\\
&&+ \frac{i\sigma^{-n}}{2k^-_2} U(-\infty, \infty)P_n(\infty) \Big) |z'_\perp) v_{s_2}(k_2) - (B=0)\Big\}\,,\nonumber
\end{eqnarray}
where we used Eq. (\ref{eq:com-gen-UP}).

Imposing the boundary condition $\lim_{x^-\to\infty}B_i(x) = 0$ and using
\begin{eqnarray}
&&\int d^2z_\perp e^{-ik_{1\perp}z_\perp} (z_\perp| p_m = \int d^2z_\perp e^{-ik_{1\perp}z_\perp} (z_\perp| k_{1m}\,,
\end{eqnarray}
we can further rewrite this as
\begin{eqnarray}
&&iM^\nu_{s_1,s_2}(q, k_1, k_2) \Big|_{\rm eik.}
= \int d^4y e^{-iqy} \theta(-y^-) \int d^2z_\perp e^{-ik_{1\perp}z_\perp} \int d^2z'_\perp e^{-ik_{2\perp}z'_\perp}\\
&&\times \Big\{\bar{u}_{s_1}(k_1) \Big( (1 + \frac{ik_{1m} \sigma^{-m}}{2k^-_1}  ) U(\infty, -\infty;z_\perp)
(z_\perp|e^{i\frac{ P^2_\perp(-\infty)}{2k^-_1}y^- } |y_\perp) - \frac{i\sigma^{-m}}{2k^-_1}U(\infty, -\infty;z_\perp)\nonumber\\
&&\times (z_\perp| P_m(-\infty) e^{i\frac{ P^2_\perp(-\infty)}{2k^-_1}y^- } |y_\perp)\Big)  e^{ik^-_1 y^+}\gamma^\nu e^{ik^-_2 y^+}
\Big( (1 - \frac{ik_{2n}\sigma^{-n}}{2k^-_2} )
(y_\perp|e^{i\frac{P^2_\perp(-\infty)}{2k^-_2}y^-}|z'_\perp)\nonumber\\
&&\times U(-\infty, \infty;z'_\perp)
- \frac{i\sigma^{-n}}{2k^-_2} (y_\perp|e^{i\frac{P^2_\perp(-\infty)}{2k^-_2}y^-} P_n(-\infty) |z'_\perp)U(-\infty, \infty;z'_\perp) \Big) v_{s_2}(k_2) 
- (B=0)\Big\}\,.\nonumber
\end{eqnarray}

Using the operator expansion~(\ref{eq:tr-WL-insertionPileft}) and assuming a pure gauge field at past infinity, we further rewrite the amplitude as
\begin{eqnarray}
&&iM^\nu_{s_1,s_2}(q, k_1, k_2) \Big|_{\rm eik.}
= \int d^4y e^{-iqy} e^{i(k^-_1+k^-_2) y^+} \theta(-y^-) \int d^2z_\perp e^{-ik_{1\perp}z_\perp} \int d^2z'_\perp e^{-ik_{2\perp}z'_\perp}\\ &&\times 
 (z_\perp|  e^{i\frac{ p^2_\perp}{2k^-_1}y^- } |y_\perp) 
(y_\perp|e^{i\frac{p^2_\perp}{2k^-_2}y^-}|z'_\perp) \Big\{\bar{u}_{s_1}(k_1) \Big( 1 + \frac{ik_{1m}\sigma^{-m} }{2k^-_1}
+ \frac{i(z-y)_m \sigma^{-m} }{2y^-}  \Big) \gamma^\nu 
  \nonumber\\
  &&\times \Big( 1- \frac{i k_{2n} \sigma^{-n}}{2k^-_2}  +  \frac{i(y-z')_n \sigma^{-n} }{2y^-}  \Big) v_{s_2}(k_2) \Big\}
\Big(U(\infty, -\infty;z_\perp) U(z_\perp, z'_\perp;-\infty) U(-\infty, \infty;z'_\perp) - 1\Big)\,.\nonumber
\end{eqnarray}

In the reference frame~(\ref{eq:ref-fr}), using Eq.~(\ref{eq:gauge-vec}) and integrating over $y^-$ and $y_\perp$, we obtain
\begin{eqnarray}
&&iM^\nu_{s_1,s_2}(q, k_1, k_2) \Big|_{\rm eik.}
= 2\pi\delta(q^- - k^-_1- k^-_2) \int d^2z_\perp e^{-ik_{1\perp}z_\perp} \int d^2z'_\perp e^{-ik_{2\perp}z'_\perp}  
\\
&&\times \bar{u}_{s_1}(k_1) \Big\{ \Big( i\gamma^\nu - \frac{k_{1m} }{2k^-_1}\sigma^{-m}\gamma^\nu + \frac{k_{2m} }{2k^-_2} \gamma^\nu \sigma^{-m} \Big) (z_\perp| \frac{1}{ q^+ - \frac{ p^2_\perp}{2k^-_1} - \frac{p^2_\perp}{2k^-_2} } |z'_\perp) + \Big( \frac{ \sigma^{-m}\gamma^\nu }{2k^-_1} + \frac{ \gamma^\nu \sigma^{-m} }{2k^-_2} \Big)
\nonumber\\
&&\times (z_\perp| \frac{p_m}{q^+ - \frac{ p^2_\perp}{2k^-_1} - \frac{p^2_\perp}{2k^-_2}}  |z'_\perp) \Big\} v_{s_2}(k_2) \Big(U(\infty, -\infty;z_\perp) U(z_\perp, z'_\perp;-\infty) U(-\infty, \infty;z'_\perp) - 1\Big)\,.
\nonumber
\end{eqnarray}

Using Eq. (\ref{eq:fourier1}) and
\begin{eqnarray}
&&(z_\perp| \frac{p_m}{ p^2_\perp + z\bar{z}Q^2 } |z'_\perp) = \frac{i\sqrt{z\bar{z}Q^2}}{2\pi}  K_1(\sqrt{z\bar{z}Q^2}| z_\perp - z'_\perp |) \frac{z_m - z'_m }{| z_\perp - z'_\perp |}\,,
\label{eq:fourier2}
\end{eqnarray}
it is straightforward to rewrite this as
\begin{eqnarray}
&&iM^\nu_{s_1,s_2}(q, k_1, k_2) \Big|_{\rm eik.}
= \delta(q^- - k^-_1- k^-_2) 2z\bar{z}q^-\int d^2z_\perp e^{-ik_{1\perp}z_\perp} \int d^2z'_\perp e^{-ik_{2\perp}z'_\perp} \label{eq:amp-gen-eik-forCGC}\\
&&\times \bar{u}_{s_1}(k_1) \Big\{ \Big( -i\gamma^\nu + \frac{k_{1m} }{2k^-_1}\sigma^{-m}\gamma^\nu - \frac{k_{2m} }{2k^-_2}\gamma^\nu \sigma^{-m} \Big)  K_0(\sqrt{z\bar{z}Q^2}| z_\perp - z'_\perp |) - i\sqrt{z\bar{z}Q^2} \Big( \frac{ \sigma^{-m}\gamma^\nu }{2k^-_1}\nonumber\\
&&+ \frac{ \gamma^\nu \sigma^{-m} }{2k^-_2}  \Big)
\frac{z_m - z'_m }{| z_\perp - z'_\perp |}  K_1(\sqrt{z\bar{z}Q^2}| z_\perp - z'_\perp |)  \Big\} v_{s_2}(k_2) \Big(U(\infty, -\infty;z_\perp) U(z_\perp, z'_\perp;-\infty) U(-\infty, \infty;z'_\perp) - 1\Big)\,.
\nonumber
\end{eqnarray}
This form of the amplitude is equivalent to Eq.~(\ref{eq:amp-gen-eik-form1}). The distinction is that in Eq.~(\ref{eq:amp-gen-eik-forCGC}) the transverse background field component is attributed entirely to the transverse gauge link, whereas in Eq.~(\ref{eq:amp-gen-eik-form1}) it is distributed between the transverse gauge link and the Pauli vertex insertions $\sim \sigma^{-k}F_{-k}$ at finite coordinate-space positions.


Within the high-energy power counting~(\ref{eq:pw-couting-sx}), both forms of the transverse field component contribution are parametrically eikonal, $\sim \lambda^0$. Our formal operator transformations preserve the eikonality of each contribution. We emphasize that physically both forms correspond to the same interaction of the (anti)quark with the transverse background field at a finite position in coordinate space, see Fig.~\ref{fig:gif6}.


A similar computation for the complex conjugated amplitude (\ref{eq:ampcc-gen-eik}) yields
\begin{eqnarray}
&&-iM^{\dag\mu}_{s_1,s_2}(q, k_1, k_2)\Big|_{\rm eik.} =  \delta(q^- - k^-_1 - k^-_2) 2z\bar{z} q^- \int d^2\omega_\perp e^{ik_{1\perp}\omega_\perp} \int d^2\omega'_\perp  e^{ik_{2\perp}\omega'_\perp} 
\label{eq:ampcc-gen-eik-forCGC}\\
&&\times \bar{v}_{s_2}(k_2) \Big\{ \Big( i\gamma^\mu +  \frac{k_{1k}}{2k^-_1} \gamma^\mu\sigma^{-k} - \frac{k_{2k}}{2k^-_2} \sigma^{-k}\gamma^\mu  \Big) K_0(\sqrt{z\bar{z}Q^2}| \omega_\perp - \omega'_\perp |) + i\sqrt{z\bar{z}Q^2} \Big( \frac{\gamma^\mu\sigma^{-k}}{2k^-_1} + \frac{\sigma^{-k}\gamma^\mu}{2k^-_2} \Big)
\nonumber\\
&&\times \frac{ \omega_k - \omega'_k }{| \omega_\perp - \omega'_\perp |} K_1(\sqrt{z\bar{z}Q^2}| \omega_\perp - \omega'_\perp |)   \Big\}  u_{s_1}(k_1) \Big(U(\infty, -\infty;\omega'_\perp) U(\omega'_\perp, \omega_\perp;-\infty)U(-\infty, \infty;\omega_\perp) - 1\Big)
\nonumber
\end{eqnarray}
Substituting Eqs.~(\ref{eq:amp-gen-eik-forCGC}) and~(\ref{eq:ampcc-gen-eik-forCGC}) into Eq.~(\ref{eq:hadronic-bfm-eik}), the dijet production cross section in the high-energy power counting~(\ref{eq:pw-couting-sx}) at the eikonal order can be computed straightforwardly. We now list the results for each virtual photon polarization.

\subsection{The dijet production cross section at the eikonal order}
For the longitudinally polarized photon we obtain the following result
\begin{eqnarray}
&&d\sigma^{\gamma^\ast p}_{00}\Big|_{\rm eik.} = \sum_f \frac{2\pi e^2_f \alpha_{\rm EM}x}{V_3Q^2}   \Big(\frac{Q}{q^-}\Big)^2 \frac{dk^-_1d^2k_{1\perp}}{(2\pi)^3 2k^-_1} \Big|_{k^2_1=0}\frac{dk^-_2d^2k_{2\perp}}{(2\pi)^3 2k^-_2} \Big|_{k^2_2=0} \delta(q^-- k^-_1 - k^-_2) 32 z^3\bar{z}^3 (q^-)^4\nonumber\\
&&\times \int d^2\omega_\perp e^{ik_{1\perp}\omega_\perp}
\int d^2\omega'_\perp  e^{ik_{2\perp}\omega'_\perp}
K_0(\sqrt{z\bar{z}Q^2}| \omega_\perp - \omega'_\perp |) \int d^2z_\perp e^{-ik_{1\perp}z_\perp} \int d^2z'_\perp e^{-ik_{2\perp}z'_\perp}\nonumber\\
&&\times K_0(\sqrt{z\bar{z}Q^2}| z_\perp - z'_\perp |) \mathcal{U}(z_\perp, z'_\perp;\omega_\perp, \omega'_\perp)\,.
\label{eq:cc00-f1}
\end{eqnarray}
In this case, the dijet production cross section is defined by a matrix element of the background field operator
\begin{eqnarray}
&&\mathcal{U}(z_\perp, z'_\perp;\omega_\perp, \omega'_\perp)\equiv \langle P, S| {\rm tr_c} \Big(U(\infty, -\infty;\omega'_\perp) U(\omega'_\perp, \omega_\perp;-\infty) U(-\infty, \infty;\omega_\perp) - 1\Big)U(\omega_\perp, z_\perp ;\infty) 
 \nonumber\\
 &&\times \Big(U(\infty, -\infty;z_\perp) U(z_\perp, z'_\perp;-\infty) U(-\infty, \infty;z'_\perp) - 1\Big)U(z'_\perp, \omega'_\perp;\infty)
|P, S\rangle\,.
\label{op-hepc-trlink}
\end{eqnarray}
Here, we restored the gauge invariance of the operator by adding transverse gauge links at the future infinity.

The same operator defines the cross sections with the transversely polarized photon. We obtain the following equations
\begin{eqnarray}
&&d\sigma^{\gamma^\ast p}_{0\lambda'=\pm1}\Big|_{\rm eik.} 
= \sum_f \frac{2\pi e^2_f \alpha_{\rm EM}x}{V_3Q^2}   \frac{Q}{q^-} \epsilon^{\lambda'=\pm1}_l \frac{dk^-_1d^2k_{1\perp}}{(2\pi)^3 2k^-_1} \Big|_{k^2_1=0}\frac{dk^-_2d^2k_{2\perp}}{(2\pi)^3 2k^-_2} \Big|_{k^2_2=0} \delta(q^-- k^-_1 - k^-_2)\nonumber\\
&&\times 16(q^-)^3( 2z - 1 ) z^2\bar{z}^2
\int d^2\omega_\perp e^{ik_{1\perp}\omega_\perp} \int d^2\omega'_\perp  e^{ik_{2\perp}\omega'_\perp} K_0(\sqrt{z\bar{z}Q^2}| \omega_\perp - \omega'_\perp |) i\sqrt{z\bar{z}Q^2}\nonumber\\
&&\times\int d^2z_\perp e^{-ik_{1\perp}z_\perp}
\int d^2z'_\perp e^{-ik_{2\perp}z'_\perp} 
\frac{z_l - z'_l }{| z_\perp - z'_\perp |} K_1(\sqrt{z\bar{z}Q^2}| z_\perp - z'_\perp |)\mathcal{U}(z_\perp, z'_\perp;\omega_\perp, \omega'_\perp)
\label{eq:cc01-f1}
\end{eqnarray}
and
\begin{eqnarray}
&&d\sigma^{\gamma^\ast p}_{\lambda=\pm10}\Big|_{\rm eik.} 
= \sum_f \frac{2\pi e^2_f \alpha_{\rm EM}x}{V_3Q^2}   \epsilon^{\lambda=\pm 1\ast}_s \frac{Q}{q^-} \frac{dk^-_1d^2k_{1\perp}}{(2\pi)^3 2k^-_1} \Big|_{k^2_1=0}\frac{dk^-_2d^2k_{2\perp}}{(2\pi)^3 2k^-_2} \Big|_{k^2_2=0} \delta(q^-- k^-_1 - k^-_2)\label{eq:cc10-f1}\\
&&\times 16 (q^-)^3 z^2 \bar{z}^2 ( 1  - 2z )
i\sqrt{z\bar{z}Q^2} \int d^2\omega_\perp e^{ik_{1\perp}\omega_\perp} \int d^2\omega'_\perp  e^{ik_{2\perp}\omega'_\perp}  \frac{ \omega_s - \omega'_s }{| \omega_\perp - \omega'_\perp |} K_1(\sqrt{z\bar{z}Q^2}| \omega_\perp - \omega'_\perp |)
\nonumber\\
&&\times \int d^2z_\perp e^{-ik_{1\perp}z_\perp} \int d^2z'_\perp e^{-ik_{2\perp}z'_\perp}   K_0(\sqrt{z\bar{z}Q^2}| z_\perp - z'_\perp |)  \mathcal{U}(z_\perp, z'_\perp;\omega_\perp, \omega'_\perp)\,.
\nonumber
\end{eqnarray}

The same matrix element (\ref{op-hepc-trlink}) appears in the dijet production cross section
\begin{eqnarray}
&&d\sigma^{\gamma^\ast p}_{\lambda=\pm1\lambda'=\pm1}\Big|_{\rm eik.} = \sum_f \frac{2\pi e^2_f \alpha_{\rm EM}x}{V_3}   \epsilon^{\lambda=\pm1\ast}_s \epsilon^{\lambda'=\pm1}_l \frac{dk^-_1d^2k_{1\perp}}{(2\pi)^3 2k^-_1} \Big|_{k^2_1=0}\frac{dk^-_2d^2k_{2\perp}}{(2\pi)^3 2k^-_2} \Big|_{k^2_2=0}\\
&&\times \delta(q^-- k^-_1 - k^-_2) 8(q^-)^2z^2\bar{z}^2
\big( (1 - 2z )^2 g^{sk}g^{l m} - \big( g^{sm}g^{kl} - g^{sl}g^{km}\big) \big) \int d^2\omega_\perp e^{ik_{1\perp}\omega_\perp} \nonumber\\
&&\times \int d^2\omega'_\perp  e^{ik_{2\perp}\omega'_\perp} \frac{ \omega_k - \omega'_k }{| \omega_\perp - \omega'_\perp |}
 K_1(\sqrt{z\bar{z}Q^2}| \omega_\perp - \omega'_\perp |) 
\int d^2z_\perp e^{-ik_{1\perp}z_\perp} \int d^2z'_\perp e^{-ik_{2\perp}z'_\perp}\nonumber\\
&&\times \frac{z_m - z'_m }{| z_\perp - z'_\perp |} K_1(\sqrt{z\bar{z}Q^2}| z_\perp - z'_\perp |) \mathcal{U}(z_\perp, z'_\perp;\omega_\perp, \omega'_\perp) \,.\nonumber
\end{eqnarray}

The last result can be further simplified by using the following identity for the polarization vector (\ref{eq:gauge-vec}):
\begin{eqnarray}
&&\epsilon^{\lambda=\pm 1\ast}_s \epsilon^{\lambda'=\pm1}_l \big( (1 - 2z )^2 g^{sk}g^{l m} - \big( g^{sm}g^{kl} - g^{sl}g^{km}\big) \big) \frac{ \omega_k - \omega'_k }{| \omega_\perp - \omega'_\perp |} \frac{z_m - z'_m }{| z_\perp - z'_\perp |} 
\nonumber\\
&&= \delta_{\lambda=\lambda'} ( -1 + 2z\bar{z} ) ( g^{km}  + i \epsilon^{km} ) \frac{ \omega_k - \omega'_k }{| \omega_\perp - \omega'_\perp | } \frac{ z_m - z'_m }{| z_\perp - z'_\perp | } +2 z \bar{z} \delta_{\lambda=-\lambda'} e^{i\lambda \phi}\,,
\end{eqnarray}
where anti-symmetric $\epsilon_{12} = 1$, and $\phi$ is defined as
\begin{eqnarray}
&&\cos\phi = \frac{ \omega_1 - \omega'_1 }{| \omega_\perp - \omega'_\perp |} \frac{z_1 - z'_1 }{| z_\perp - z'_\perp |}  - \frac{ \omega_2 - \omega'_2 }{| \omega_\perp - \omega'_\perp |} \frac{z_2 - z'_2 }{| z_\perp - z'_\perp |} \,.
\end{eqnarray}

This yields
\begin{eqnarray}
&&d\sigma^{\gamma^\ast p}_{\lambda=\pm1\lambda'=\pm1}\Big|_{\rm eik.} = \sum_f \frac{2\pi e^2_f \alpha_{\rm EM}x}{V_3}    \frac{dk^-_1d^2k_{1\perp}}{(2\pi)^3 2k^-_1} \Big|_{k^2_1=0}\frac{dk^-_2d^2k_{2\perp}}{(2\pi)^3 2k^-_2} \Big|_{k^2_2=0}\label{eq:cc11-f1}\\
&&\times \delta(q^-- k^-_1 - k^-_2) 8(q^-)^2z^2\bar{z}^2
\int d^2\omega_\perp e^{ik_{1\perp}\omega_\perp}
\int d^2\omega'_\perp  e^{ik_{2\perp}\omega'_\perp} 
 K_1(\sqrt{z\bar{z}Q^2}| \omega_\perp - \omega'_\perp |)\nonumber\\
&&\times \int d^2z_\perp e^{-ik_{1\perp}z_\perp} \int d^2z'_\perp e^{-ik_{2\perp}z'_\perp}
K_1(\sqrt{z\bar{z}Q^2}| z_\perp - z'_\perp |) \Big(\delta_{\lambda\lambda'} ( -1 + 2z\bar{z} )
( g^{km}  + i \epsilon^{km} ) \nonumber\\
&&\times \frac{ \omega_k - \omega'_k }{| \omega_\perp - \omega'_\perp | } \frac{ z_m - z'_m }{| z_\perp - z'_\perp | } +2 z \bar{z} \delta_{\lambda(-\lambda')} e^{i\lambda \phi} \Big)
\mathcal{U}(z_\perp, z'_\perp;\omega_\perp, \omega'_\perp) \,.
\nonumber
\end{eqnarray}
\end{widetext}
To some extend, Eqs. (\ref{eq:cc00-f1}), (\ref{eq:cc01-f1}), (\ref{eq:cc10-f1}), and (\ref{eq:cc11-f1}) represent the final result for the dijet production cross section at the eikonal order within the high-energy power counting (\ref{eq:pw-couting-sx}). Note that the cross section is defined by a matrix element of the operator (\ref{op-hepc-trlink}) with the transverse gauge links contribution. While the transverse gauge links in Eq. (\ref{op-hepc-trlink}) are formally written as operators at the spacial infinities, these operators appeared as a result of our formal manipulations with the operators in the dijet production cross section (\ref{eq:hadronic-bfm-general}). The physical content of these links corresponds to interactions of the (anti)quark with the transverse component of the background field at a finite position. As a result, contribution of the transverse gauge factors to the dijet production cross section is not trivial and cannot be neglected. For instance, one can derive another form of the dijet production cross section using an alternative representation of the amplitudes (\ref{eq:amp-gen-eik}) and (\ref{eq:ampcc-gen-eik}), which corresponds to another choice of the operator basis. We, however, leave analysis of this operator basis for another publication.

\section{Dijet production at the eikonal order in CGC approach\label{sec:cgc-eik}}
In the previous section, we derived the dijet production cross section at the eikonal order within the high-energy power counting (\ref{eq:pw-couting-sx}). The goal of this section is to relate our results to the ones derived within CGC approach (\ref{eq:cgc-frm}).

Previously, we found that the choice of the operator basis defining the cross section at the eikonal order in the power counting (\ref{eq:pw-couting-sx}) is not unique. Using formal operator transformations, one can rearrange contribution of the transverse field component to either a strength tensor insertion, see Eq. (\ref{eq:amp-gen-eik-form1}), or a transverse gauge link contribution at the past infinity. Both choices of the operator basis are equivalent to each other and physically correspond to interactions of the quantum (anti)quarks with the transverse component of the background field at a finite position in the coordinate space, which provides an eikonal contribution to the dijet production cross section.

We find that the choice of the operator basis without the strength tensor contribution, as in Eqs. (\ref{eq:cc00-f1}), (\ref{eq:cc01-f1}), (\ref{eq:cc10-f1}), and (\ref{eq:cc11-f1}), to be more convenient for the purpose of matching our results with the CGC framework. Indeed, from these equations we see that our results are different from the cross sections computed in CGC framework up to the transverse field component contribution. Neglecting the transverse field component, i.e. removing the transverse gauge links at the past infinity in Eqs. (\ref{eq:cc00-f1}), (\ref{eq:cc01-f1}), (\ref{eq:cc10-f1}), and (\ref{eq:cc11-f1}), we immediately obtain the dijet production cross sections, which are in complete agreement with the cross sections derived within CGC approach.\footnote{Note, however, that even at $B_i=0$, the background field within the high-energy power counting (\ref{eq:pw-couting-sx}) is different from the CGC background field (\ref{eq:cgc-frm}). Our matching implies comparison of the operator structures and kinematic coefficients arising in two frameworks.} We will provide the corresponding equations below in this section.

At the same time, while matching of our results with CGC framework seems to be straightforward, one shouldn't make light of the transverse gauge link contribution. The transverse gauge link provides a non-trivial effect, which is more obvious in the choice of the operator basis with the strength tensor contribution as in the amplitudes (\ref{eq:amp-gen-eik}) and (\ref{eq:ampcc-gen-eik}). In this case, the transverse background field component contributes directly to the field-strength tensor $F_{-i}$ at a finite position in the coordinate space providing a non-trivial contribution to the cross-section at the eikonal order. For completeness of our analysis, let us match our results in this operator basis to the cross sections computed in CGC framework.

Neglecting contribution of the transverse component of the background field in the amplitude (\ref{eq:amp-gen-eik}) computed within the high-energy power counting (\ref{eq:pw-couting-sx}), we write the corresponding CGC matching of our result as
\begin{widetext}
\begin{eqnarray}
&&iM^\nu_{s_1,s_2}(q, k_1, k_2) \Big|^{\rm CGC}_{\rm eik.}
= \int d^4y e^{-iqy} \theta(-y^-) \int d^2z_\perp e^{-ik_{1\perp}z_\perp} \int d^2z'_\perp e^{-ik_{2\perp}z'_\perp} \Big\{\bar{u}_{s_1}(k_1)\nonumber\\
&&\times \Big(U(\infty, -\infty;z_\perp)
- \frac{ig}{2k^-_1}\int^{\infty}_{-\infty} dz^- U(\infty, z^-;z_\perp) \sigma^{-m}\partial_m B_{-}(z^-,z_\perp)  U(z^-, -\infty;z_\perp)\Big)\nonumber\\
&&\times (z_\perp| e^{i\frac{ p^2_\perp}{2k^-_1}y^- } |y_\perp) U(-\infty, y^-;y_\perp)
e^{ik^-_1 y^+}\gamma^\nu e^{ik^-_2 y^+} U(y^-, -\infty;y_\perp) (y_\perp|e^{i\frac{p^2_\perp}{2k^-_2}y^-} |z'_\perp)\Big(U(-\infty, \infty;z'_\perp)\nonumber\\ 
&&+ \frac{ig}{2k^-_2}\int^{-\infty}_{\infty} dz'^- U(-\infty, z'^-;z'_\perp)
\sigma^{-n}\partial_nB_-(z'^-, z'_\perp) U(z'^-, \infty;z'_\perp)\Big) v_{s_2}(k_2) - (B=0)\Big\}\,.
\label{eq:amp-gen}
\end{eqnarray}

Using Eq. (\ref{eq:comm-WL-b0}) we can rewrite this equation as\footnote{Here the background fields are understood as operators $\hat{B}|x_\perp) = B(x_\perp)|x_\perp)$ acting on states, see Eq. (\ref{eq:sch-eihen}) etc.}
\begin{eqnarray}
&&iM^\nu_{s_1,s_2}(q, k_1, k_2)\Big|^{\rm CGC}_{\rm eik.}
= \int d^4y e^{-iqy} \theta(-y^-) \Big\{\bar{u}_{s_1}(k_1) (k_{1\perp}| \Big(U(\infty, -\infty) + \frac{i\sigma^{-m}}{2k^-_1} k_{1m} U(\infty, -\infty)
\nonumber\\
&& - \frac{i\sigma^{-m}}{2k^-_1}U(\infty, -\infty)p_m \Big) e^{i\frac{ p^2_\perp}{2k^-_1}y^- } |y_\perp) 
 e^{ik^-_1 y^+}\gamma^\nu e^{ik^-_2 y^+} (y_\perp| e^{i\frac{p^2_\perp}{2k^-_2}y^-} \Big(U(-\infty, \infty) - \frac{i\sigma^{-n}}{2k^-_2} p_n U(-\infty, \infty)
\nonumber\\
&& - \frac{i\sigma^{-n}}{2k^-_2} U(-\infty, \infty) k_{2n} \Big) |-k_{2\perp})v_{s_2}(k_2) - (B=0)\Big\}\,.
\label{eq:amp-CGC}
\end{eqnarray}
Note that in CGC approximation all covariant transverse derivatives in Eq. (\ref{eq:amp-gen-eik}) are replaced with simple transverse momenta.

For the polarization vector (\ref{eq:gauge-vec}), in the reference frame (\ref{eq:ref-fr}), we can further rewrite the amplitude (\ref{eq:amp-CGC}) as\footnote{Here we insert a complete set of transverse states, i.e. $\int d^2z_\perp |z_\perp)(z_\perp| = 1$, etc.}
\begin{eqnarray}
&&iM^\nu_{s_1,s_2}(q, k_1, k_2)\Big|^{\rm CGC}_{\rm eik.}
= \delta(q^-- k^-_1 - k^-_2) 2q^-z\bar{z} \int d^2z_\perp \int d^2z'_\perp \bar{u}_{s_1}(k_1) e^{-ik_{1\perp}z_\perp}
\label{eq:CGC-amp-fin1}\\
&&\times \Big\{  \Big( -i\gamma^\nu + \frac{ k_{1m}}{2k^-_1}\sigma^{-m}\gamma^\nu - \frac{k_{2m} }{2k^-_2} \gamma^\nu \sigma^{-m}\Big) K_0(\sqrt{z\bar{z}Q^2}| z_\perp - z'_\perp |)  - i\sqrt{z\bar{z}Q^2} \Big( \frac{\gamma^\nu \sigma^{-m}}{2k^-_2}  + \frac{\sigma^{-m}\gamma^\nu}{2k^-_1}\Big)\nonumber\\
&&\times \frac{z_m - z'_m }{| z_\perp - z'_\perp |}
K_1(\sqrt{z\bar{z}Q^2}| z_\perp - z'_\perp |) \Big\} e^{-ik_{2\perp}z'_\perp} v_{s_2}(k_2) \Big(U(\infty, -\infty;z_\perp) U(-\infty, \infty;z'_\perp) - 1\Big) \,,
 \nonumber
\end{eqnarray}
where we used Eqs. (\ref{eq:fourier1})
and (\ref{eq:fourier2}).
Note that terms $\sim\sigma^{-m}$ in this result originate in the $\sim \sigma^{\mu\nu}F_{\mu\nu}$ terms of the amplitude (\ref{eq:amp-gen-eik}) and the (anti)quark propagator in the background field (\ref{eq:prop-def}), where they correspond to the interaction term defined by the Pauli vertex.

Similarly, for the complex conjugated amplitude (\ref{eq:ampcc-gen-eik}) computed in the high-energy power counting (\ref{eq:pw-couting-sx}) at the eikonal order, in CGC approximations (\ref{eq:cgc-frm}) for the background field we find:
\begin{eqnarray}
&&-iM^{\dag\mu}_{s_1,s_2}(q, k_1, k_2)\Big|^{\rm CGC}_{\rm eik.} =  \delta(q^- - k^-_1 - k^-_2) 2q^-z\bar{z} \int d^2\omega_\perp \int d^2\omega'_\perp \bar{v}_{s_2}(k_2) e^{ik_{2\perp}\omega'_\perp} 
\label{eq:CGC-amp-fin2}\\
&&\times \Big\{ \Big( i\gamma^\mu +  \frac{k_{1k}}{2k^-_1} \gamma^\mu\sigma^{-k} - \frac{k_{2k}}{2k^-_2} \sigma^{-k}\gamma^\mu  \Big) K_0(\sqrt{z\bar{z}Q^2}| \omega_\perp - \omega'_\perp |) + i\sqrt{z\bar{z}Q^2} \Big( \frac{\gamma^\mu\sigma^{-k}}{2k^-_1} + \frac{\sigma^{-k}\gamma^\mu}{2k^-_2} \Big)\nonumber\\
&&\times \frac{ \omega_k - \omega'_k }{| \omega_\perp - \omega'_\perp |}
K_1(\sqrt{z\bar{z}Q^2}| \omega_\perp - \omega'_\perp |)   \Big\} e^{ik_{1\perp}\omega_\perp} u_{s_1}(k_1) \Big(U(\infty, -\infty;\omega'_\perp) U(-\infty, \infty;\omega_\perp) - 1\Big)\,.
\nonumber
\end{eqnarray}

Substituting Eqs. (\ref{eq:CGC-amp-fin1}) and (\ref{eq:CGC-amp-fin2}) into the cross section (\ref{eq:hadronic-bfm-eik}), we obtain the following CGC matching of our results. For the longitudinally polarized photon we have
\begin{eqnarray}
&&d\sigma^{\gamma^\ast p}_{00}\Big|^{\rm CGC}_{\rm eik.} = \sum_f \frac{2\pi e^2_f \alpha_{\rm EM}x}{V_3Q^2}   \Big(\frac{Q}{q^-}\Big)^2 \frac{dk^-_1d^2k_{1\perp}}{(2\pi)^3 2k^-_1} \Big|_{k^2_1=0}\frac{dk^-_2d^2k_{2\perp}}{(2\pi)^3 2k^-_2} \Big|_{k^2_2=0} \delta(q^-- k^-_1 - k^-_2) 32 z^3\bar{z}^3 (q^-)^4
\nonumber\\
&&\times \int d^2\omega_\perp e^{ik_{1\perp}\omega_\perp}
 \int d^2\omega'_\perp  e^{ik_{2\perp}\omega'_\perp}  K_0(\sqrt{z\bar{z}Q^2}| \omega_\perp - \omega'_\perp |) \int d^2z_\perp e^{-ik_{1\perp}z_\perp} \int d^2z'_\perp  e^{-ik_{2\perp}z'_\perp}
 \nonumber\\
&&\times K_0(\sqrt{z\bar{z}Q^2}| z_\perp - z'_\perp |) 
\mathcal{U}(z_\perp, z'_\perp;\omega_\perp, \omega'_\perp)\Big|_{\rm CGC} \,,
 \label{eq:CGC-00res}
\end{eqnarray}
where the background field operator
\begin{equation}
\mathcal{U}(z_\perp, z'_\perp;\omega_\perp, \omega'_\perp)\Big|_{\rm CGC}
\equiv \langle P, S| {\rm tr} \Big(U(\infty, -\infty;\omega'_\perp) U(-\infty, \infty;\omega_\perp) - 1\Big) \Big(U(\infty, -\infty;z_\perp) U(-\infty, \infty;z'_\perp) - 1\Big)
|P, S\rangle
\end{equation}
is constructed from the light-cone gauge factors, c.f. Eq. (\ref{op-hepc-trlink}).

The same matrix element defines cross sections
\begin{eqnarray}
&&d\sigma^{\gamma^\ast p}_{0\lambda'=\pm1}\Big|^{\rm CGC}_{\rm eik.} = \sum_f \frac{2\pi e^2_f \alpha_{\rm EM}x}{V_3Q^2}   \frac{Q}{q^-} \epsilon^{\lambda'=\pm1}_l \frac{dk^-_1d^2k_{1\perp}}{(2\pi)^3 2k^-_1} \Big|_{k^2_1=0}\frac{dk^-_2d^2k_{2\perp}}{(2\pi)^3 2k^-_2} \Big|_{k^2_2=0} \delta(q^-- k^-_1 - k^-_2)
\nonumber\\
&&\times 16z^2\bar{z}^2( z - \bar{z} )(q^-)^3 
\int d^2\omega_\perp e^{ik_{1\perp}\omega_\perp} \int d^2\omega'_\perp  e^{ik_{2\perp}\omega'_\perp} K_0(\sqrt{z\bar{z}Q^2}| \omega_\perp - \omega'_\perp |) \int d^2z_\perp e^{-ik_{1\perp}z_\perp}
\nonumber\\
&&\times \int d^2z'_\perp e^{-ik_{2\perp}z'_\perp} i\sqrt{z\bar{z}Q^2} \frac{z_l - z'_l }{| z_\perp - z'_\perp |}
K_1(\sqrt{z\bar{z}Q^2}| z_\perp - z'_\perp |) \mathcal{U}(z_\perp, z'_\perp;\omega_\perp, \omega'_\perp)\Big|_{\rm CGC}
\label{eq:int-term1-CGC}
\end{eqnarray}
and
\begin{eqnarray}
&&d\sigma^{\gamma^\ast p}_{\lambda=\pm10}\Big|^{\rm CGC}_{\rm eik.} = \sum_f \frac{2\pi e^2_f \alpha_{\rm EM}x}{V_3Q^2}   \epsilon^{\lambda=\pm 1\ast}_s \frac{Q}{q^-} \frac{dk^-_1d^2k_{1\perp}}{(2\pi)^3 2k^-_1} \Big|_{k^2_1=0}\frac{dk^-_2d^2k_{2\perp}}{(2\pi)^3 2k^-_2} \Big|_{k^2_2=0} \delta(q^-- k^-_1 - k^-_2)\nonumber\\
&&\times 16 z^2 \bar{z}^2 ( \bar{z}  - z ) (q^-)^3 
\int d^2\omega_\perp e^{ik_{1\perp}\omega_\perp} \int d^2\omega'_\perp  e^{ik_{2\perp}\omega'_\perp} i\sqrt{z\bar{z}Q^2} \frac{ \omega_s - \omega'_s }{| \omega_\perp - \omega'_\perp |}
 K_1(\sqrt{z\bar{z}Q^2}| \omega_\perp - \omega'_\perp |)\nonumber\\
&&\times \int d^2z_\perp e^{-ik_{1\perp}z_\perp} \int d^2z'_\perp  e^{-ik_{2\perp}z'_\perp}  
K_0(\sqrt{z\bar{z}Q^2}| z_\perp - z'_\perp |)  \mathcal{U}(z_\perp, z'_\perp;\omega_\perp, \omega'_\perp)\Big|_{\rm CGC}\,.
\label{eq:int-term2-CGC}
\end{eqnarray}
Again, these results were obtained from our general expression for the dijet production cross section derived within the high-energy power counting (\ref{eq:pw-couting-sx}), see Eq. (\ref{eq:hadronic-bfm-eik}), by setting the transverse component of the background field $B_i = 0$, which corresponds to the CGC approximations (\ref{eq:cgc-frm}).

Finally, for the transversly polarized photon we have
\begin{eqnarray}
&&d\sigma^{\gamma^\ast p}_{\lambda=\pm1\lambda'=\pm1}\Big|^{\rm CGC}_{\rm eik.} = \sum_f \frac{2\pi e^2_f \alpha_{\rm EM}x}{V_3}   \epsilon^{\lambda=\pm1\ast}_s \epsilon^{\lambda'=\pm1}_l \frac{dk^-_1d^2k_{1\perp}}{(2\pi)^3 2k^-_1} \Big|_{k^2_1=0}\frac{dk^-_2d^2k_{2\perp}}{(2\pi)^3 2k^-_2} \Big|_{k^2_2=0}\label{eq:CGC-11res}\\
&&\times \delta(q^-- k^-_1 - k^-_2)
8(q^-)^2z^2\bar{z}^2
\int d^2\omega_\perp e^{ik_{1\perp}\omega_\perp} \int d^2\omega'_\perp  e^{ik_{2\perp}\omega'_\perp} 
 K_1(\sqrt{z\bar{z}Q^2}| \omega_\perp - \omega'_\perp |)\nonumber\\
 &&\times \int d^2z_\perp e^{-ik_{1\perp}z_\perp} \int d^2z'_\perp  e^{-ik_{2\perp}z'_\perp}
 K_1(\sqrt{z\bar{z}Q^2}| z_\perp - z'_\perp |)
\Big(\delta_{\lambda=\lambda'} ( -1 + 2z\bar{z} ) ( g^{km}  + i \epsilon^{km} )\nonumber\\
&&\times \frac{ \omega_k - \omega'_k }{| \omega_\perp - \omega'_\perp | } \frac{ z_m - z'_m }{| z_\perp - z'_\perp | } +2 z \bar{z} \delta_{\lambda=-\lambda'} e^{i\lambda \phi} \Big) \mathcal{U}(z_\perp, z'_\perp;\omega_\perp, \omega'_\perp)\Big|_{\rm CGC} \,.
 \nonumber
\end{eqnarray}
\end{widetext}

Eqs. (\ref{eq:CGC-00res}), (\ref{eq:int-term1-CGC}), (\ref{eq:int-term2-CGC}), and (\ref{eq:CGC-11res}) coincide with the known results for the dijet production cross section derived in the CGC framework, see e.g. Ref. \cite{Kovchegov:2024wjs}. Therefore, we conclude that our expression for the cross section, see Eqs. (\ref{eq:hadronic-bfm-eik}), (\ref{eq:amp-gen-eik}), and (\ref{eq:ampcc-gen-eik}), is in agreement with the CGC predictions, when contribution of the transverse component of the background field is trivial $B_i=0$.

As we noticed before, contribution of the transverse field component within the high-energy power counting (\ref{eq:pw-couting-sx}) is not trivial and appears at the eikonal order of scattering. To reveal this contribution in a more transparent way, it is instructive to match our results for the dijet production cross section at the eikonal order, Eqs. (\ref{eq:hadronic-bfm-eik}), (\ref{eq:amp-gen-eik}), and (\ref{eq:ampcc-gen-eik}), with the cross section computed in the back-to-back limit, see Eq. (\ref{eq:btb-ltwist-pr}). As we previously discussed, the back-to-back power counting of terms is based on the ordering of transverse momenta between the quantum and background fields. This is rather different from the eikonal counting of the background fields (\ref{eq:pw-couting-sx}). As we will see in the next section, contribution of the transverse field component plays an essential role in the matching of two results, and explicitly appears as a non-trivial contribution in operators of our eikonal result, Eqs. (\ref{eq:hadronic-bfm-eik}), (\ref{eq:amp-gen-eik}), and (\ref{eq:ampcc-gen-eik}), written in the back-to-back approximation. Note that this non-trivial contribution does not appear in the back-to-back limit of the eikonal results computed in CGC framework, Eqs. (\ref{eq:CGC-00res}), (\ref{eq:int-term1-CGC}), (\ref{eq:int-term2-CGC}), and (\ref{eq:CGC-11res}), where contribution of the transverse field component is trivial.

To illustrate this more explicitly, let us consider the back-to-back limit of the CGC result at the eikonal order. To construct this limit, it is convenient to start with the back-to-back limit of the eikonal dijet production cross section computed within the high-energy power counting (\ref{eq:pw-couting-sx}). We will construct this limit in Sec. \ref{sec:btb-eik}. For now, however, let us borrow the final result of that section (\ref{eq:btb-ltwist-pr-ek}). 

It is easy to see that the eikonal cross section computed within the high-energy power counting (\ref{eq:pw-couting-sx}) in the back-to-back limit (\ref{eq:btb-ltwist-pr-ek}) essentially coincides with the dijet production cross section (\ref{eq:btb-ltwist-pr}) computed within the back-to-back power counting up to an exponential factor in the gluon TMD operator, which is a result of the eikonal approximations in Eq. (\ref{eq:btb-ltwist-pr-ek}).

As we explained above, the back-to-back limit of the CGC result can be read from these equations by setting the transverse component of the background field $B_i=0$. Specifically, from Eqs. (\ref{eq:lead-order-col-btb}) and (\ref{eq:lead-order-col-btb-TT})\footnote{The cross sections (\ref{eq:interf-first}) and (\ref{eq:interf-first}) yield a trivial contribution after symmetrization with respect to an interchange of the quark and antiquark.}
\begin{widetext}
\begin{eqnarray}
&&d\sigma^{\gamma^\ast p}_{00}\Big|^{\rm CGC}_{\rm back-to-back} = \sum_f \frac{2\pi e^2_f g^2 \alpha_{\rm EM}x}{Q^2} \Big(\frac{Q}{q^-}\Big)^2\frac{dk^-_1d^2k_{1\perp}}{(2\pi)^3 2k^-_1} \Big|_{k^2_1=0}\frac{dk^-_2d^2k_{2\perp}}{(2\pi)^3 2k^-_2} \Big|_{k^2_2=0}\label{eq:lead-order-col-btb-CGC}\\
&&\times 2\pi\delta(q^--k^-_1-k^-_2)
2\bar{P}_k \bar{P}_m \frac{32 z^3 \bar{z}^3 (q^-)^4}{(\bar{P}^2_\perp + z\bar{z}Q^2)^4}  \int d^2x_\perp e^{i \Delta_{\perp} x_\perp} \int_{-\infty}^{\infty} dz^- \langle P, S|
    F^a_{-k}(z^- , x_\perp)\nonumber\\
&&\times U^{ac}(z^-, \infty; x_\perp)
 U^{cb}(\infty, 0^-;0_\perp) F^b_{-m}(0^-, 0_\perp) |P, S\rangle\,.
\nonumber
\end{eqnarray}
and
\begin{eqnarray}
&&d\sigma^{\gamma^\ast p}_{\lambda=\pm1\lambda'=\pm1}\Big|^{\rm CGC}_{\rm back-to-back} =   \sum_f \frac{2\pi e^2_f g^2 \alpha_{\rm EM}x}{Q^2} \frac{dk^-_1d^2k_{1\perp}}{(2\pi)^3 2k^-_1} \Big|_{k^2_1=0}\frac{dk^-_2d^2k_{2\perp}}{(2\pi)^3 2k^-_2} \Big|_{k^2_2=0}\label{eq:lead-order-col-btb-TT-CGC}\\
&&\times 2\pi\delta(q^--k^-_1-k^-_2)
\frac{4z^2\bar{z}^2 (q^-)^2}{(\bar{P}^2_\perp + z\bar{z}Q^2)^2} \Big[ -\delta_{\lambda\lambda'} \Big( \frac{z}{\bar{z}} + \frac{\bar{z}}{z}\Big) ( g_{sl} + i\lambda \epsilon_{sl} ) +  2 \delta_{\lambda(-\lambda')} \Big( g_{1s}g_{1l} -  g_{2s}g_{2 l}\nonumber\\
&&+ i \lambda ( g_{1s}g_{2 l} +  g_{2s} g_{1l}) \Big) \Big]
\Big(  g^{ks}g^{ml} + \frac{2g^{ks}\bar{P}^m \bar{P}^l }{\bar{P}^2_\perp + z\bar{z}Q^2} + \frac{2g^{ml} \bar{P}^k \bar{P}^s }{\bar{P}^2_\perp + z\bar{z}Q^2} + \frac{4 \bar{P}^k \bar{P}^s \bar{P}^m \bar{P}^l }{(\bar{P}^2_\perp + z\bar{z}Q^2)^2}\Big) \int d^2x_\perp e^{i \Delta_\perp x_\perp}\nonumber\\
&&\times \int_{-\infty}^{\infty} dz^-
\langle P, S| F^a_{-k}(z^- , x_\perp) U^{ac}(z^-, \infty; x_\perp) U^{cb}(\infty, 0^-;0_\perp) F^b_{-m}(0^-, 0_\perp) |P, S\rangle\,.
\nonumber
\end{eqnarray}

In Eqs. (\ref{eq:lead-order-col-btb-CGC}) and (\ref{eq:lead-order-col-btb-TT-CGC}) the strength tensors should be understood as $F_{-k} = -\partial_k A_-$, within the CGC framework (\ref{eq:cgc-frm}). Because of that, parametrization of the matrix elements in  CGC cross sections (\ref{eq:lead-order-col-btb-CGC}) and (\ref{eq:lead-order-col-btb-TT-CGC}) is different from the parametrization (\ref{eq:TMD-gluon-param}) of the matrix elements in the back-to-back cross sections (\ref{eq:lead-order-col-btb}) and (\ref{eq:lead-order-col-btb-TT}). Note that in the latter case the strength tensors include contribution of the transverse background field component: $F_{-k} = \partial_- A_k - \partial_k A_- - ig [A_-, A_k]$.

Specifically, in the small-x limit one expects various relations between the spin-dependent distribution functions, see e.g. Ref. \cite{Boer:2016fqd}, which leads to modification of the general parametrization formula (\ref{eq:TMD-gluon-param}). Since the relations between the distribution functions in the small-x limit are not fully understood, let us limit our analysis to the case of an unpolarized target, which implies the following parametrization of the matrix element:
\begin{eqnarray}
&& \int \frac{d^2x_\perp}{(2\pi)^2} e^{i \Delta_\perp x_\perp} \int_{-\infty}^{\infty} \frac{dz^-}{2\pi} \langle P| F^a_{-k}(z^- , x_\perp) U^{ac}(z^-, \infty; x_\perp) U^{cb}(\infty, 0^-;0_\perp) F^b_{-m}(0^-, 0_\perp) |P\rangle
\nonumber\\
&&= \frac{P^+}{2}\Big[ - g^{km} \mathcal{D}(\Delta^2_\perp) + \frac{\Delta^{km}_\perp}{M^2} \mathcal{H}(\Delta^2_\perp) \Big]\,,
\label{eq:parapsmx}
\end{eqnarray}
where
\begin{eqnarray}
&&\mathcal{D}(\Delta^2_\perp)=\lim_{x\to0}x f_1(x, \Delta^2_\perp);\ \ \ \mathcal{H}(\Delta^2_\perp) =\lim_{x\to0}x h^\perp_1(x, \Delta^2_\perp)\,.
\end{eqnarray}

Substituting Eq. (\ref{eq:parapsmx}) into Eqs. (\ref{eq:lead-order-col-btb-CGC}) and (\ref{eq:lead-order-col-btb-TT-CGC}) we obtain
\begin{eqnarray}
&&d\sigma^{\gamma^\ast p}_{00}\Big|^{\rm CGC}_{\rm back-to-back} = \sum_f (2\pi)^4 e^2_f g^2 \alpha_{\rm EM}x \frac{dk^-_1d^2k_{1\perp}}{(2\pi)^3 2k^-_1} \Big|_{k^2_1=0}\frac{dk^-_2d^2k_{2\perp}}{(2\pi)^3 2k^-_2} \Big|_{k^2_2=0} 2\pi\delta(q^--k^-_1-k^-_2)q^-
\label{eq:lead-order-col-btb-CGC-param}\\
&&\times \frac{32 z^3 \bar{z}^3 \bar{P}^2_\perp (P\cdot q)}{(\bar{P}^2_\perp + z\bar{z}Q^2)^4}
 \Big[ \mathcal{D}(\Delta^2_\perp) + \frac{\Delta^2_\perp}{2M^2} \cos\big(2(\phi_{\bar{P}} - \phi_{\Delta})\big)\mathcal{H}(\Delta^2_\perp) \Big]
\nonumber
\end{eqnarray}
and\footnote{Here, for brevity, we again present only $\lambda=\lambda'$ contribution.}
\begin{eqnarray}
&&d\sigma^{\gamma^\ast p}_{\lambda=\lambda'=\pm1}\Big|^{\rm CGC}_{\rm back-to-back} =   \sum_f (2\pi)^4 e^2_f g^2 \alpha_{\rm EM}x \frac{dk^-_1d^2k_{1\perp}}{(2\pi)^3 2k^-_1} \Big|_{k^2_1=0}\frac{dk^-_2d^2k_{2\perp}}{(2\pi)^3 2k^-_2} \Big|_{k^2_2=0} 2\pi\delta(q^--k^-_1-k^-_2)q^-
\label{eq:lead-order-col-btb-TT-CGC-param}\\
&&\times 
\frac{8z^2\bar{z}^2 ( z^2 + \bar{z}^2 ) (P\cdot q)}{(\bar{P}^2_\perp + z\bar{z}Q^2)^4} \Big[ \frac{ \bar{P}^4_\perp + z^2\bar{z}^2Q^4 }{2z\bar{z} Q^2} \mathcal{D}(\Delta^2_\perp) - \frac{\Delta^2_\perp P^2_\perp}{2M^2} \cos \big(2 (\phi_{\bar{P}} - \phi_{\Delta})\big) \mathcal{H}(\Delta^2_\perp) \Big]\,.
\nonumber
\end{eqnarray}
\end{widetext}

Comparing the dijet production cross section in the back-to-back limit computed in CGC framework (\ref{eq:lead-order-col-btb-CGC-param}) and (\ref{eq:lead-order-col-btb-TT-CGC-param}) with the corresponding equations computed within the back-to-back power counting (\ref{eq:00-btb-param}) and (\ref{eq:11-btb-param}), we find agreement in the structure of angular modulations between two results for the unpolarized scattering.

At the same time, we expect a difference in the structure of angular modulations for the scattering on a polarized target. As we mentioned above, this difference is anticipated due to various relations between the spin-dependent distribution functions appearing the small-x limit, see Ref. \cite{Boer:2016fqd}. The difference in angular modulations can be utilized in analysis of the experimental data as a signature of the saturation of the dense QCD medium described by the CGC framework. Quantitative analysis of this difference and its origin goes beyond the scope of this paper, and is left for a future publication.

In the next section, we will complete our analysis by deriving the back-to-back limit of the eikonal cross section computed within the high-energy power counting (\ref{eq:pw-couting-sx}). We will demonstrate a non-trivial contribution of the transverse background field component, which appears in the field-strength tensors $F_{-k} = \partial_- A_k - \partial_k A_- - ig [A_-, A_k]$ of the gluon TMD operator of the dijet production cross section in the back-to-back limit at the eikonal order.









\section{Dijet production at the eikonal order in the back-to-back limit\label{sec:btb-eik}}
One of the main advantages of our framework is that it allows efficient matching of results obtained in different kinematic approximations. In this section, we demonstrate this explicitly by comparing the dijet production cross section computed within the high-energy power counting~(\ref{eq:pw-couting-sx}) at the eikonal order with the same cross section computed in the back-to-back limit.


In general, distinct kinematic limits are dominated by different parton dynamics. This underlying dynamics gives rise to a unique expansion parameter, from which a factorization formula for the physical observable in a given limit can be derived.


For instance, our derivation of the dijet production in the back-to-back limit is based on the expansion parameter $\Delta_\perp/\bar{P}_\perp \ll 1$, originating from the ordering of parton emission in transverse momenta that dominates this kinematic regime. Such ordering typically occurs in scattering at large Bjorken $x_B = Q^2/(2P\cdot q)$ and effectively leads to a twist counting, with the dijet production cross section determined by TMD operators of different power.


The high-energy power counting~(\ref{eq:pw-couting-sx}), by contrast, describes scattering in the high-energy limit of small Bjorken $x_B$. This power counting implies ordering of parton emission in longitudinal momenta, while the transverse momenta remain unordered, as seen from Eq.~(\ref{eq:pw-couting-sx}).


Physical observables in different kinematic limits are thus dominated by different aspects of the parton dynamics. The dominant effects are not obviously related, making the comparison of the same observable computed in different regimes far from straightforward.


In our approach, however, the analysis of physical observables in different kinematic limits is carried out within a single computational framework, enabling us to study relations between the results and understand transitions between different limits. We begin with the general expression~(\ref{eq:hadronic-bfm-general}) for the dijet production cross section, valid in arbitrary kinematics. Kinematic assumptions are imposed by expanding the (anti)quark propagators onto different contours consistent with the expansion parameters of the respective regimes.


While expanding the cross section and applying the power counting allows one to estimate the size of different terms and isolate the dominant contribution in a given limit, the expansion procedure itself does not alter the physical content of the equations. In this regard, Eq.~(\ref{eq:hadronic-bfm-general}) is complete and general, as are its various expanded forms derived in this paper.


To isolate the dominant contribution in a given limit, we expand the path-ordered exponents in Eq.~(\ref{eq:hadronic-bfm-general}) using formal operator transformations. This expansion introduces no new physical information; it merely reorganizes the terms of the path-ordered exponents in the (anti)quark propagators. Different expansions introduced in our analysis are therefore completely equivalent to each other. The sole purpose of the expansion is to represent the physical content of Eq.~(\ref{eq:hadronic-bfm-general}) in a form suitable for extracting the dominant contribution within a given power counting scheme. This enables efficient mapping between results computed in different kinematic limits by comparing the corresponding expansion series.


Once the path-ordered exponents in Eq.~(\ref{eq:hadronic-bfm-general}) are expanded and their terms organized according to the expansion parameter, our method allows one to map each term to a different power counting scheme by re-expanding it onto a different contour. This effectively builds a dictionary between different power counting schemes for the dijet production in different kinematic limits.


To illustrate the matching procedure, we consider the amplitudes~(\ref{eq:amp-gen-eik}) and~(\ref{eq:ampcc-gen-eik}) derived within the high-energy power counting~(\ref{eq:pw-couting-sx}) at the eikonal order and match them to the back-to-back limit by re-expanding onto the light-cone contour in Fig.~\ref{fig:fig13}.


\begin{widetext}
We first rewrite the amplitude~(\ref{eq:amp-gen-eik}) as
\begin{eqnarray}
&&iM^\nu_{s_1,s_2}(q, k_1, k_2) \Big|_{\rm eik.}
= \int d^4y e^{-iqy} \theta(-y^-) \Big\{\bar{u}_{s_1}(k_1) (k_{1\perp}| \Big(U(\infty, -\infty)\label{eq:amp-eik-simp-btb}\\
&&+ \frac{ig}{2k^-_1}\int^{\infty}_{-\infty} dz^- U(\infty, z^-) \sigma^{-m}F_{-m}(z^-)
U(z^-, -\infty) \Big) e^{i\frac{ P^2_\perp(-\infty)}{2k^-_1}y^- } U(-\infty, y^-) |y_\perp) e^{ik^-_1 y^+}\gamma^\nu e^{ik^-_2 y^+}\nonumber\\
&&\times (y_\perp|  U(y^-, -\infty) e^{i\frac{P^2_\perp(-\infty)}{2k^-_2}y^-} \Big(U(-\infty, \infty) 
- \frac{ig}{2k^-_2}\int^{-\infty}_{\infty} dz'^- U(-\infty, z'^-)\nonumber\\
&&\times \sigma^{-n}F_{-n}(z'^-) U(z'^-, \infty)\Big)|-k_{2\perp})v_{s_2}(k_2) - (B=0)\Big\}\,.
\nonumber
\end{eqnarray}
To match this result with the back-to-back power counting, we need to re-expand the operators in a way consistent with the back-to-back limit, reorganizing the terms of Eq.~(\ref{eq:amp-eik-simp-btb}) according to their size within the back-to-back power counting. This is accomplished by re-expanding the operators onto the light-cone contour in Fig.~\ref{fig:fig13}, which can be constructed by commuting all transverse covariant operators $P_k$ to the final (anti)quark state.


For the exponential factors~(\ref{eq:exp-fact-cova}), this commutation takes the form, cf.\ Eq.~(\ref{eq:commutation-identity}),
\begin{eqnarray}
&&U(\infty, -\infty) e^{i\frac{ P^2_\perp(-\infty)}{2k^-_1}y^-}
\label{eq:com-PperpU}\\
&&= e^{i\frac{ P^2_\perp(\infty)}{2k^-_1}y^-} \Big[U(\infty, -\infty) +\Big( \frac{-iy^-}{2k^-_1}\Big) [ P^2_\perp , U(\infty, -\infty) + \frac{1}{2!}\Big(\frac{-iy^-}{2k^-_1}\Big)^2[P^2_\perp , [P^2_\perp , U(\infty, -\infty)]] + \dots\Big] \,,
\nonumber
\end{eqnarray}
where the ellipsis denotes higher-order terms. On the right-hand side, we adopt the convention $P_\perp\to P_\perp(\infty)$ when the covariant momentum stands to the left of the gauge factor, and $P_\perp\to P_\perp(-\infty)$ when it stands to the right.


In Eq.~(\ref{eq:com-PperpU}), the covariant transverse momenta can be further commuted to the left, yielding
 \begin{eqnarray}
&&U(\infty, -\infty)e^{i\frac{ P^2_\perp(-\infty)}{2k^-_1}y^-}\label{eq:comm-rel}\\
&&= e^{i\frac{ P^2_\perp(\infty)}{2k^-_1}y^-}\Big[ U(\infty, -\infty) -\frac{iy^-}{2k^-_1} \Big(2g \int^{\infty}_{-\infty}dz^- P_m(\infty) U(\infty, z^-) F_{-m}(z^-)U(z^-, -\infty)
\nonumber\\
&&+ ig \int^{\infty}_{-\infty}dz^- U(\infty, z^-)  D^m F_{-m}(z^-) U(z^-, -\infty)
- 2g^2 \int^{\infty}_{-\infty}dz^- \int^{z^-}_{-\infty}dz'^-   U(\infty, z^-) F_{-m}(z^-)\nonumber\\
&&\times U(z^-, z'^-)
F_{-m}(z'^-)U(z'^-, -\infty)\Big)  + \dots \Big]\,.
\nonumber
\end{eqnarray}
The higher-order terms of this commutation relation can be constructed straightforwardly. This identity represents the original operator on the left-hand side as an infinite series on the right-hand side, corresponding to an expansion onto the light-cone direction. The expansion~(\ref{eq:comm-rel}) is consistent with the back-to-back power counting: higher-order terms are parametrically suppressed by higher powers of the expansion parameter, corresponding to additional field-strength tensor insertions.


To isolate the leading-power back-to-back contribution in Eq.~(\ref{eq:amp-eik-simp-btb}), it therefore suffices to retain only terms linear in the background field-strength tensors. At leading power, we use the replacement
 \begin{equation}
U(\infty, -\infty)e^{i\frac{ P^2_\perp(-\infty)}{2k^-_1}y^-}
\to e^{i\frac{ P^2_\perp(\infty)}{2k^-_1}y^-}\Big[ U(\infty, -\infty) -\frac{igy^-}{k^-_1} \int^{\infty}_{-\infty}dz^- P_m(\infty) U(\infty, z^-) F_{-m}(z^-)U(z^-, -\infty)  \Big]\,.
\end{equation}

The remaining terms in Eq.~(\ref{eq:amp-eik-simp-btb}) can be similarly re-expanded onto the light-cone contour. For the leading-power matching, it suffices to retain terms linear in the background field-strength tensors. For instance, at leading order we use the replacement
\begin{eqnarray}
&&\Big(\frac{ig}{2k^-_1} \int^{\infty}_{-\infty} dz^- U(\infty, z^-) \sigma^{-m}F_{-m}(z^-)   U(z^-, -\infty)\Big) e^{i\frac{ P^2_\perp(-\infty)}{2k^-_1}y^- } 
\nonumber\\
&&\to e^{i\frac{ P^2_\perp(\infty)}{2k^-_1}y^- } \Big( \frac{ig}{2k^-_1}\int^{\infty}_{-\infty} dz^- U(\infty, z^-) \sigma^{-m}F_{-m}(z^-)   U(z^-, -\infty)\Big)\,.
\label{eq:simp-singFbtb}
\end{eqnarray}

Re-expanding the operators in Eq.~(\ref{eq:amp-eik-simp-btb}) onto the light-cone contour in Fig.~\ref{fig:fig13}, we can match our eikonal result with the leading-power contribution in the back-to-back limit. The result reads
\begin{eqnarray}
&&iM^\nu_{s_1,s_2}(q, k_1, k_2) \Big|^{\rm back-to-back}_{\rm eik.}
= \int d^4y e^{-iqy} \theta(-y^-) \Big\{\bar{u}_{s_1}(k_1) e^{i\frac{ k^2_{1\perp}}{2k^-_1}y^- } e^{-ik_{1\perp}y_\perp} \Big(U(\infty, y^-;y_\perp)\nonumber\\
&&-\frac{igk_{1m} y^-}{k^-_1} \int^{\infty}_{-\infty}dz^-  
U(\infty, z^-;y_\perp) F_{-m}(z^-,y_\perp) U(z^-, y^-; y_\perp) + \frac{ig}{2k^-_1}\int^{\infty}_{-\infty} dz^- U(\infty, z^-;y_\perp)\nonumber\\
&&\times \sigma^{-m}F_{-m}(z^-, y_\perp) U(z^-, y^-;y_\perp) \Big)
e^{ik^-_1 y^+}\gamma^\nu e^{ik^-_2 y^+}    \Big(U(y^-, \infty;y_\perp) - \frac{ig k_{2n} y^-}{k^-_2} \int^{-\infty}_{\infty}dz^- U(y^-, z^-;y_\perp)\nonumber\\
&&\times F_{-n}(z^-,y_\perp)U(z^-, \infty;y_\perp)
- \frac{ig}{2k^-_2}\int^{-\infty}_{\infty} dz'^- U(y^-, z'^-;y_\perp) \sigma^{-n}F_{-n}(z'^-,y_\perp) U(z'^-, \infty;y_\perp)\Big)\nonumber\\
&&\times e^{i\frac{k^2_{2\perp}}{2k^-_2}y^-} e^{-ik_{2\perp}y_\perp} v_{s_2}(k_2) - (B=0)\Big\}\,,
\label{eq:amp-eik-simp-btb-exp1}
\end{eqnarray}
where we used a supplementary boundary condition for the background fields: $\lim_{x^-\to\infty}B_i(x) = 0$. The operators in Eq. (\ref{eq:amp-eik-simp-btb-exp1}) are along the light-cone contour in Fig. \ref{fig:fig13}. The expression for the amplitude can be further simplified as
\begin{eqnarray}
&&iM^\nu_{s_1,s_2}(q, k_1, k_2) \Big|^{\rm back-to-back}_{\rm eik.}
= 2\pi\delta(q^- - k^-_1 - k^-_2) \int d^2y_\perp e^{i(q_\perp-k_{1\perp}-k_{2\perp}) y_\perp} \bar{u}_{s_1}(k_1)~~~~~~~~~~~~\label{eq:amp-eik-simp-btb-expfin}\\
&&\times \Big\{ \Big( \frac{\sigma^{-m}\gamma^\nu}{2k^-_1} + \frac{\gamma^\nu \sigma^{-m}}{2k^-_2}  \Big) 
\frac{i}{q^+- k^+_1- k^+_2} - \Big( \frac{k_{1m} \gamma^\nu}{k^-_1} - \frac{k_{2m}\gamma^\nu }{k^-_2} \Big)\frac{1}{(q^+- k^+_1- k^+_2)^2}
 \Big\} t^d v_{s_2}(k_2)\nonumber\\
 &&\times\Big( ig \int^{\infty}_{-\infty} dz^-   U^{db}( \infty, z^-;y_\perp) F^b_{-m}(z^-, y_\perp) \Big) \,.
\nonumber
\end{eqnarray}

Similarly, for the complex conjugated amplitude (\ref{eq:ampcc-gen-eik}) we obtain
\begin{eqnarray}
&&-iM^{\dag\mu}_{s_1,s_2}(q, k_1, k_2)\Big|^{\rm back-to-back}_{\rm eik.} = 2\pi\delta(q^--k^-_1-k^-_2)  \int d^2x_\perp e^{-i(q_\perp - k_{1\perp}-k_{2\perp}) x_\perp}\label{eq:ampcc-eik-simp-btb-expfin}\\
&&\times\bar{v}_{s_2}(k_2) \Big\{  \Big(\frac{\gamma^\mu\sigma^{-k}}{2k^-_1}
+ \frac{\sigma^{-k} \gamma^\mu}{2k^-_2} \Big)
\frac{i}{q^+-k^+_1-k^+_2} + \Big(\frac{k_{1k}\gamma^\mu }{k^-_1}  - \frac{k_{2k}\gamma^\mu}{k^-_2} \Big) \frac{1}{(q^+-k^+_1-k^+_2)^2}
\Big\}  t^c  u_{s_1}(k_1)\nonumber\\
&&\times \Big(ig \int^{\infty}_{-\infty} d\omega^- U^{ca}(\infty, \omega^-;x_\perp) F^a_{-k}(\omega^-,x_\perp)\Big) \,.\nonumber
\end{eqnarray}

Substituting Eqs.~(\ref{eq:amp-eik-simp-btb-expfin}) and~(\ref{eq:ampcc-eik-simp-btb-expfin}) into the cross section~(\ref{eq:hadronic-bfm-eik}), we obtain
\begin{eqnarray}
&&d\sigma^{\gamma^\ast p}_{\lambda\lambda'}\Big|^{\rm back-to-back}_{\rm eik.} = \sum_f\sum_{s_1,s_2}\frac{2\pi e^2_f \alpha_{\rm EM}x}{Q^2}   \epsilon^{\lambda\ast}_\mu \epsilon^{\lambda'}_\nu \frac{dk^-_1d^2k_{1\perp}}{(2\pi)^3 2k^-_1} \Big|_{k^2_1=0}\frac{dk^-_2d^2k_{2\perp}}{(2\pi)^3 2k^-_2} \Big|_{k^2_2=0}\label{eq:btb-ltwist-pr-ek}\\
&&\times 2\pi\delta(q^--k^-_1-k^-_2) \bar{v}_{s_2}(k_2)
(ig)^2 \Big\{  \Big(\frac{\gamma^\mu\sigma^{-k}}{2k^-_1}
+ \frac{\sigma^{-k} \gamma^\mu}{2k^-_2} \Big) \frac{i}{q^+-k^+_1-k^+_2}
+ \Big(\frac{k_{1k}\gamma^\mu }{k^-_1}  - \frac{k_{2k}\gamma^\mu}{k^-_2} \Big)\nonumber\\
&&\times \frac{1}{(q^+-k^+_1-k^+_2)^2}
\Big\}  t^c  u_{s_1}(k_1) \bar{u}_{s_1}(k_1)
\Big\{ \Big( \frac{\sigma^{-m}\gamma^\nu}{2k^-_1}
+ \frac{\gamma^\nu \sigma^{-m}}{2k^-_2}  \Big) \frac{i}{q^+- k^+_1- k^+_2} 
- \Big( \frac{k_{1m} \gamma^\nu}{k^-_1} \nonumber\\
&&- \frac{k_{2m}\gamma^\nu }{k^-_2} \Big)\frac{1}{(q^+- k^+_1- k^+_2)^2}
 \Big\} t^d v_{s_2}(k_2)
 \int d^2x_\perp e^{-i(q_\perp - k_{1\perp}-k_{2\perp}) x_\perp}
\langle P, S| \int^{\infty}_{-\infty} dz^- F^a_{-k}(z^-,x_\perp)\nonumber\\
&&\times U^{ac}(z^-, \infty;x_\perp)  U^{db}( \infty, 0^-;0_\perp) F^b_{-m}(0^-, 0_\perp)
|P, S\rangle \,,
\nonumber
\end{eqnarray}

Let us compare this result with Eq.~(\ref{eq:btb-ltwist-pr}), derived within the back-to-back power counting. The two expressions are nearly identical. The only difference resides in the matrix element: Eq.~(\ref{eq:btb-ltwist-pr-ek}) contains
\begin{eqnarray}
&&\langle P, S| \int^{\infty}_{-\infty} dz^- F^a_{-k}(z^-,x_\perp) U^{ac}(z^-, \infty;x_\perp)
 U^{db}( \infty, 0^-;0_\perp) F^b_{-m}(0^-, 0_\perp)
|P, S\rangle\,,
\label{eq:full-TMD-no-phase}
\end{eqnarray}
whereas the back-to-back cross section~(\ref{eq:btb-ltwist-pr}) involves
\begin{eqnarray}
&&\langle P, S|
   \int_{-\infty}^{\infty} dz^- e^{i(q^+-k^+_1-k^+_2)z^-}  F^a_{-k}(z^- , x_\perp) U^{ac}(z^-, \infty; x_\perp)
 U^{db}(\infty, 0^-;0_\perp) F^b_{-m}(0^-, 0_\perp) |P, S\rangle\,.
   \label{eq:full-TMD-phase}
\end{eqnarray}
\end{widetext}
The origin of this difference is clear: the exponential factor in Eq.~(\ref{eq:full-TMD-phase}) provides sub-eikonal corrections that were neglected in the eikonal amplitudes~(\ref{eq:amp-gen-eik}) and~(\ref{eq:ampcc-gen-eik}).

Up to this exponential factor, the cross sections~(\ref{eq:btb-ltwist-pr}) and~(\ref{eq:btb-ltwist-pr-ek}) are in full agreement, confirming that the eikonal cross section~(\ref{eq:hadronic-bfm-eik}) is consistent with the leading-power result in the back-to-back limit.


It is important to note that the TMD operator in Eq.~(\ref{eq:btb-ltwist-pr-ek}) includes field-strength tensors with contributions from the transverse background field component, systematically accounted for in our analysis. The transverse component enters in two distinct ways.


First, it appears in the Pauli vertex insertions $\sim \sigma^{-m}F_{-m}$ of the initial eikonal amplitudes~(\ref{eq:amp-gen-eik}) and~(\ref{eq:ampcc-gen-eik}). These operators originate from the chromomagnetic interaction $\sim \sigma^{\mu\nu} F_{\mu\nu}$ in the (anti)quark propagator~(\ref{eq:prop-def}). This contribution enters the operator~(\ref{eq:full-TMD-no-phase}) directly through the $\sim \sigma^{-k}$ terms in Eq.~(\ref{eq:btb-ltwist-pr-ek}), and similarly the operator~(\ref{eq:full-TMD-phase}) in Eq.~(\ref{eq:btb-ltwist-pr}).

Second, the transverse field component contributes to the field-strength tensors of Eq.~(\ref{eq:full-TMD-no-phase}) through the transverse gauge links at past infinity in the initial eikonal amplitudes~(\ref{eq:amp-gen-eik}) and~(\ref{eq:ampcc-gen-eik}). These gauge factors, see Eq.~(\ref{eq:amp-eik-simp-btb}), correspond to the exponential factor~(\ref{eq:exp-fact-cova}). Commuting this factor through the light-cone gauge factors, see Eq.~(\ref{eq:comm-rel}), generates an infinite series of field-strength tensor insertions containing the transverse field component. At leading power in the back-to-back limit, it suffices to retain terms linear in the field-strength tensor, see Eq.~(\ref{eq:simp-singFbtb}). These linear terms from Eq.~(\ref{eq:comm-rel}) contribute to the operator~(\ref{eq:full-TMD-no-phase}) in the final result~(\ref{eq:btb-ltwist-pr-ek}).


We thus find that the transverse background field component provides a non-trivial contribution to the dijet production cross section already at the eikonal order --- one of the main results of this paper. In the CGC framework, the transverse field component is neglected, reducing the background field-strength tensors to transverse derivatives of the light-cone gauge factors, see Eq.~(\ref{eq:comm-WL-b0}). However, our analysis of dijet production demonstrates that this is not always sufficient for describing physical observables at the eikonal order. The transverse background field component can provide an eikonal contribution that must be taken into account, either through an appropriate parametrization of the matrix elements or through computation of the operators within an effective model.


\section{Conclusions}
In this paper, we developed a formalism for computing physical observables within the background field method. In this approach, one begins by deriving a general expression for the physical observable in terms of quark and gluon propagators in the background field. While exact and valid in general kinematics, these expressions are not directly amenable to phenomenological analysis, since the propagators entangle the dynamics of the quantum partons with the interactions with the background fields. In the QCD factorization approach, one needs to disentangle these two elements: the quantum parton dynamics, encoded in perturbative coefficients, and the interactions of the partons with the background fields, giving rise to background field operators that characterize the non-perturbative QCD medium of the target.


This disentanglement is not straightforward, especially when the gauge-covariant form of the background field operators must be derived. Conventionally, it is performed order by order in the coupling constant, leading to a laborious and opaque analysis that often cannot be unambiguously resolved. Our approach provides a complete solution to this problem through a gauge-covariant procedure that allows one to unambiguously derive the full hierarchy of background field operators defining the physical observable.


A key element of our approach is the representation of parton propagators as path-ordered exponents. We developed a procedure for constructing this representation and derived the corresponding equations. At this level, the path-ordered exponent representation is exact and valid in arbitrary kinematics. Similarly to the propagators in the general expression, the path-ordered exponents entangle the parton dynamics and the background field contributions. We developed an efficient method for disentangling these contributions by expanding the fields in the path-ordered exponents onto a fixed contour of appropriate choice in coordinate space.


Our methods allow expansion of the path-ordered exponent representation onto an arbitrary linear piecewise contour. This involves expansion onto the light-cone direction and parallel shifts of the exponents in the transverse direction. The identities derived for these manipulations are completely general, enabling application in different kinematic regimes and approximations.


Once the path-ordered exponents are expanded onto a given contour, the expressions for physical observables assume a factorized form amenable to phenomenological analysis, once the operators obtained in the expansion are parametrized in terms of distribution functions. The methods developed for the path-ordered exponents are explicitly gauge-covariant and allow one to systematically derive the hierarchy of QCD operators defining the physical observable to any required order.


As noted above, our approach allows one to construct the expansion of the path-ordered exponents onto an arbitrary contour, enabling application in different kinematic regimes. In various limits, the physical observables are characterized by different expansion parameters defined by the kinematics of the problem. The choice of expansion contour should be consistent with the expansion parameter, meaning that higher-order terms of the expansion should be suppressed by higher powers of the expansion parameter relative to the leading-order terms. When this condition is met, the terms of the expansion are ordered by construction according to their contribution to the physical observable, and computation of the dominant contribution requires only a finite number of terms.


The expansion of the path-ordered exponents onto a contour is a purely formal procedure --- exact and independent of the kinematic aspects of the scattering process. Different contour expansions are therefore completely equivalent and should be understood as different ways to organize terms in the general expression. This makes our approach an efficient method for matching results computed in different kinematic limits. Within our procedure, one can re-expand the result from one expansion contour to another, enabling quantitative comparison of results across different kinematic limits. This re-expansion procedure effectively provides a dictionary between computations of the same physical observable in different regimes, both in terms of matching the corresponding operator hierarchies and the kinematic coefficients.


Our approach can be applied to various physical observables, provided they are represented in terms of quantum propagators in the background field within the QCD factorization framework. To illustrate the method, we considered the specific case of DIS dijet production. Following our procedure, we first derived a general expression for the dijet production valid in arbitrary kinematics, constructed from the (anti)quark propagators in the background field. Using our framework, we rewrote the propagators in their path-ordered exponent representation --- an essential step enabling computation of the dijet production in different kinematic limits. In this paper, we considered two kinematic limits.


We began with the back-to-back limit of dijet production, in which the ratio of the transverse momentum imbalance between the jets to a typical jet momentum is small, $\Delta_\perp/ \bar{P}_\perp \to 0$. This ratio serves as the expansion parameter. We demonstrated that the expansion contour for the path-ordered exponents corresponds to the operator contour of the TMD operators, and showed how our procedure constructs the expansion onto this contour in an explicitly gauge-covariant manner. We explained how to reconstruct the full hierarchy of gluon TMD operators defining dijet production in the back-to-back limit to any required order. To illustrate, we explicitly derived the dijet production cross section at leading power. Our final expressions agree with known results.


We then considered a different kinematic regime: high-energy dijet production at small $x$. Going beyond the standard CGC framework, we analyzed the dijet production within the high-energy power counting~\eqref{eq:pw-couting-sx}. The boost parameter $\lambda$ in this power counting serves as the expansion parameter, $\lambda^{-1}\to0$. In the context of small-$x$ computations, this power counting has been previously applied to sub-eikonal corrections in high-energy scattering and spin effects at small $x$, see e.g. Refs. \cite{Balitsky:2015qba,Chirilli:2018kkw,Altinoluk:2021lvu,Cougoulic:2022gbk}. We found that within the high-energy power counting~(\ref{eq:pw-couting-sx}), the appropriate expansion contour is a staple-like contour. Expanding the background fields of the propagators onto this contour, we found that at the eikonal order the background field operators consist of gauge links along the expansion contour with Pauli vertex insertions of the field-strength tensors $F_{-i}$.


In our analysis, we fully reconstructed the contribution of the transverse background field component $B_i$. We found that this component provides a non-trivial contribution already at the eikonal order --- one of the main results of this paper. Within the high-energy power counting~(\ref{eq:pw-couting-sx}), the transverse component is suppressed, $B_i\ll B_- \sim \lambda$; however, its longitudinal derivative is enhanced by the Lorentz boost of the background fields, $\partial_-B_i\sim \partial_i B_- \sim \lambda$. As a result, the transverse component generates an eikonal contribution to the field-strength tensor $F_{-i}$ comparable to the standard contribution of the longitudinal component $\partial_i B_-$.


Furthermore, the transverse background field component contributes to the transverse gauge links in the operators constructed through our expansion. While the presence of transverse gauge links is always implicitly assumed, systematic methods for analyzing their contribution are scarce; very few examples are available in the literature, see e.g.\ Ref.~\cite{Belitsky:2002sm}. In this paper, we performed a systematic derivation of the transverse gauge link contribution. We found this contribution --- and the contribution of the transverse background field component in general --- to be non-trivial at the eikonal order, and argued that it must be taken into account either through an appropriate parametrization of the operator matrix elements or through effective model computations.


When the transverse background field component is set to zero, $B_i = 0$, our results for the dijet production cross section at the eikonal order reduce to the corresponding CGC results.


Using formal operator transformations, we also found that the choice of operator basis for the dijet production is not unique. At the eikonal order, one can attribute the transverse background field contribution entirely to the transverse gauge links, or alternatively employ an operator basis with field-strength tensor insertions into the light-cone gauge links. Both choices are completely equivalent, with no physical distinction between them. Regardless of how the transverse field component appears at the operator level, it physically corresponds to the interaction of the quantum (anti)quark with the transverse background field of the target at a finite coordinate-space position. Within the high-energy power counting~(\ref{eq:pw-couting-sx}), this interaction is non-trivial already at the eikonal order.


Our approach provides an efficient method for matching results computed in different kinematic limits, making it particularly instructive for analyzing interplay effects between large and small $x$. To illustrate the matching procedure, we compared our results for dijet production in the back-to-back and high-energy limits. Starting with the dijet production cross section within the high-energy power counting~(\ref{eq:pw-couting-sx}) at the eikonal order, we re-expanded the background field operators onto the TMD operator contour defining the back-to-back expansion. This re-expansion generates an infinite series of terms corresponding to different powers of the back-to-back expansion parameter $\Delta_\perp/ \bar{P}_\perp \to 0$. The expansion of the eikonal result onto the TMD contour is consistent with this parameter. We explained how the matching can be performed to any required order and provided an explicit result by matching the eikonal result with the leading-power contribution in the back-to-back limit. For this purpose, only the leading term of the re-expansion series is needed. At this order, we found complete agreement between the eikonal and back-to-back results for the dijet production.


Since our formalism is explicitly gauge-covariant, derivation of the gauge-covariant operators defining the scattering is straightforward and can be carried out to any given order. We outlined how to construct the hierarchy of TMD operators defining the scattering in the back-to-back limit. Similarly, our analysis can be extended to the sub-eikonal order in the high-energy limit.


In this regard, we find our approach advantageous compared to currently available computation schemes. Existing methods typically rely on the shock-wave approximation with rigid boundaries of the background potential $B_\mu$, distinguishing quantum emission before, after, and inside these boundaries. We find this step to be rather artificial and leading to complications in deriving the operator hierarchy at the sub-eikonal order. Our approach instead assumes a general form of the background field without rigid boundaries, making the derivation of sub-eikonal corrections more direct. We plan to present this derivation in a separate publication.


The application of our approach extends beyond the dijet production cross section. It can be applied to other high-energy scattering observables within the QCD factorization framework. Once a given observable is represented in terms of quantum propagators in the background fields, the analysis largely follows the procedure presented here.


As an outlook, we note that our analysis focused primarily on the (anti)quark propagators in the background field. For some observables, a similar computation of gluon propagators is necessary. This extension is straightforward, since the gluon propagator in the background-Feynman gauge has a similar structure to the quark propagators considered here, and our methods generalize readily.


We also plan to extend our formalism to include the quark background field. While this will require some additional technical steps, see e.g.\ Ref.~\cite{Mukherjee:2025aiw}, the general procedure remains unchanged.


Although we limited our analysis to the leading order, we expect that our methods can be extended to quantum loop computations with only a few technical modifications related to operator manipulations within the loops. The general procedure at loop order follows the framework presented here. Once loop diagrams for the quantum partons are expressed in terms of propagators in the background fields, the expansion techniques developed in this paper can be applied. After the background fields are expanded and the operators constructed, the expressions assume a factorized form, and the computation is completed by deriving the perturbative coefficients through loop integration. We plan to present such a computation for the dijet production in a separate paper.


While we expect the extension to gluon contributions and quantum loops to be straightforward, several aspects of our approach require further investigation. One concerns the dependence on the form of the background field. We restricted our analysis to $B_\mu(x) = B_{\mu}(x^-, x_\perp)$, independent of the $x^+$ coordinate. While this dependence appears to be irrelevant at the leading order of the expansion, it could potentially introduce operators $\sim F_{+-}$ at higher orders. We find the systematic inclusion of the $x^+$ dependence to be quite involved and leave it for a future publication.


Finally, it would be interesting to extend our approach to the case of multiple background fields, as is generally necessary for computing observables involving different participating hadrons and deriving the corresponding factorization theorems. Available examples of such computations in the background field method suggest that a general formalism can be constructed. Nevertheless, such a derivation is by no means trivial, and we leave it for future research.


\section{Acknowledgments}
We are grateful to Ian Balitsky, Swagato Mukherjee, Shaswat Tiwari, and Fei Yao for inspiring discussions. This work is supported by the U.S. Department of Energy, Office of Science, Office of Nuclear Physics through Contract No. DE-SC0020081, and within the framework of Saturated Glue (SURGE) Topical Collaboration in Nuclear Theory. 
We also acknowledge the United States-Israel Binational Science Foundation grant 2022132.

A.T. thanks the Aspen Center for Physics, which is supported by National Science Foundation grant PHY-2210452, where part of this work was performed.

\appendix



\begin{widetext}
\section{Path-ordered exponent representation of the scalar propagator\label{app:tr-prop}}
Let us consider the propagator (\ref{eq:transv-prop}) in more details. We can formally expand it as
\begin{eqnarray}
&&(x| \frac{1}{p^+  - \frac{P^2_\perp}{2p^-} + i\epsilon p^-}|y)
\label{eq:sc-prop-tr-exp}\\
&&= (x| \frac{1}{p^+ + i\epsilon p^-} + \frac{1}{p^+ + i\epsilon p^-}\frac{P^2_\perp}{2p^-} \frac{1}{p^+ + i\epsilon p^-} + \frac{1}{p^+ + i\epsilon p^-}\frac{P^2_\perp}{2p^-} \frac{1}{p^+ + i\epsilon p^-} \frac{P^2_\perp}{2p^-} \frac{1}{p^+ + i\epsilon p^-} + \dots |y)\,,
\nonumber
\end{eqnarray}
which represents an infinite sum of $P^2_\perp$ insertions connected with a free propagator
\begin{eqnarray}
&&(x| \frac{1}{p^+ + i\epsilon p^-}|y)\label{eq:scprop-free}\\
&&= \Big( \frac{-i}{2\pi} \theta(x^- - y^-) \int^\infty_0 dp^- + \frac{i}{2\pi} \theta(y^- - x^-) \int^0_{-\infty} dp^- \Big) e^{-ip^- (x^+ - y^+)}(2\pi)^2\delta^2(x_\perp-y_\perp)\,.\nonumber
\end{eqnarray}

Introducing intermediate coordinates and using Eq. (\ref{eq:scprop-free}), one can easily rewrite Eq. (\ref{eq:sc-prop-tr-exp}) as
\begin{eqnarray}
&&(x| \frac{1}{p^+  - \frac{P^2_\perp}{2p^-} + i\epsilon p^-}|y)
= \Big( \frac{-i}{2\pi} \theta(x^- - y^-) \int^\infty_0 dp^- + \frac{i}{2\pi} \theta(y^- - x^-) \int^0_{-\infty} dp^- \Big) e^{-ip^- (x^+ - y^+)} 
\nonumber\\
&&\times (x_\perp| \Big\{1 + \int^{x^-}_{y^-} dz^- \frac{-iP^2_\perp(z^-)}{2p^-} + \int^{x^-}_{y^-} dz^-_1 \frac{-i P^2_\perp(z^-_1)}{2p^-} \int^{z^-_1}_{y^-} dz^-_2 \frac{-iP^2_\perp(z^-_2)}{2p^-}
+ \dots\Big\} |y_\perp)\,,
\label{eq:scal-prop-pp-only}
\end{eqnarray}
where $P^2_\perp(z^-)\equiv p^2_\perp + \{p_k, A_k(z^-)\} + A_k(z^-) A_k(z^-)$.
\end{widetext}

Eq. (\ref{eq:scal-prop-pp-only}) contains an infinite sum of $P^2_\perp(z^-)$ insertions ordered in the longitudinal direction. To understand the structure of this result, let us group the terms based on the corresponding power of $1/p^-$. In the following derivation we will explicitly work out terms up to the $1/(p^-)^2$ order, which is sufficient to see the general pattern. Note, that our procedure can be applied to terms of any power of $1/p^-$, which leads to the same results.\footnote{We explicitly checked this up to the $1/(p^-)^3$ order.}

Our overall strategy is to resum the infinite number of terms in Eq. (\ref{eq:scal-prop-pp-only}) by exponentiating $P^2_\perp$ insertions. The philosophy behind this approach is that $P^2_\perp$ insertions describe transition of the scalar quark in the transverse direction. Indeed, without these insertions the scalar quark does not propagate in the transverse space which is manifested by a $\delta$-function in Eq. (\ref{eq:scprop-free}), c.f. Eq. (\ref{eq:sc-prop-tr-exp}). As we will find, exponentiation of these insertions allows to efficiently describe these transitions and, in particular, introduce the associated transverse Wilson links.

To observe exponentiation of the $P^2_\perp$ component, let's first rewrite Eq. (\ref{eq:scal-prop-pp-only}) in a way that each power of $1/p^-$ is compensated with the same power of $z^-$ in the numerator. Specifically, in the first two terms of Eq. (\ref{eq:scal-prop-pp-only}) this can be achieved by inserting $1 = dz^-/dz^-$:
\begin{eqnarray}
&&1 + \int^{x^-}_{y^-} dz^- \frac{-iP^2_\perp(z^-)}{2p^-} \nonumber\\
&&= 1 + \int^{x^-}_{y^-} dz^- \Big(\frac{dz^-}{dz^-}\Big)\frac{-iP^2_\perp(z^-)}{2p^-}\,,
\end{eqnarray}
and integrate by parts with respect to $z^-$ in the second term, which yields
\begin{eqnarray}
&&1 + \int^{x^-}_{y^-} dz^- \frac{-iP^2_\perp(z^-)}{2p^-} = 1 + \frac{-ix^-P^2_\perp(x^-)}{2p^-}\nonumber\\
&& + \frac{iy^-P^2_\perp(y^-)}{2p^-} - \int^{x^-}_{y^-} dz^- z^-\frac{-i\partial_{z^-} P^2_\perp(z^-)}{2p^-}\,.
\label{eq:int-by-parts-proc}
\end{eqnarray}

A similar procedure can be applied to all terms in Eq. (\ref{eq:scal-prop-pp-only}),\footnote{In the higher order terms in $1/p^-$, instead of $1 = \frac{d}{dz^-}z^-$ one has to use $z^- = \frac{1}{2}\frac{d}{dz^-}(z^-)^2$ etc.} which yields
\begin{widetext}
\begin{eqnarray}
&&(x| \frac{1}{p^+  - \frac{P^2_\perp}{2p^-} + i\epsilon p^-}|y)\nonumber\\
&&= \Big( \frac{-i}{2\pi} \theta(x^- - y^-) \int^\infty_0 dp^- + \frac{i}{2\pi} \theta(y^- - x^-) \int^0_{-\infty} dp^- \Big) e^{-ip^- (x^+ - y^+)} 
\nonumber\\
&&\times (x_\perp| \Big\{1 + \frac{-i x^- P^2_\perp(x^-)}{2p^-} + \frac{1}{2} \frac{-i x^- P^2_\perp(x^-)}{2p^-} \frac{-i x^- P^2_\perp(x^-)}{2p^-} + \frac{iy^-P^2_\perp(y^-)}{2p^-}\nonumber\\
&&+ \frac{1}{2} \frac{i y^- P^2_\perp(y^-)}{2p^-} \frac{i y^- P^2_\perp(y^-)}{2p^-}
+ \frac{-i x^- P^2_\perp(x^-)}{2p^-}  \frac{iy^-P^2_\perp(y^-)}{2p^-} - \int^{x^-}_{y^-} dz^- \Big( z^- \frac{-i\partial_{z^-}P^2_\perp(z^-)}{2p^-}\Big)\nonumber\\
&&- \frac{-i x^- P^2_\perp(x^-)}{2p^-} \int^{x^-}_{y^-} dz^- \Big( z^- \frac{-i\partial_{z^-}P^2_\perp(z^-)}{2p^-} \Big)
- \int^{x^-}_{y^-} dz^- \Big(z^- \frac{-i \partial_{z^- }P^2_\perp(z^-)}{2p^-}\Big)  \frac{iy^-P^2_\perp(y^-)}{2p^-}\nonumber\\
&&- \frac{1}{2}\int^{x^-}_{y^-} dz^- \Big( z^- \frac{-i \partial_{z^-} P^2_\perp(z^-)}{2p^-} \Big) \frac{-i z^- P^2_\perp(z^-)}{2p^-}
+ \frac{1}{2}\int^{x^-}_{y^-} dz^- \frac{-i z^- P^2_\perp(z^-)}{2p^-} \Big(z^- \frac{-i\partial_{z^-} P^2_\perp(z^-)}{2p^-}\Big)\nonumber\\
&&+ \int^{x^-}_{y^-} dz^-_1 \Big(z^-_1 \frac{-i \partial_{z^-_1}P^2_\perp(z^-_1)}{2p^-}\Big) \int^{z^-_1}_{y^-} dz^-_2 \Big( z^-_2 \frac{-i\partial_{z^-_2}P^2_\perp(z^-_2)}{2p^-} \Big) 
+ \mathcal{O}\Big(\frac{1}{(p^-)^3}\Big)\Big\} |y_\perp)
\label{eq:scal-part-before-exp}
\end{eqnarray}

Eq. (\ref{eq:scal-part-before-exp}) makes one pattern immediately obvious: the boundary terms\footnote{These terms appear after the integration by parts with respect to $z^-$ in Eq. (\ref{eq:int-by-parts-proc}) etc.} proportional to $-i x^- P^2_\perp(x^-)/2p^-$ and  $i y^- P^2_\perp(y^-)/2p^-$ can be exponentiated, such that the equation takes the following form
\begin{eqnarray}
&&(x| \frac{1}{p^+  - \frac{P^2_\perp}{2p^-} + i\epsilon p^-}|y)\nonumber\\
&&= \Big( \frac{-i}{2\pi} \theta(x^- - y^-) \int^\infty_0 dp^- + \frac{i}{2\pi} \theta(y^- - x^-) \int^0_{-\infty} dp^- \Big) e^{-ip^- (x^+ - y^+)} 
\nonumber\\
&&\times (x_\perp| e^{-i\frac{ P^2_\perp(x^-)}{2p^-}x^- }\Big\{1 - \int^{x^-}_{y^-} dz^- \Big( z^- \frac{-i\partial_{z^-}P^2_\perp(z^-)}{2p^-}\Big) - \frac{1}{2}\int^{x^-}_{y^-} dz^- \Big( z^- \frac{-i \partial_{z^-} P^2_\perp(z^-)}{2p^-} \Big)\nonumber\\
&&\times \frac{-i z^- P^2_\perp(z^-)}{2p^-}
+ \frac{1}{2}\int^{x^-}_{y^-} dz^- \frac{-i z^- P^2_\perp(z^-)}{2p^-} \Big(z^- \frac{-i\partial_{z^-} P^2_\perp(z^-)}{2p^-}\Big)
 + \int^{x^-}_{y^-} dz^-_1 \Big(z^-_1 \frac{-i \partial_{z^-_1}P^2_\perp(z^-_1)}{2p^-}\Big)\nonumber\\
&&\times \int^{z^-_1}_{y^-} dz^-_2 \Big( z^-_2 \frac{-i\partial_{z^-_2}P^2_\perp(z^-_2)}{2p^-} \Big) 
+ \mathcal{O}\Big(\frac{1}{(p^-)^3}\Big) \Big\} e^{i\frac{P^2_\perp(y^-)}{2p^-}y^-} |y_\perp)\,.
\label{eq:scal-transv-with-der}
\end{eqnarray}

It is easy to check that terms in the curly brackets can be further exponentiated as
\begin{eqnarray}
&&(x| \frac{1}{p^+  - \frac{P^2_\perp}{2p^-} + i\epsilon p^-}|y)\label{eq:tr-prop-ins}\\
&&= \Big( \frac{-i}{2\pi} \theta(x^- - y^-) \int^\infty_0 dp^- + \frac{i}{2\pi} \theta(y^- - x^-) \int^0_{-\infty} dp^- \Big) e^{-ip^- (x^+ - y^+)} 
\nonumber\\
&&\times (x_\perp| e^{-i\frac{ P^2_\perp(x^-)}{2p^-}x^- }\Big\{1 + i\int^{x^-}_{y^-} dz^- \Big(e^{i\frac{ P^2_\perp(z^-)}{2p^-}z^-}i\partial_- e^{-i\frac{ P^2_\perp(z^-)}{2p^-}z^-} - \frac{P^2_\perp(z^-)}{2p^-}\Big)\nonumber\\
&&+ (i)^2\int^{x^-}_{y^-} dz^-_1 \int^{z^-_1}_{y^-} dz^-_2
\Big(e^{i\frac{ P^2_\perp(z^-_1)}{2p^-}z^-_1}i\partial_- e^{-i\frac{ P^2_\perp(z^-_1)}{2p^-}z^-_1} - \frac{P^2_\perp(z^-_1)}{2p^-}\Big) \Big(e^{i\frac{ P^2_\perp(z^-_2)}{2p^-}z^-_2}i\partial_- e^{-i\frac{ P^2_\perp(z^-_2)}{2p^-}z^-_2}\nonumber\\
&&- \frac{P^2_\perp(z^-_2)}{2p^-}\Big) + \dots \Big\} e^{i\frac{P^2_\perp(y^-)}{2p^-}y^-} |y_\perp)\,,
 \nonumber
\end{eqnarray}
where ellipsis stands for terms of the higher powers of the
\begin{eqnarray}
&&i\int^{x^-}_{y^-} dz^- \Big(e^{i\frac{ P^2_\perp(z^-)}{2p^-}z^-}i\partial_- e^{-i\frac{ P^2_\perp(z^-)}{2p^-}z^-} - \frac{P^2_\perp(z^-)}{2p^-}\Big)
\label{eq-general-ins}
\end{eqnarray}
insertions.

Since all insertions (\ref{eq-general-ins}) in Eq. (\ref{eq:tr-prop-ins}) are ordered along a path connecting $x^-$ and $y^-$, Eq. (\ref{eq:tr-prop-ins}) can be written as a path-ordered exponent:  
\begin{eqnarray}
&&(x| \frac{1}{p^+  - \frac{P^2_\perp}{2p^-} + i\epsilon p^-}|y)\label{eq:tr-prop-Pexp}\\
&&= \Big( \frac{-i}{2\pi} \theta(x^- - y^-) \int^\infty_0 dp^- + \frac{i}{2\pi} \theta(y^- - x^-) \int^0_{-\infty} dp^- \Big) e^{-ip^- (x^+ - y^+)}\nonumber 
\\
&&\times (x_\perp| e^{-i\frac{ P^2_\perp(x^-)}{2p^-}x^- }\mathcal{P}\exp\Big\{ i\int^{x^-}_{y^-} dz^- \Big(e^{i\frac{ P^2_\perp(z^-)}{2p^-}z^-}i\partial_- e^{-i\frac{ P^2_\perp(z^-)}{2p^-}z^-} - \frac{P^2_\perp(z^-)}{2p^-}\Big) \Big\} e^{i\frac{P^2_\perp(y^-)}{2p^-}y^-} |y_\perp)\,.
 \nonumber
\end{eqnarray}
Note that this form of the propagator (\ref{eq:transv-prop}) is exact. It can be checked that in any given order of the $1/p^-$ expansion Eq. (\ref{eq:tr-prop-Pexp}) reproduces corresponding terms of a general result (\ref{eq:scal-prop-pp-only}).

The path-ordered exponent in Eq. (\ref{eq:tr-prop-Pexp}) has one important property, which makes this form of the propagator especially suitable for our analysis of the structure of QCD operators appearing in the dijet production. For an arbitrary shift $\xi^-$, the path-ordered exponent satisfies
\begin{eqnarray}
&&e^{-i\frac{ P^2_\perp(x^-)}{2p^-}x^- }\mathcal{P}\exp\Big\{ i\int^{x^-}_{y^-} dz^- \Big(e^{i\frac{ P^2_\perp(z^-)}{2p^-}z^-}i\partial_- e^{-i\frac{ P^2_\perp(z^-)}{2p^-}z^-} - \frac{P^2_\perp(z^-)}{2p^-}\Big) \Big\} e^{i\frac{P^2_\perp(y^-)}{2p^-}y^-}
\label{eq:parallel-shift-scalarmin}\\
&&= e^{-i\frac{ P^2_\perp(x^-)}{2p^-}(x^--\xi^-) }\mathcal{P}\exp\Big\{ i\int^{x^-}_{y^-} dz^- \Big(e^{i\frac{ P^2_\perp(z^-)}{2p^-}(z^--\xi^-)}i\partial_- e^{-i\frac{ P^2_\perp(z^-)}{2p^-}(z^--\xi^-)} - \frac{P^2_\perp(z^-)}{2p^-}\Big) \Big\} \nonumber\\
&&\times e^{i\frac{P^2_\perp(y^-)}{2p^-}(y^--\xi^-)}\,,
\nonumber
\end{eqnarray}
which can be easily checked by expanding both sides of the equation in powers of $1/p^-$ and comparing the corresponding terms. As we will discuss later, this result will allow us to modify the path of Wilson lines appearing in the dijet production, and project the result of our computation to a set of the quadrupole operators, or TMD operators in the back-to-back limit.

\section{Computation of the scalar propagator in the mixed representation\label{qpp-LSZ}}
In this section, we will present derivation of Eq. (\ref{eq:LSZ-fin}) starting with the form of the scalar propagator in the background field (\ref{eq:scalar-exp-Amin-exp-form-full-S}), which can be rewritten as
\begin{eqnarray}
&&\lim_{k^2\to0}k^2(k| \frac{1}{P^2 + i\epsilon}|y)\\
&&= \lim_{k^2\to0} \int d^4z e^{ikz} (2i k^- \frac{\partial}{\partial z^-}  - k^2_\perp) \frac{1}{2k^-} \Big( \frac{-i}{2\pi} \theta(z^- - y^-) \int^\infty_0 dp^-
\nonumber\\
&& + \frac{i}{2\pi} \theta(y^- - z^-) \int^0_{-\infty} dp^- \Big) e^{-ip^- (z^+ - y^+)} (z_\perp| e^{-i\frac{ P^2_\perp(z^-)}{2p^-}z^- } \mathcal{S}(z^-, y^-) e^{i\frac{P^2_\perp(y^-)}{2p^-}y^-}|y_\perp)\,.
\nonumber
\end{eqnarray}

Integrating over $z^+$ variable and $p^-$, it is straightforward to obtain
\begin{eqnarray}
&&\lim_{k^2\to0}k^2(k| \frac{1}{P^2 + i\epsilon}|y)\label{eqApa:basic-form}\\
&&= \lim_{k^2\to0} \int dz^- e^{i\frac{k^2_\perp}{2k^-}z^-} (i \frac{\partial}{\partial z^-}  - \frac{k^2_\perp}{2k^-})   \Big( -i \theta (z^--y^-)\theta(k^-) + i \theta (y^- - z^-)\theta(-k^-) \Big)
\nonumber\\
&&\times (k_\perp|  e^{-i\frac{P^2_\perp(z^-)}{2k^-} z^-} \mathcal{S}(z^-, y^-) e^{i\frac{P^2_\perp(y^-)}{2k^-}y^-} |y_\perp) e^{ik^- y^+}\,.
\nonumber
\end{eqnarray}

It is easy to see that in this equation the derivative with respect to $z^-$ acting on the $\theta$-functions yields a trivial contribution $e^{iky}|_{k^2 = 0}$, which can be explicitly written as
\begin{eqnarray}
&&\lim_{k^2\to0}k^2(k| \frac{1}{P^2 + i\epsilon}|y)\\
&&= e^{iky}\Big|_{k^2 = 0}
+ \lim_{k^2\to0} \Big( -i \theta(k^-) \int^\infty_{y^-} dz^-  + i \theta(-k^-) \int^{y^-}_{-\infty} dz^-\Big) e^{i\frac{k^2_\perp}{2k^-}z^-}
\nonumber\\
&&\times (i \frac{\partial}{\partial z^-}  - \frac{k^2_\perp}{2k^-}) (k_\perp|  e^{-i\frac{P^2_\perp(z^-)}{2k^-} z^-} \mathcal{S}(z^-, y^-) e^{i\frac{P^2_\perp(y^-)}{2k^-}y^-} |y_\perp) e^{ik^- y^+}\,.\nonumber
\end{eqnarray}

Differentiating further, we can write this result as
\begin{eqnarray}
&&\lim_{k^2\to0}k^2(k| \frac{1}{P^2 + i\epsilon}|y)\label{eqApa:after-dif}\\
&&= e^{iky}\Big|_{k^2 = 0} + \lim_{k^2\to0} \Big( -i \theta(k^-) \int^\infty_{y^-} dz^-  + i \theta(-k^-) \int^{y^-}_{-\infty} dz^-\Big) e^{i\frac{k^2_\perp}{2k^-}z^-} (k_\perp|
\nonumber\\
&&\times   \Big[ (i \frac{\partial}{\partial z^-}  - \frac{k^2_\perp}{2k^-}) e^{-i\frac{P^2_\perp(z^-)}{2k^-} z^-}\Big]  \mathcal{S}(z^-, y^-) e^{i\frac{P^2_\perp(y^-)}{2k^-}y^-} |y_\perp) e^{ik^- y^+} + \lim_{k^2\to0} (k_\perp| \Big( i \theta(k^-) \int^\infty_{y^-} dz^-\nonumber\\
&&+ i \theta(-k^-) \int_{y^-}^{-\infty} dz^-\Big)
\Big(e^{i\frac{k^2_\perp}{2k^-}z^-} iD_- e^{-i\frac{ P^2_\perp(z^-)}{2k^-}z^-} - e^{i\frac{k^2_\perp}{2k^-}z^-} e^{-i\frac{P^2_\perp(z^-)}{2k^-} z^-} \frac{P^2_\perp(z^-)}{2k^-}\Big) \mathcal{S}(z^-, y^-)\nonumber\\
&&\times e^{i\frac{P^2_\perp(y^-)}{2k^-}y^-} |y_\perp)e^{ik^- y^+}\,,
\nonumber
\end{eqnarray}
where in the second term the derivative acts only on the exponential factor, and in the last two terms we take into account:
\begin{eqnarray}
&&-i\frac{\partial}{\partial z^-}\mathcal{S}(z^-, y^-)= \Big(e^{i\frac{ P^2_\perp(z^-)}{2p^-}z^-}iD_- e^{-i\frac{ P^2_\perp(z^-)}{2p^-}z^-} - \frac{P^2_\perp(z^-)}{2p^-}\Big) \mathcal{S}(z^-, y^-)\,.
\end{eqnarray}

Now, let us consider the second term of Eq. (\ref{eqApa:after-dif}). Taking into account
\begin{eqnarray}
&&(i \frac{\partial}{\partial z^-}  - \frac{k^2_\perp}{2k^-}) e^{-i\frac{P^2_\perp(z^-)}{2k^-} z^-} = i e^{-i\frac{k^2_\perp}{2k^-} z^-} \Big(\frac{\partial}{\partial z^-}e^{i\frac{ k^2_\perp}{2k^-} z^-} e^{-i\frac{P^2_\perp(z^-) }{2k^-} z^-} \Big)
\end{eqnarray}
and expanding the path-ordered exponent $\mathcal{S}(z^-, y^-)$, we write 
\begin{eqnarray}
&&\lim_{k^2\to0} \Big( -i \theta(k^-) \int^\infty_{y^-} dz^-  + i \theta(-k^-) \int^{y^-}_{-\infty} dz^-\Big) e^{i\frac{k^2_\perp}{2k^-}z^-} (k_\perp| \Big[ (i \frac{\partial}{\partial z^-}  - \frac{k^2_\perp}{2k^-}) e^{-i\frac{P^2_\perp(z^-)}{2k^-} z^-}\Big]\nonumber\\
&&\times \mathcal{S}(z^-, y^-) 
e^{i\frac{P^2_\perp(y^-)}{2k^-}y^-} |y_\perp)  e^{ik^- y^+}\\
&&=  \lim_{k^2\to0} \Big(  \theta(k^-) \int^\infty_{y^-} dz^-  - \theta(-k^-) \int^{y^-}_{-\infty} dz^-\Big) (k_\perp| \Big(\frac{\partial}{\partial z^-}e^{i\frac{ k^2_\perp}{2k^-} z^-} e^{-i\frac{P^2_\perp(z^-) }{2k^-} z^-} \Big) 
 \nonumber\\
&&\times \Big(1 + i\int^{z^-}_{y^-} dz^-_1 \Big(e^{i\frac{ P^2_\perp(z^-_1)}{2p^-}z^-_1}iD_- e^{-i\frac{ P^2_\perp(z^-_1)}{2p^-}z^-_1} - \frac{P^2_\perp(z^-_1)}{2p^-}\Big) + \dots\Big) e^{i\frac{P^2_\perp(y^-)}{2k^-}y^-} |y_\perp) e^{ik^- y^+}\,,\nonumber
\end{eqnarray}
where ellipsis stands for the higher order terms of the expansion.\footnote{While we only explicitly show computation of the first two terms of the expansion, computation of the higher order terms is identical and can be performed without any loss of generality.} Separating the first term of the expansion and changing the limits of integration in the higher order terms we get
\begin{eqnarray}
&&\lim_{k^2\to0} \Big( -i \theta(k^-) \int^\infty_{y^-} dz^-  + i \theta(-k^-) \int^{y^-}_{-\infty} dz^-\Big) e^{i\frac{k^2_\perp}{2k^-}z^-} (k_\perp| \Big[ (i \frac{\partial}{\partial z^-}  - \frac{k^2_\perp}{2k^-}) e^{-i\frac{P^2_\perp(z^-)}{2k^-} z^-}\Big]\nonumber\\
&&\times \mathcal{S}(z^-, y^-) e^{i\frac{P^2_\perp(y^-)}{2k^-}y^-} |y_\perp)
e^{ik^- y^+} \\
&&= \lim_{k^2\to0} (k_\perp| \Big(  \theta(k^-) \int^\infty_{y^-} dz^- \Big(\frac{\partial}{\partial z^-}e^{i\frac{ k^2_\perp}{2k^-} z^-} e^{-i\frac{P^2_\perp(z^-) }{2k^-} z^-} \Big)  + \theta(-k^-) \int_{y^-}^{-\infty} dz^- \nonumber\\
&&\times \Big(\frac{\partial}{\partial z^-}e^{i\frac{ k^2_\perp}{2k^-} z^-} e^{-i\frac{P^2_\perp(z^-) }{2k^-} z^-} \Big) \Big)
e^{i\frac{P^2_\perp(y^-)}{2k^-}y^-} |y_\perp) e^{ik^- y^+} + \lim_{k^2\to0} (k_\perp| \Big\{ \Big(  i\theta(k^-) \int^{\infty}_{y^-} dz^-_1 \int^\infty_{z^-_1} dz^-\nonumber\\
&&\times \Big(\frac{\partial}{\partial z^-}e^{i\frac{ k^2_\perp}{2k^-} z^-} e^{-i\frac{P^2_\perp(z^-) }{2k^-} z^-} \Big) + i \theta(-k^-) \int^{-\infty}_{y^-} dz^-_1
\int_{z^-_1}^{-\infty} dz^- \Big(\frac{\partial}{\partial z^-}e^{i\frac{ k^2_\perp}{2k^-} z^-} e^{-i\frac{P^2_\perp(z^-) }{2k^-} z^-} \Big) \Big)\nonumber\\
&&\times\Big(e^{i\frac{ P^2_\perp(z^-_1)}{2p^-}z^-_1}iD_- e^{-i\frac{ P^2_\perp(z^-_1)}{2p^-}z^-_1} - \frac{P^2_\perp(z^-_1)}{2p^-}\Big) + \dots\Big\}  e^{i\frac{P^2_\perp(y^-)}{2k^-}y^-} |y_\perp) e^{ik^- y^+}\,.\nonumber
\end{eqnarray}

Integrating over $z^-$ variable, we obtain
\begin{eqnarray}
&&\lim_{k^2\to0} \Big( -i \theta(k^-) \int^\infty_{y^-} dz^-  + i \theta(-k^-) \int^{y^-}_{-\infty} dz^-\Big) e^{i\frac{k^2_\perp}{2k^-}z^-} (k_\perp| \Big[ (i \frac{\partial}{\partial z^-}  - \frac{k^2_\perp}{2k^-}) e^{-i\frac{P^2_\perp(z^-)}{2k^-} z^-}\Big]\nonumber\\
&&\times\mathcal{S}(z^-, y^-) e^{i\frac{P^2_\perp(y^-)}{2k^-}y^-} |y_\perp)
e^{ik^- y^+}\\
&&= \lim_{k^2\to0} (k_\perp| \Big(  \theta(k^-) e^{i\frac{ k^2_\perp}{2k^-} z^-} e^{-i\frac{P^2_\perp(z^-) }{2k^-} z^-} \Big|^\infty_{y^-}  + \theta(-k^-) e^{i\frac{ k^2_\perp}{2k^-} z^-} e^{-i\frac{P^2_\perp(z^-) }{2k^-} z^-} \Big|_{y^-}^{-\infty} \Big)\nonumber\\
&&\times e^{i\frac{P^2_\perp(y^-)}{2k^-}y^-} |y_\perp) e^{ik^- y^+}
+ \lim_{k^2\to0} (k_\perp| \Big\{ \Big(  i\theta(k^-) \int^{\infty}_{y^-} dz^-_1 e^{i\frac{ k^2_\perp}{2k^-} z^-} e^{-i\frac{P^2_\perp(z^-) }{2k^-} z^-} \Big|^\infty_{z^-_1}\nonumber\\
&&+ i \theta(-k^-) \int^{-\infty}_{y^-} dz^-_1 e^{i\frac{ k^2_\perp}{2k^-} z^-} e^{-i\frac{P^2_\perp(z^-) }{2k^-} z^-} \Big|_{z^-_1}^{-\infty} \Big)
\Big(e^{i\frac{ P^2_\perp(z^-_1)}{2p^-}z^-_1}iD_- e^{-i\frac{ P^2_\perp(z^-_1)}{2p^-}z^-_1} - \frac{P^2_\perp(z^-_1)}{2p^-}\Big) + \dots\Big\}\nonumber\\
&&\times e^{i\frac{P^2_\perp(y^-)}{2k^-}y^-} |y_\perp) e^{ik^- y^+}\,,\nonumber
\end{eqnarray}
which can be rewritten as
\begin{eqnarray}
&&\lim_{k^2\to0} \Big( -i \theta(k^-) \int^\infty_{y^-} dz^-  + i \theta(-k^-) \int^{y^-}_{-\infty} dz^-\Big) e^{i\frac{k^2_\perp}{2k^-}z^-} (k_\perp| \Big[ (i \frac{\partial}{\partial z^-}  - \frac{k^2_\perp}{2k^-}) e^{-i\frac{P^2_\perp(z^-)}{2k^-} z^-}\Big]\nonumber\\
&&\times \mathcal{S}(z^-, y^-) e^{i\frac{P^2_\perp(y^-)}{2k^-}y^-} |y_\perp) 
e^{ik^- y^+}\\
&&= \lim_{k^2\to0} (k_\perp| \Big(  \theta(k^-) e^{i\frac{ k^2_\perp}{2k^-} z^-} e^{-i\frac{P^2_\perp(z^-) }{2k^-} z^-} \Big|^\infty  + \theta(-k^-) e^{i\frac{ k^2_\perp}{2k^-} z^-} e^{-i\frac{P^2_\perp(z^-) }{2k^-} z^-} \Big|^{-\infty} \Big)\nonumber\\
&&\times e^{i\frac{P^2_\perp(y^-)}{2k^-}y^-} |y_\perp) e^{ik^- y^+}
- e^{iky}\Big|_{k^2 = 0} + \lim_{k^2\to0} (k_\perp| \Big\{ \Big(  \theta(k^-) e^{i\frac{ k^2_\perp}{2k^-} z^-} e^{-i\frac{P^2_\perp(z^-) }{2k^-} z^-} \Big|^\infty i \int^{\infty}_{y^-} dz^-_1 \nonumber\\
&&+ \theta(-k^-) e^{i\frac{ k^2_\perp}{2k^-} z^-} e^{-i\frac{P^2_\perp(z^-) }{2k^-} z^-} \Big|^{-\infty} i \int^{-\infty}_{y^-} dz^-_1  \Big)
\Big(e^{i\frac{ P^2_\perp(z^-_1)}{2p^-}z^-_1}iD_- e^{-i\frac{ P^2_\perp(z^-_1)}{2p^-}z^-_1} - \frac{P^2_\perp(z^-_1)}{2p^-}\Big) + \dots\Big\}\nonumber\\
&&\times e^{i\frac{P^2_\perp(y^-)}{2k^-}y^-} |y_\perp) e^{ik^- y^+} 
- \lim_{k^2\to0} (k_\perp| \Big\{ \Big(  i\theta(k^-) \int^{\infty}_{y^-} dz^-_1 + i \theta(-k^-)
\int^{-\infty}_{y^-} dz^-_1 \Big)\nonumber\\
&&\times \Big( e^{i\frac{ k^2_\perp}{2k^-} z^-_1} iD_- e^{-i\frac{ P^2_\perp(z^-_1)}{2p^-}z^-_1}
- e^{i\frac{ k^2_\perp}{2k^-} z^-_1} e^{-i\frac{P^2_\perp(z^-_1) }{2k^-} z^-_1} \frac{P^2_\perp(z^-_1)}{2p^-}\Big) + \dots\Big\}  e^{i\frac{P^2_\perp(y^-)}{2k^-}y^-} |y_\perp) e^{ik^- y^+}\,.\nonumber
\end{eqnarray}

It is easy to see that the infinite number of terms in this result can be combined into the path-ordered exponents as
\begin{eqnarray}
&&\lim_{k^2\to0} \Big( -i \theta(k^-) \int^\infty_{y^-} dz^-  + i \theta(-k^-) \int^{y^-}_{-\infty} dz^-\Big) e^{i\frac{k^2_\perp}{2k^-}z^-} (k_\perp| \Big[ (i \frac{\partial}{\partial z^-}  - \frac{k^2_\perp}{2k^-}) e^{-i\frac{P^2_\perp(z^-)}{2k^-} z^-}\Big]\nonumber\\
&&\times \mathcal{S}(z^-, y^-)e^{i\frac{P^2_\perp(y^-)}{2k^-}y^-} |y_\perp) 
 e^{ik^- y^+}\label{eqApp-LSZint}\\
 &&= 
- e^{iky}\Big|_{k^2 = 0} + \lim_{k^2\to0} (k_\perp| \Big(  \theta(k^-) e^{i\frac{ k^2_\perp}{2k^-} z^-} e^{-i\frac{P^2_\perp(z^-) }{2k^-} z^-} \Big|^\infty \mathcal{S}(\infty, y^-)
+ \theta(-k^-) e^{i\frac{ k^2_\perp}{2k^-} z^-} e^{-i\frac{P^2_\perp(z^-) }{2k^-} z^-} \Big|^{-\infty}\nonumber\\
&&\times \mathcal{S}(-\infty, y^-) \Big) e^{i\frac{P^2_\perp(y^-)}{2k^-}y^-} |y_\perp) e^{ik^- y^+} - \lim_{k^2\to0} (k_\perp| \Big(  i\theta(k^-) \int^{\infty}_{y^-} dz^- 
+ i \theta(-k^-) \int^{-\infty}_{y^-} dz^- \Big)\nonumber\\
&&\Big( e^{i\frac{ k^2_\perp}{2k^-} z^-} iD_- e^{-i\frac{ P^2_\perp(z^-)}{2p^-}z^-}
- e^{i\frac{ k^2_\perp}{2k^-} z^-} e^{-i\frac{P^2_\perp(z^-) }{2k^-} z^-} \frac{P^2_\perp(z^-)}{2p^-}\Big) \mathcal{S}(z^-, y^-) e^{i\frac{P^2_\perp(y^-)}{2k^-}y^-} |y_\perp) e^{ik^- y^+}\,.
 \nonumber
\end{eqnarray}

Substituting Eq. (\ref{eqApp-LSZint}) into Eq. (\ref{eqApa:after-dif}), we finally obtain the following form of the scalar propagator
\begin{eqnarray}
&&\lim_{k^2\to0}k^2(k| \frac{1}{P^2 + i\epsilon}|y) = \lim_{k^2\to0} (k_\perp| \Big(  \theta(k^-) e^{i\frac{ k^2_\perp}{2k^-} z^-} e^{-i\frac{P^2_\perp(z^-) }{2k^-} z^-} \Big|^\infty \mathcal{S}(\infty, y^-)
 \nonumber\\
 &&+ \theta(-k^-) e^{i\frac{ k^2_\perp}{2k^-} z^-} e^{-i\frac{P^2_\perp(z^-) }{2k^-} z^-} \Big|^{-\infty} \mathcal{S}(-\infty, y^-) \Big) e^{i\frac{P^2_\perp(y^-)}{2k^-}y^-} |y_\perp) e^{ik^- y^+} \,.
\end{eqnarray}
This completes our derivation of Eq. (\ref{eq:LSZ-fin}).
\end{widetext}

\bibliographystyle{unsrt}
\bibliography{ref}

\end{document}